\documentclass[a4paper,11pt]{article}
\pdfoutput=1 

\usepackage{jcappub} 

\usepackage[T1]{fontenc} 

\usepackage{xcolor}

\usepackage{caption, subcaption}

\usepackage{graphicx}

\usepackage{amsmath}

\usepackage{footmisc}

\usepackage{booktabs,siunitx}

\usepackage[toc]{appendix}

\newcommand{\len}{\phi}
\newcommand{\bn}{\hat{\bf n}}

\newcommand{\cmb}{\Theta}

\newcommand{\lm}[1]{l_{#1}m_{#1}} 

\newcommand{\thrj}[6] { \begin{pmatrix} 
                       #1 & #2 & #3 \\
		       #4 & #5 & #6
                      \end{pmatrix} }

\title{\boldmath Primordial Power Spectrum from Lensed CMB Temperature Spectrum using Iterative Delensing.}

\author[a,b,1]{Rajorshi Sushovan Chandra,\note{Corresponding author.}}
\author[b,a]{and Tarun Souradeep}

\affiliation[a]{Inter-University Centre for Astronomy and Astrophysics, Post Bag 4, Ganeshkhind, \\Pune 411 007, India}
\affiliation[b]{Raman Research Institute, C. V. Raman Avenue, Sadashivanagar, \\Bengaluru , 560 080, India}

\abstract{We address a current caveat in the deconvolution of  the Primordial Power Spectrum (PPS), from observed Cosmic Microwave Background (CMB) temperature  anisotropy, in the presence of weak lensing of the CMB by the large scale structure (LSS) in the Universe. Richardson-Lucy (RL) deconvolution algorithm has been used in the context of reconstructing a free-form PPS, $P_R(k)$ from the observed lensed CMB temperature anisotropy power spectrum $\widetilde{C}_{\ell}^{TT}$. We propose and demonstrate that the RL algorithm works in the context of a non-linear convolution where the non-linear contribution is small, such as the effect of weak lensing of the $\widetilde{C}_{\ell}^{TT}$, for the deconvolution of the PPS from it. The Non-Linear Iterative Richardson-Lucy (NIRL) algorithm is successful at both convergence, as well as fidelity, in reconstructing features in some underlying PPS. This makes PPS reconstruction efforts more robust in accounting for the weak lensing effect in the CMB temperature observations. No prior assumptions on the PPS are involved during the iterative delensing process, and distinct improvement is noted over a power-law template based delensing approach used earlier, at the cost of moderately increased computational cost due to the NIRL reconstruction kernel. }

\begin{document}
\maketitle
\flushbottom

\section{Introduction}
\label{sec:intro}
There have been purported anomalies in recent cosmological results, some of which are statistically significant, such as the Hubble constant discrepancy from the $\Lambda$CDM standard model validation using Planck data \cite{Aghanim:2019ame,Aghanim:2018eyx,Aghanim:2018oex,Akrami:2018odb} versus local measurements \cite{Riess:2019cxk}, which now stands at a $4.4\sigma$ deviation \cite{Keeley:2019esp,Keeley:2020rmo}. Some of such anomalies are less clear as to if they are statistical artefacts, or actual observational outliers, such as the $A_L$ anomaly, where deviations from the theoretical smoothing of the weak lensing of CMB power spectra shows a 10\% excess, albeit at only $2\sigma$ deviation \cite{Aghanim:2018eyx,ACT:2020goa,Millea:2020iuw}. Ironically, the theoretical framework of weak lensing, under General Relativity, is highly quantified and well established by a plethora of independent and diverse observational data. Yet it is the weak lensing of CMB power spectra that still presents a primary challenge in the analysis of the relevant observables, both as an observable in its own right, as well as a contaminant. This has led to a burgeoning field of research which seeks to reconcile the strong validity of the evolutionary physics that lead to present day observables, with these possible observed anomalies, by seeking modifications to the early Universe observables, such as in the field of inflationary models, particularly the structure of the Primordial Power Spectrum (PPS) $P_R(k)$ \cite{Starobinsky:1982ee,1982PhRvL..49.1110G,Starobinsky:1992ts,Beutler:2019ojk}. There have been two classes of approaches in this regard, which are complementary to each other, a model based and a free form reconstruction approach. There have been model construction based approaches, such as \cite{Hazra:2014goa,Domenech:2019cyh}, both to explain low ${\ell}$ features as well as address possible high ${\ell}$ features such as lensing excess in $C_{\ell}^{TT}$. Recent work has focussed on analysing the the $H_0$ tension in context of a free-form $P_R(k)$ reconstruction using the improved Richardson-Lucy (IRL) deconvolution algorithm \cite{Hazra:2018opk,Keeley:2020rmo} as well as for general feature hunting. There are also other free-form and other reconstruction methods that have been developed\cite{Nicholson:2009zj,2010JCAP...01..016N,2014A&A...566A..77P,Paykari:2014zua,Lanusse:2014sra,Tegmark:2002cy,Meerburg:2013cla}. Results have shown promise within the bounds and assumptions of this method, and produce a relatively high frequency (with respect to the reconstruction bounds) oscillatory feature at high $k$ regions, while considering a prior assumption of the $H_0$, $\Omega_m$ and $\sigma_8$ evaluated from local Universe observations and remaining parameters from $\Lambda$CDM inference. It is interesting to note that such a $P_R(k)$ fit does solve the purported tensions while improving the precision of those parameters upon inference. Since a free-form fitting is likely to risk the possibility of over-fitting i.e., fitting of noise, this necessitates a level of precision over the reconstruction algorithm that is capable of discerning and filtering out such effects while being able to probe actual data. The effect of weak lensing acoustic peak smoothing of the $C_L^{TT}$ power spectrum has been been demonstrated and discussed in context of IRL based $P_R(k)$ reconstruction earlier \cite{Hazra:2014jwa,Hazra:2018opk}. Weak lensing when not considered, can easily produce artefacts at high $k$ in the $P_R(k)$ and is degenerate with any physical origin deviations from power law in $P_R(k)$ and has to be accounted for. It is a worthwhile exercise to consider the non linear effects of the PPS convolution under weak lensing when the PPS itself is not limited to a power law. In this work we present a formal expression of the weak lensing effect on the $C_{\ell}^{TT}$ power spectrum transport kernel and provide a modified version of the RL deconvolution algorithm that can account for this non linear convolution with the end goal of improving deconvolution accuracy. Such an exercise has tangible implications when validating the results of PPS based approaches to address anomalies like $H_0$ discordance and lensing excess, especially when applying on high precision CMB data from current and future missions.

The layout of the paper is as follows, We present the weak lensing correction kernel to the temperature anisotropy power spectrum in section \ref{sec:cltt_lens_harmonic}. Then we propose our modified NIRL algorithm for addressing such non-linear convolutions and provide the numerical algorithm for its operation in section \ref{sec:NIRLdeconv}. We then provide a brief overview and description of the transfer kernel as well as the numerical details involved in simulating the lensing correction kernel to be used in the NIRL estimator in section \ref{sec:cltt_kernelsim}. Finally we move on to the results where we perform a null test to determine convergence and a feature based reconstruction to determine estimator accuracy, in section \ref{sec:NIRL_ressimtest}. In the end we round it off with a discussion and possible future work goals in section \ref{sec:discuss}.

\section{Temperature Power Spectrum $C_{\ell}^{TT}$ Weak Lensing Correction}
\label{sec:cltt_lens_harmonic}

A key observation when deconvolving the $P_R(k)$ from the observed $C_{\ell}^{TT}$ is the effect of weak lensing on the $C_{\ell}^{TT}$ power spectrum and it's consequent effects on the recovered PPS $P_R(k)$. It is shown that when utilizing the full $C_{\ell}^{TT}$ data, a reconstruction of the PPS that does not account for weak lensing acoustic damping, produces a distinct spurious feature similar to a wave packet at the $k$ regions of $3\times10^{-2}$ to $2\times10^{-1}$. The current approach to dealing with this is to generate a lensing template using fiducial cosmology and pre-process the $C_{\ell}^{TT}$ data by subtracting said template before IRL deconvolution. While practical and efficient with respect to computing resources, the issue with this approach is that the template based delensing process inherently assumes a fiducial $C_L^{\phi\phi}$ weak lensing power spectrum. It has been demonstrated earlier that the weak lensing power spectrum $C_L^{\phi\phi}$ can be expressed as a convolution of the primordial power spectrum $P_R(k)$ with the transport kernel $G_L^{\phi\phi}(k)$ and the RL estimator has been successfully employed in deconvolution with relevant statistical analyses and optimization algorithms \cite{Chandra:2021ydm}. This naturally poses a circular problem of having to assume a form of $C_L^{\phi\phi}$, based on a Power Law $P_R(k)$, when we are trying to reconstruct the $P_R(k)$ itself. Hence, our attempt to reconstruct the $P_R(k)$ directly without having to make any assumptions in the secondary corrections, at the cost of some reasonable computation resources, provides a more rigorous basis for the reconstruction and produces an estimator robust to future increases in measurement precision of the observed power spectrum and future model deviations.

In order to proceed towards a clean delensed reconstruction of the $P_R(k)$ from the lensed $\widetilde{C}_{\ell}^{TT}$, we lay out the mathematical formalism for the lensing effect on the power spectrum in this section. Using the formalism developed in Hu et. al. \cite{Hu:2000ee}, we write the Taylor expansion of the lensed temperature anisotropy field expanded around the projected gradient order lensing potential. By taking terms upto the second order in $\phi_{LM}$, such that we can obtain a full expression of the lensed $\widetilde{C}_{\ell}^{TT}$ upto first order in ${C}_L^{\phi\phi}$, we obtain

\begin{equation}
\begin{split}
\widetilde{\Theta}_{{\ell}m} &= {\Theta}_{{\ell}m}     						
+ \sum_{\lm{1}} \sum_{\lm{2}} {\Theta}_{\lm{2}} \phi_{\lm{1}} I_{{\ell} l_1 l_2}^{m m_1 m_2}     						
+ {1 \over 2} \sum_{\substack{l_1m_1 \\ l_2m_2}} \sum_{l_3 m_3} \len_{l_1 m_1} \len_{l_3 m_3}^* \cmb_{l_2 m_2} J_{{\ell} l_1 l_2 l_3}^{m m_1 m_2 m_3} \text{,} \\
\end{split}
\label{eqn:delens_start1}
\end{equation}
where the terms $I_{{\ell} l_1 l_2}^{m m_1 m_2}$ and $J_{{\ell} l_1 l_2 l_3}^{m m_1 m_2 m_3}$ correspond to geometric shape factors which incorporate information about the spherical harmonic basis in terms of the spherical harmonic functions 

\begin{equation}
\begin{split}
I_{{\ell} l_1 l_2}^{m m_1 m_2} & = \int d\bn Y_{\ell m}^* 
( \nabla_i Y_{l_1 m_1} ) ( \nabla^i Y_{l_2 m_2} ) \\
J_{{\ell} l_1 l_2 l_3}^{m m_1 m_2 m_3} & = \int d\bn Y_{\ell m}^* 
( \nabla_i Y_{l_1 m_1} ) ( \nabla_j Y_{l_3 m_3} ) 
\nabla^i \nabla^j Y_{l_2 m_2}  
\end{split}
\end{equation}
Assuming the Gaussianity and statistical isotropy conditions for both the temperature anisotropy and weak lensing field hold, we obtain the 2-point correlation function 

\begin{equation}
\begin{split}
\widetilde{C}_{\ell}^{TT} &= {C}_{\ell}^{TT} + {C}_{\ell}^{TT} \sum_{l_1} {C}_{l_1}^{\phi\phi} S^{(b)}_{{\ell}l_1} + \sum_{l_1l_2} {C}_{l_1}^{\phi\phi} {C}_{l_2}^{TT} S^{(a)}_{{\ell}l_1l_2}
\text{.}
\end{split}
\end{equation}
Using spherical harmonic operations and identities and following the procedure developed in \cite{Hu:2000ee,Lewis:2006fu,2011PhRvD..83d3005H}, we can simplify the shape factor sums into the following relations 

\begin{equation}
\begin{split}
S^{(a)}_{{\ell}l_1l_2} &= \frac{1}{2{\ell}+1} (F_{{\ell}l_1l_2})^2 \\
S^{(b)}_{{\ell}l_1} &= -\frac{1}{2} {\ell}({\ell}+1) l_1(l_1+1) \frac{2l_1+1}{4\pi} \\
\widetilde{C}_{\ell}^{TT} &= {C}_{\ell}^{TT} + \Delta{C}_{\ell}^{(b)TT} + \Delta{C}_{\ell}^{(a)TT}  \text{,}
\end{split}
\label{eqn:lencltt_deriv}
\end{equation}
where $F_{l_1l_2l_3}$ is given by.

\begin{equation}
\begin{split}
F_{l_1l_2l_3} = \frac{1}{2} [l_2(l_2+1)+l_3(l_3+1)-l_1(l_1+1)]
\sqrt{\frac{(2l_1+1)(2l_2+1)(2l_3+1)}{4\pi}} 
\thrj{l_1}{l_2}{l_3}{0}{0}{0}  \text{.}
\end{split}
\label{Fll'Leven}
\end{equation}
This final result given in equation \ref{eqn:lencltt_deriv} corresponds to the now commonly known result of the damping of the peak and filling of the valleys of the $C_{\ell}^{TT}$ power spectrum, which expresses the effect of weak lensing on the temperature anisotropy. It can be seen that weak lensing becomes a significant effect, especially at high ${\ell}$, which is also where the observational precision is higher with respect to the cosmic variance limit. Therefore it becomes imperative to account for this effect when performing any kind of parameter or free from estimation from the temperature power spectrum. In this paper we present a novel methodology that addresses the weak lensing contribution in the context of a free form PPS reconstruction using the Richardson-Lucy deconvolution algorithm.

\section{Method: Non-Linear Richardson-Lucy Deconvolution}
\label{sec:NIRLdeconv}

A broad body of work has been performed on the Richardson-Lucy (RL) deconvolution in existing literature \cite{1974AJ.....79..745L,1972JOSA...62...55R,Shafieloo:2003gf,Hazra:2014jwa, Shafieloo:2006hs, Shafieloo:2007tk, Nicholson:2009pi,2011NJPh...13j3024S,Hazra:2013xva, 2010JCAP...04..010H, Shaikh:2016hrf,Hazra:2017joc,Debono:2020emh}. We will follow the notation, formalism and groundwork laid out in \cite{Chandra:2021ydm} in this work. We recall that the MRL algorithm, which is the RL algorithm modified to incorporate observed covariance matrix information, based upon the condition of the positive definite nature of the power spectra and radiative transport kernel \cite{1993MNRAS.265..145B,1994MNRAS.267..323B}, as well as the linear convolution functional form, is given by 

\begin{equation}
C_{\ell}^{XX} = \sum_{k} P_k G_{{\ell}k}^{XX} \text{.}
\end{equation}
In our formulation of the lensing effect on $C_{\ell}^{TT}$ we can write the two power spectra of the CMB temperature and Weak Lensing field as 

\begin{equation}
\begin{split}
& {C}_{\ell}^{TT} = \sum_{k} P_{k} G_{{\ell}k}^{TT} \\
& {C}_L^{\phi\phi} = \sum_{k_1} P_{k_1} G_{Lk_1}^{\phi\phi} \text{.}
\end{split}
\end{equation}
Substituting these expressions into equation \ref{eqn:lencltt_deriv} we can express the lensed temperature anisotropy power spectrum as 

\begin{equation}
\begin{split}
\tilde{C}_{\ell}^{TT} &= \sum_{k} P_{k} \tilde{G}_{{\ell}k}^{TT} \\
& = \sum_{k} P_{k} 
\bigg[ 
G_{{\ell}k}^{TT} \bigg(1 + \sum_{k_1} P_{k_1} \sum_{l_1} G_{l_1k_1}^{\phi\phi} S^{(b)}_{{\ell}l_1} \bigg)  + \bigg( \sum_{k_1} P_{k_1} \sum_{l_1l_2} G_{l_1k_1}^{\phi\phi} G_{l_2k}^{TT} S^{(a)}_{{\ell}l_1l_2} \bigg) 
\bigg] \text{.} \\
\end{split}
\end{equation}
In this form the lensed temperature spectrum becomes a non-linear function of the PPS, $P_{k}$. The challenge is that the Richardson-Lucy algorithm is in principle designed to work on a linear convolution operation. At this juncture, we propose a way out; owing to the fact that the lensing effect is a secondary higher order correction and an order of magnitude weaker than the primary power spectrum at high $\ell$, one can approximate the higher order lensing contribution from the $P_{k}$ using the previous iteration estimate when being reconstructed by the Richardson-Lucy algorithm. There are two ways to implement this; we can either add the lensing correction to the probability/convolution kernel after each iteration, or we can delens the data using the template constructed from the previous iteration $P_{k}$. We prefer not to modify the data as much as possible and consider the kernel update as being a numerically 'cleaner' way to analyse the data. For observational data the template approach could result in negative value of $C_{\ell}^{TT}$ at high ${\ell}$'s due to realisation noise and so on, which add a layer of needless complexity. However in principle, both methods should yield the same result in the context of a power law form PPS, when applied on ideal simulated data. Given the original MRL algorithm used till date, in equation \ref{eq:IRL_eqns}

\begin{equation}
\begin{split}
P_{k}^{(i+1)} &= P_{k}^{(i)} \bigg{[} 1 + 
\sum_{\ell} \tilde{G}_{\ell k}^{TT} \bigg{(} \frac{\hat{C}_{\ell}^{TT}}{C_{\ell}^{TT(i)}} - 1 \bigg{)}
\text{tanh}^2 \big{(} 
[\hat{C}_{\ell}^{TT} - {C}_{\ell}^{TT(i)}]
\Sigma^{-1}
[\hat{C}_{\ell}^{TT} - {C}_{\ell}^{TT(i)}]^T 
\big{)} \bigg{]} \\
&= P_{k}^{(i)} \bigg{[} 1 + 
\sum_{\ell} \tilde{G}_{\ell k}^{TT} \bigg{(} \frac{\hat{C}_{\ell}^{TT}}{C_{\ell}^{TT(i)}} - 1 \bigg{)}
\text{tanh}^2 \bigg{(} \frac{\hat{C}_{\ell}^{TT}-C_{\ell}^{TT(i)}}{\hat{\sigma}_{\ell}} \bigg{)}^2 \bigg{]} \\
\tilde{G}_{\ell k}^{TT} &\text{ : Discretized Kernel normalised over ${\ell}$ } \\
\hat{C}_{\ell}^{TT} &\text{ : Data $C_{\ell}^{TT}$} \\
\Sigma^{-1} &: \text{Error covariance matric of the data}\ \hat{C}_{\ell}^{TT} \text{, diagonals given by}\ \hat{\sigma}_{\ell}  \\
C_{\ell}^{TT(i)} &=  \sum_{k} G_{\ell k}^{TT} P^{(i)}_k \text{.} \\
\end{split}
\label{eq:IRL_eqns}
\end{equation}
The transfer kernel ${\tilde{G}}_{\ell k}^{TT (i)}$ is corrected at each RL iteration $i$ to include the lensing contribution $\Delta G_{\ell k}^{TT(i)}$ calculated from the previous iterated estimate $P_{k}^{(i)}$. The estimator is expected to iteratively converge to the actual solution as before, if secondary corrections are accounted for by this method. This correction is expressed in the following equation, where the lensed kernel (iterated) is given by, 

\begin{equation}
\begin{split}
{\tilde{G}}_{\ell k}^{TT (i)} &= 
G_{\ell k}^{TT} \bigg[1 + \sum_{k_1} P_{k_1}^{(i)} \sum_{l_1} G_{l_1k_1}^{\phi\phi} S^{(b)}_{\ell l_1} \bigg]  + 
\bigg[ \sum_{k_1} P_{k_1}^{(i)} \sum_{l_1l_2} G_{l_1k_1}^{\phi\phi} G_{l_2k}^{TT} S^{(a)}_{\ell l_1l_2} \bigg] \\
&= G_{\ell k}^{TT} + \Delta G_{\ell k}^{TT(i)} \text{.}
\end{split}
\label{eqn:numer_lenscorr}
\end{equation}
In addition we forgo the regularising $\text{tanh}^2$ error accounting term for this project, since the goal is to validate the non-linear iterative reconstruction modification. It is straightforward to extend the error accounting for the NIRL algorithm for analysis on noise incorporated data. The final NIRL algorithm is given by

\begin{equation}
\begin{split}
P_{k}^{(i+1)} &= P_{k}^{(i)} \bigg{[} 1 + 
\sum_{\ell} \hat{\tilde{G}}_{\ell k}^{TT (i)} \bigg{(} \frac{\hat{C}_{\ell}^{TT}}{C_{\ell}^{TT(i)}} - 1 \bigg{)} \bigg{]} \\
\hat{\tilde{G}}_{\ell k}^{TT (i)} &= G_{\ell k}^{TT} + \Delta G_{\ell k}^{TT}(P_{k}^{(i)}) \text{ : Normalized over $\ell$} \\
\hat{C}_{\ell}^{TT} &\text{ : Lensed $C_{\ell}^{TT}$ Data} \\
C_{\ell}^{TT(i)} &=  \sum_{k} \hat{\tilde{G}}_{\ell k}^{TT (i)} P^{(i)}_k \text{.} \\
\end{split}
\label{eq:IRL_eqns_nonlin_notanerr}
\end{equation}
Since we do not make any prior assumptions of the lensing correction being made, and let it vary with the estimator, we expect that any features in the PPS that deviate from the power law model used in the delensing template based approach should be recovered much better when corrected using the non-linear iterative approach postulated here. In addition, we note that this method is likely to be useful for any general application of the Richardson-Lucy deconvolution method that is applicable on such deconvolution problems and expect interesting results when applied to physical effects that are more likely to involved non-linear effects, such as late time structure formation power spectra, deblurring and image processing of multi stage blur (such as, multiple lens assemblies) and so on.
The rigorous mathematical validation of the broader scope of such a method, is deferred here as an interesting open question. As an example, a series expansion derivation may be able to show that such an estimator will converge to the true solution when the non-linear corrections can be treated as a perturbation around the primary convolution.

\section{Results Ia: Kernel Simulation}
\label{sec:cltt_kernelsim}

In this section we present a preliminary proof of concept simulated test of the non-linear iterative Richardson-Lucy (NIRL) deconvolution estimator. The two key goals are to study the convergence of the NIRL estimator itself and to compare its efficacy in reconstruction compared to the template based lensing cleaning approach in previous literate. We will focus on the ideal theoretical power spectra simulations and forgo the realisation based data with cosmic variance limited error bars, as well as the Monte-Carlo sampling based error covariance matrix analysis. This is due to computation time constraints as the iterative method to account for non-linear corrections increases the time per NIRL iteration and any reasonable number of sample or error accounting would make the time infeasible for the scope of this method's paper, which is primarily to establish the validity of the NIRL algorithm itself. In addition the higher granularity and fine features in the temperature radiative transport convolution kernel means it has a capacity for larger degrees of freedom for the free form $P_R(k)$ fit, which in turn requires a larger sample set of Monte-Carlo reconstruction realisations. All this adds to computation time. However it is not algorithmically challenging and we expect to perform the comprehensive analysis in future on actual observation data.

\subsection{Kernel Simulation and Lensing Correction}
\label{sub_sec:kernel_behav_lens}

\begin{figure}
\begin{subfigure}{1\textwidth}
\includegraphics[width=1.00\linewidth]{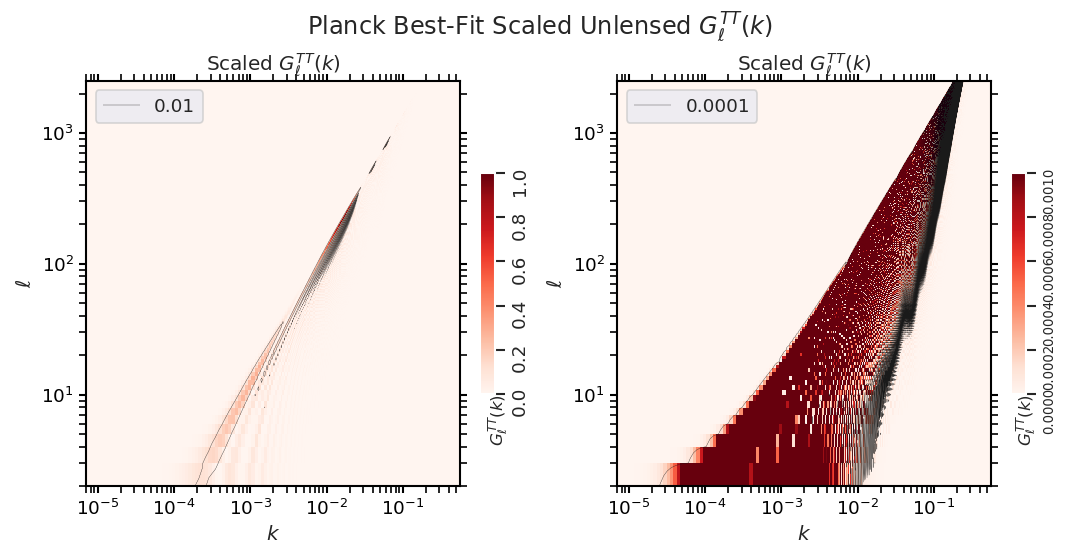}
\end{subfigure}
\caption{This plot shows a top-down view of the scaled transfer function given by $G_{\ell}^{TT}(k)/G_{\ell}^{TT}(k)_{max}$, scaled to a maximum power value of 1. The first plot shows the function with the complete power range 0 to 1 on the colorbar and a power level contour of 0.01. The second plot shows the function in a power range 0 upto 0.001 on the colorbar with a power level contour of 0.0001.}
\label{fig:glkTT_heatmap_dual}
\end{figure}

\begin{figure}
\begin{subfigure}{1\textwidth}
\includegraphics[width=1.00\linewidth]{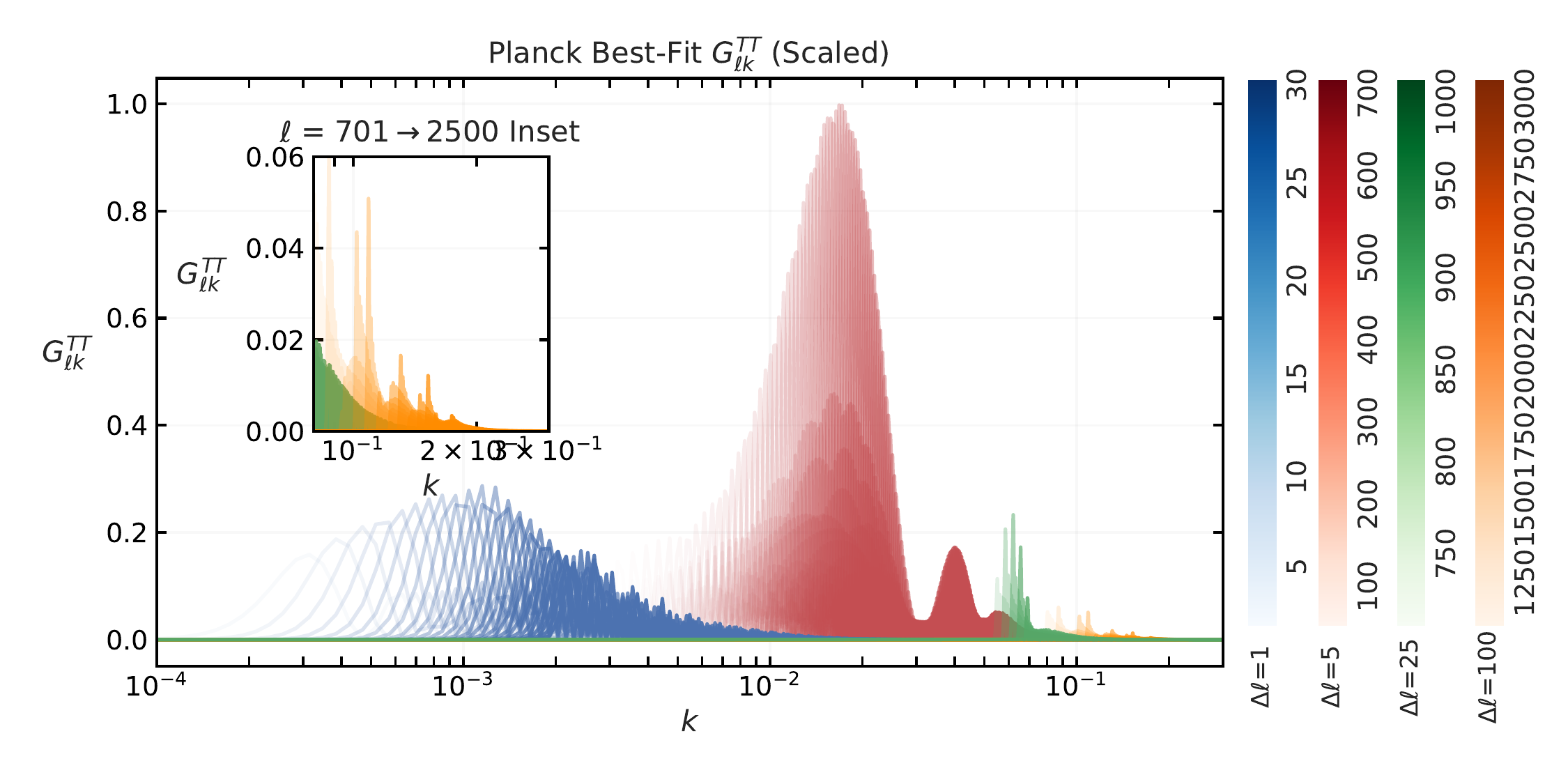}
\end{subfigure}
\caption{The plot shows the kernel with the numerical integration step size multiplied $G_{\ell k}^{TT} = G_{\ell}^{TT}(k)\Delta{k}$, projected onto the $k$ vs $G_{\ell}^{TT}(k)$ plane where $\ell$ is parametrized as a color gradient within blocks of $\ell$ with corresponding $\Delta \ell$ step sizes showing the plotted frequency of $\ell$ blocks. The $\ell$ blocks are roughly segmented by the relative amount of power they transfer.}
\label{fig:glkTT_side_paramd}
\end{figure}

In this section we provide an overview of the transfer kernel, the simulation procedure as well as the lensing correction added to it. We plot the top down heatmap of the kernel in figure \ref{fig:glkTT_heatmap_dual}. We use the Planck \cite{Aghanim:2018eyx} inferred cosmological parameters to generate the CMB temperature anisotropy transfer kernel, detailed in table \ref{table:phy_param}. The simulation parameters used are default except for the setting $\textbf{pars.set\_accuracy}(\textbf{lSampleBoost}=51)$. The numerical equivalent kernel expression $G_{\ell k}^{TT} = G_{\ell }^{TT}(k)\Delta{k}$ is shown in figure \ref{fig:glkTT_side_paramd}.

\begin{table}
\emph{\begin{center}
\begin{tabular}{ l r r l }
  \toprule
  Parameter & \multicolumn{1}{c}{Planck Best-Fit} \\
  \midrule
  $H_0$ & \num{ 67.36} \\
  $\Omega_b h^2$ & \num{ 0.02237 } \\
  $\Omega_c h^2$ & \num{ 0.1200 } \\
  $m_\nu$ & \num{0.06 } eV \\
  $\Omega_K$ & \num{ 0 } \\
  $\tau$ & \num{ 0.0544 } \\
  $A_s$ & \num{ 0.9 } \\
  $n_s$ & \num{ 0.9649 } \\
  $r$ & \num{ 0 } \\
  \bottomrule
\end{tabular}
\end{center}}
\caption{The physical parameters on which the CAMB calculated $\Lambda$CDM cosmology radiative transfer kernel is based on \cite{Aghanim:2018eyx}. }
\label{table:phy_param}
\end{table}

We present the temperature anisotropy power spectrum kernel $G_{Lk}^{TT}$ in figures \ref{fig:glkTT_heatmap_dual} and \ref{fig:glkTT_side_paramd}. The kernel is significantly sharp and the acoustic oscillation behaviour is clearly encoded in the transfer function, which allows the temperature anisotropy spectrum to be a discriminating observable for inferring cosmological parameters. Much work has already been done on this estimator in existing literature. We briefly point out that reconstruction support is broadly limited to the regions of $k \approx [10^{-4} \rightarrow 0.15]$. The initial guess sensitivity follows the support limitations, as shown in \cite{Shafieloo:2003gf,Chandra:2021ydm}.

Regarding the weak lensing correction as a function of the Primordial Power Spectrum, $P_R(k)$, we use the equation \ref{eq:IRL_eqns_nonlin_notanerr}. The key computation work involves calculating the shape factors $S^{(a)}_{\ell l_1l_2}$ and $S^{(b)}_{\ell l_1}$. Term (a), going over 3 indices is both time consuming and memory heavy, owing to the 3j symbols and indices respectively. The computation can be easily solved by a one-time calculation and storage, which costs around 30 GB of storage memory. This leads to an approximate read in and calculation time of 2 minutes for the lensing correction calculation per NIRL iteration. The lensing correction is clearly a function of the unlensed temperature and lensing potential kernel themselves and hence will be sensitive to the $\ell$ range accounted for.

\begin{table}
{\begin{center}
\begin{tabular}{ l r r l }
  \toprule
  $l_1,l_2$ Range For $S^{(a)}_{\ell l_1l_2}, S^{(b)}_{\ell l_1}$  & \multicolumn{1}{c}{Range} \\
  \midrule
  $l_1$ & (\num{2} $\rightarrow$ {2500}) \\
  $l_2$ & (\num{2} $\rightarrow$ {3000}) \\
  \bottomrule
\end{tabular}
\end{center}}
\caption{The table provides the weak lensing correction precision range for $l_1,l_2$ in $S^{(a)}_{\ell l_1l_2}, S^{(b)}_{\ell l_1}$.}
\label{table:wklns_lrange}
\end{table}
%
The $l_2$ range must be kept reasonably larger than the lensed $\widetilde{C}_{\ell}^{TT}$ range which is being probed, in order to properly account for the lensing correction power. We have considered the case when $l_2 = \ell + 500$. We will show quantify this choice later on by comparing it to CAMB precision. In the next section we move on to our simulation results.

\section{Results Ib: Simulated NIRL Reconstruction}
\label{sec:NIRL_ressimtest}

\subsection{Results i): Power Law Reconstruction with Delensing}
\label{sub_sec:pwlw_NIRL_delen}

In this section we first perform the tests where we evaluate the accuracy of our weak lensing correction mechanism by doing a null test to see if it performs convergence under our NIRL estimator algorithm. For a power law $P_R(k)$, we perform the following tests in table \ref{table:pwlw_tests}. The identifier $\textbf{(Power Law)}$ here denotes that all power spectra are constructed assuming a power law PPS.

\begin{table}
{\begin{center}
\begin{tabular}{ l r r l }
  \toprule
  Null Test Steps  \\
  \midrule
  1. Weak Lensing Kernel $\Delta C_{\ell}^{TT}$ vs CAMB $\Delta C_{\ell}^{TT}$ \\
  2. IRL Reconstruction from Unlensed CAMB $C_{\ell}^{TT} \textbf{(Power Law)}$ \\
  3. IRL Reconstruction from Lensed CAMB $\widetilde{C}_{\ell}^{TT} \textbf{(Power Law)}$ \\
  4. IRL Reconstruction from $\widetilde{C}_{\ell}^{TT} - \Delta C_{\ell}^{TT}\textbf{(Power Law)}$ \\
  5. NIRL Reconstruction from Lensed CAMB $\widetilde{C}_{\ell}^{TT} \textbf{(Power Law)}$ \\
  \bottomrule
\end{tabular}
\end{center}}
\caption{This table lists the null test steps undertaken to validate the NIRL estimator and its convergence.}
\label{table:pwlw_tests}
\end{table}

\begin{figure}
    \centering 
\begin{subfigure}{0.47\textwidth}
  \includegraphics[width=\linewidth]{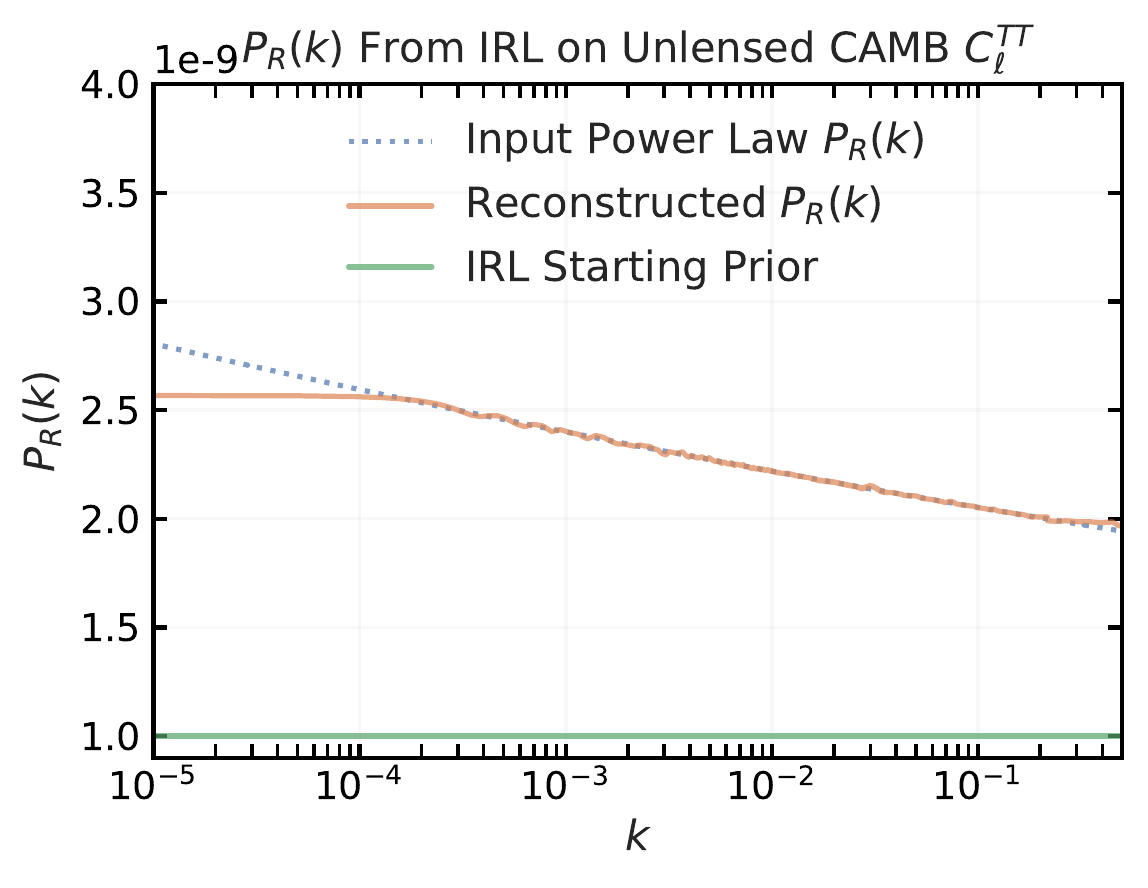}
  \caption{}
  \label{fig:pr4_kerneltest_reslt_1}
\end{subfigure}\hfil 
\begin{subfigure}{0.47\textwidth}
  \includegraphics[width=\linewidth]{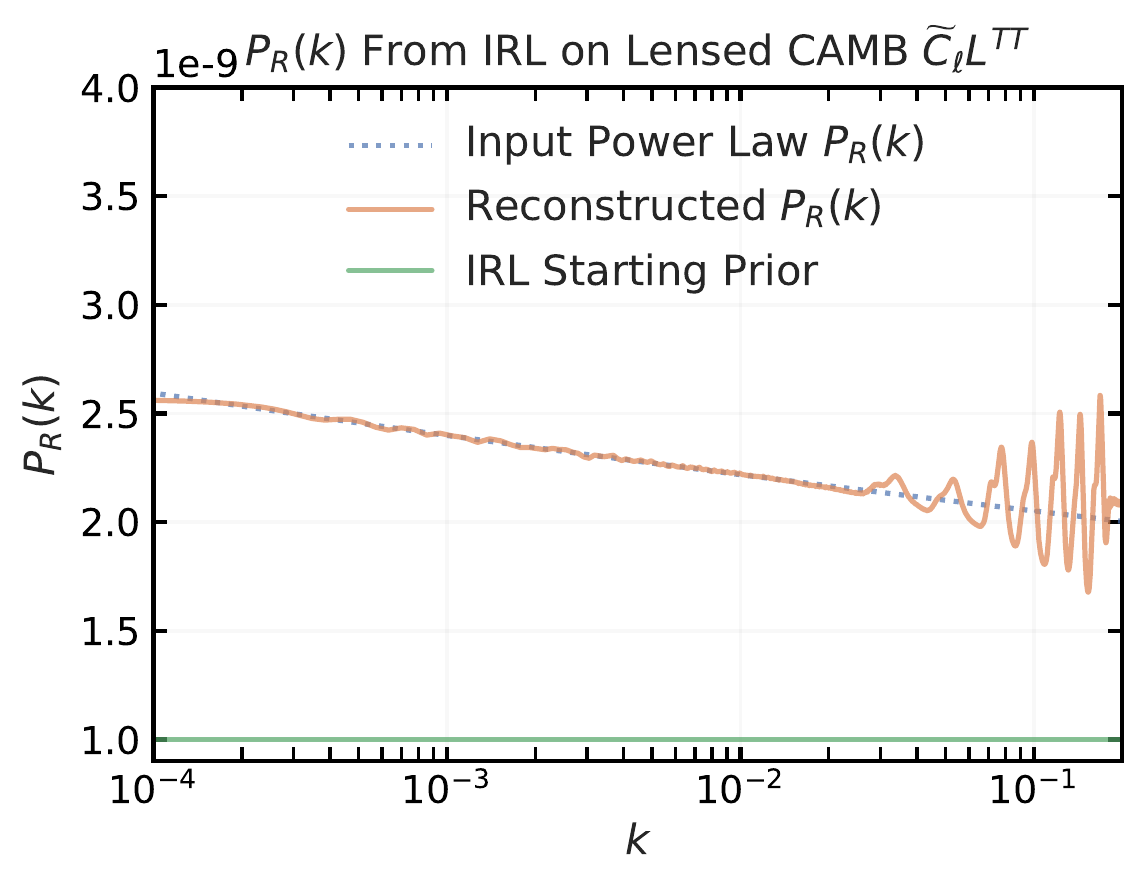}
  \caption{}
  \label{fig:pr4_kerneltest_reslt_2}
\end{subfigure}

\begin{subfigure}{0.47\textwidth}
  \includegraphics[width=\linewidth]{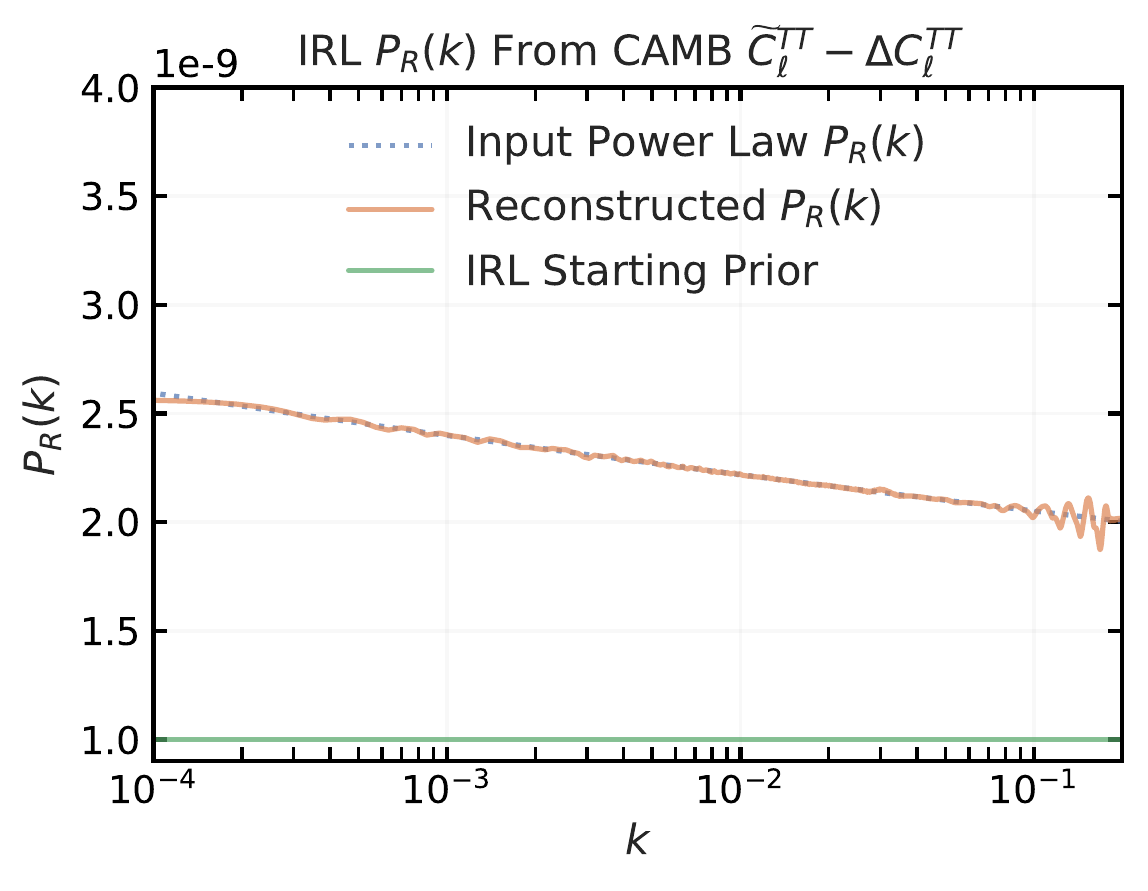}
  \caption{}
  \label{fig:pr4_kerneltest_reslt_3}
\end{subfigure}\hfil 
\begin{subfigure}{0.47\textwidth}
  \includegraphics[width=\linewidth]{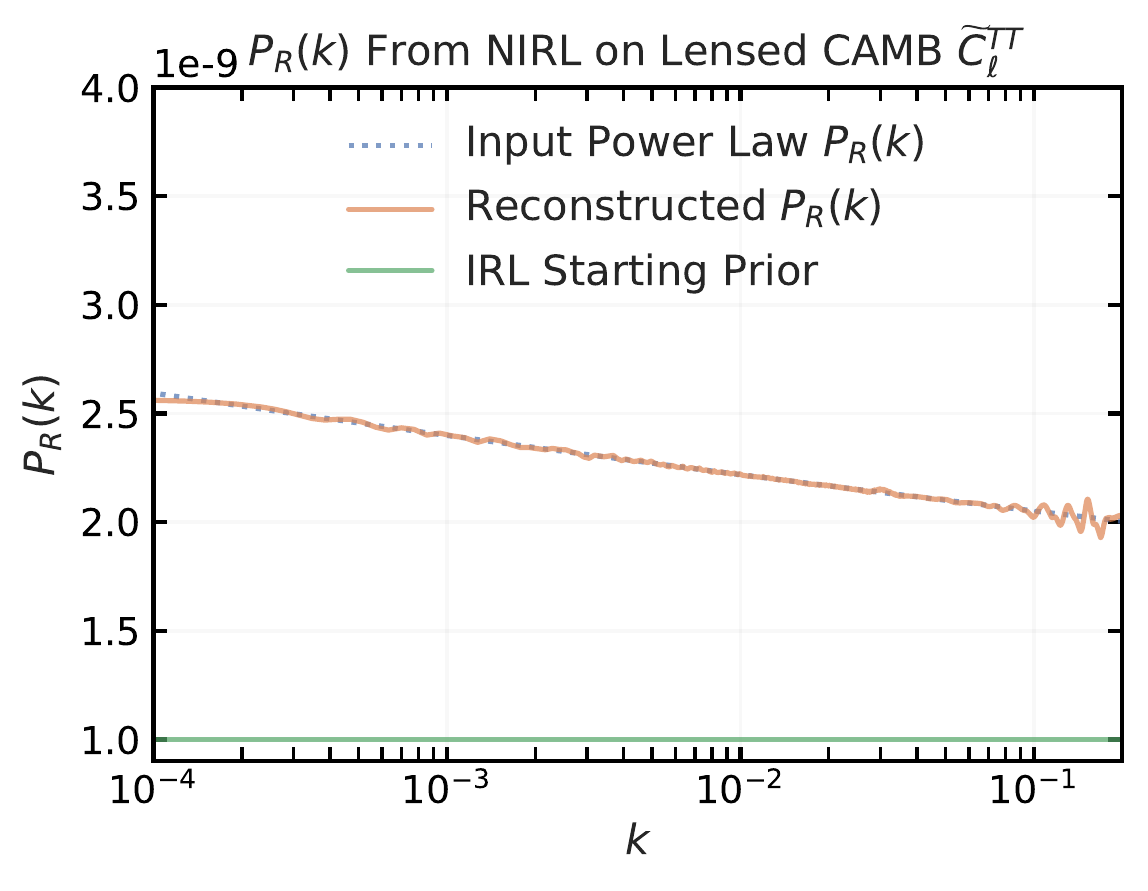}
  \caption{}
  \label{fig:pr4_kerneltest_reslt_4}
\end{subfigure}
\caption{All four plot panels show the input power law $P_R(k)$ in blue dashed lines and the reconstructed $P_R(k)$ in orange lines. The initial starting guess for the iterative reconstruction is given in green.}
\label{fig:pr4_kerneltest_reslt}
\end{figure}

The results are shown in figure \ref{fig:pr4_kerneltest_reslt}. All IRL/NIRL reconstruction have been carried out for fixed $100$ iterative loops in order to keep the comparisons consistent. 

\begin{figure}
\begin{subfigure}{1\textwidth}
\includegraphics[width=1.00\linewidth]{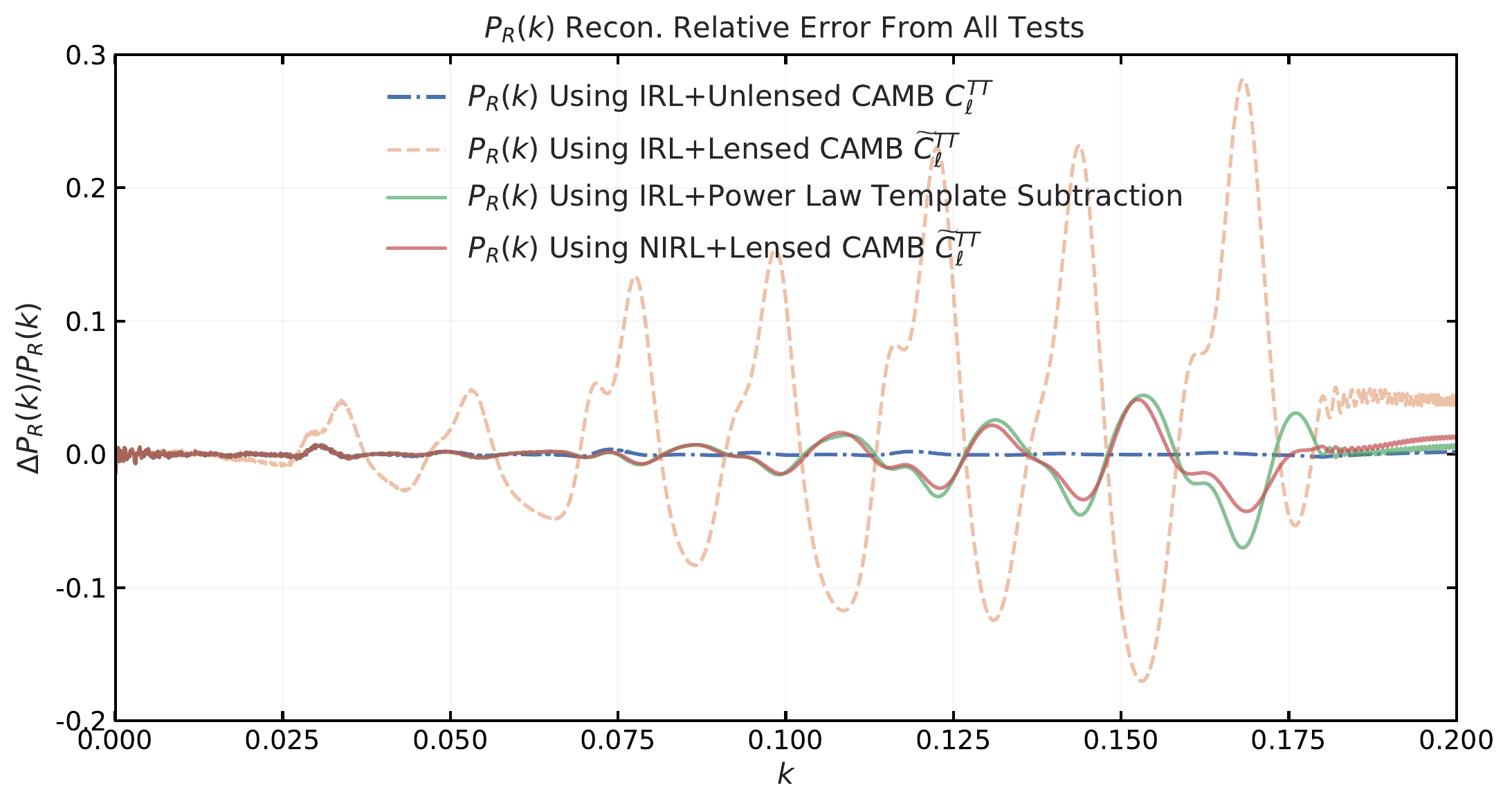}
\end{subfigure}
\caption{The figure displays the relative error of the 4 $P_R(k)$ reconstructions from figure \ref{fig:pr4_kerneltest_reslt}, with respect to \ref{fig:pr4_kerneltest_reslt_1}. }
\label{fig:pr4_krnltest_reslt_relerr}
\end{figure}

The first plot \ref{fig:pr4_kerneltest_reslt_1} shows the result of a run with the original IRL reconstructor on the unlensed ${C}_{\ell}^{TT}$ spectrum generate using CAMB. As expected the reconstruction visually recovers the input power law $P_R(k)$ accurately, with the support regions clearly within $k \in [10^{-4} \rightarrow 0.2]$. This is our null test comparison. (In addition, for future reconstructions, this support range will be used to depict all recovered $P_R(k)$ plots). In plot \ref{fig:pr4_kerneltest_reslt_2} we show the $P_R(k)$ recovered using the IRL method from the Lensed $\widetilde{C}_{\ell}^{TT}$ spectrum from CAMB. This clearly shows the generation of a spurious feature on the recovered $P_R(k)$, which is a consequence of the ignored lensing contribution. To address this, we can take two approaches; a Power Law based lensing template correction, subtracted from the lensed $\widetilde{C}_{\ell}^{TT} - \Delta C_{\ell}^{TT}$ or an NIRL approach for a non linear convolution directly on the lensed $\widetilde{C}_{\ell}^{TT}$. In figure \ref{fig:pr4_kerneltest_reslt_3} we show the template based approach. Since our simulated data plus lensing is based on the power law model only, the delensing is accurate and we recover the power law again. The small zigzag artefact is a consequence of the kernel simulation inaccuracy at high ${\ell}$ shown in figure \ref{fig:pr4_kerneltest_3}. This can be cleaned in future applications by a higher precision kernel calculation, though for the current work it was shown the discrepancy is below cosmic variance and hence can be tolerated. For the main test of the NIRL estimator, we now recover the $P_R(k)$ using the NIRL approach in equation \ref{eq:IRL_eqns_nonlin_notanerr} on the lensed $\widetilde{C}_{\ell}^{TT}$ from CAMB, without making any template based assumptions. The result is plotted in figure \ref{fig:pr4_kerneltest_reslt_4}. It is evident visually that the reconstruction is successful at recovering the power law $P_R(k)$ once again using our method. The relative difference error between the four reconstructed $P_R(k)$ versus the ideal CAMB reconstruction in \ref{fig:pr4_kerneltest_reslt_1} is given in figure \ref{fig:pr4_krnltest_reslt_relerr}. The plot clearly shows the change in relative accuracy as we move from $P_R(k)$ reconstructed from the naive lensed $\widetilde{C}_{\ell}^{TT}$ and template subtracted $\widetilde{C}_{\ell}^{TT} - \Delta C_{\ell}^{TT}$ using IRL, to the $\widetilde{C}_{\ell}^{TT}$ using NIRL. Since all template assumptions and truths are power law based, the best reconstruction is achieved by both template and NIRL methods and the NIRL estimator pass the null test of reaching reconstruction convergence even for a non-linear convolution effect. The summary of the test results are given in table \ref{table:pr4_kerneltest_reslt_summ}.

\begin{table}
{\begin{center}
\begin{tabular}{ l r r l }
  \toprule
  Data & \multicolumn{2}{c}{Algorithm}{Result} \\
  \midrule
  Unlensed ${C}_{\ell}^{TT}$ & IRL & Exact \\
  Lensed $\widetilde{C}_{\ell}^{TT}$ & IRL & Poor \\
  Lensed $\widetilde{C}_{\ell}^{TT}-\Delta C_{\ell}^{TT}$ & IRL & Good \\
  Lensed $\widetilde{C}_{\ell}^{TT}$ & NIRL & Good \\
  \bottomrule
\end{tabular}
\end{center}}
\caption{This table lists the last four Reconstruction tests from \ref{table:pwlw_tests} with their reconstruction quality.}
\label{table:pr4_kerneltest_reslt_summ}
\end{table}

\subsection{Results ii): Bump Feature Reconstruction with Delensing}
\label{sub_sec:feat_NIRL_delen}

In this section we pursue the validation process further. We have seen that the NIRL technique is quantitatively equivalent to the Template delensing approach in the case of a power law PPS, where both converge to the ideal reconstruction. This is to be expected as both the ideal and the template delensing approach are based on a power law model. The question of usefulness of NIRL arises when we consider a primordial power spectrum $P_R(k)$ which is not inherently a power law model. Any spectrum simulated on this assumption, unlensed ${C}_{\ell}^{TT}$ or lensed $\widetilde{C}_{\ell}^{TT}$, will have differences in power compared to a power law based spectrum. When it comes to the lensed $\widetilde{C}_{\ell}^{TT}$, the template weak lensing correction approach has an obvious disadvantage; it assumes a power law primordial power spectrum $P_R(k)$. As a result the lensing correction is likely to miss a complete estimate of the lensing correction. Since NIRL makes no prior assumption about the $P_R(k)$ at any step, it is important to ascertain if there is any difference between the Template versus NIRL reconstruction approach.

\begin{figure}
    \centering 
\begin{subfigure}{0.5\textwidth}
  \includegraphics[width=\linewidth]{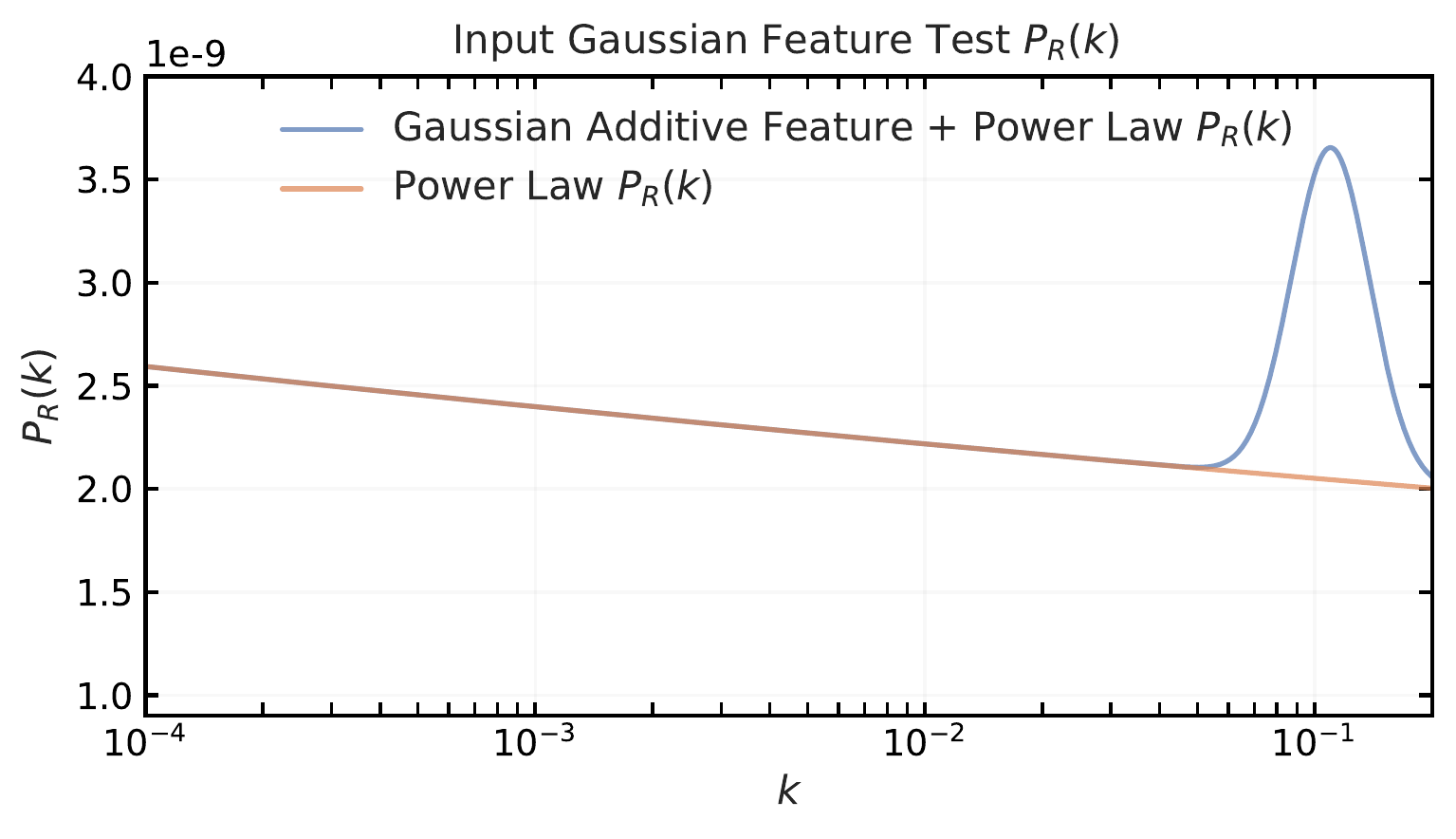}
  \caption{}
  \label{fig:pr4_feattest_1}
\end{subfigure}\hfil 
\begin{subfigure}{0.5\textwidth}
  \includegraphics[width=\linewidth]{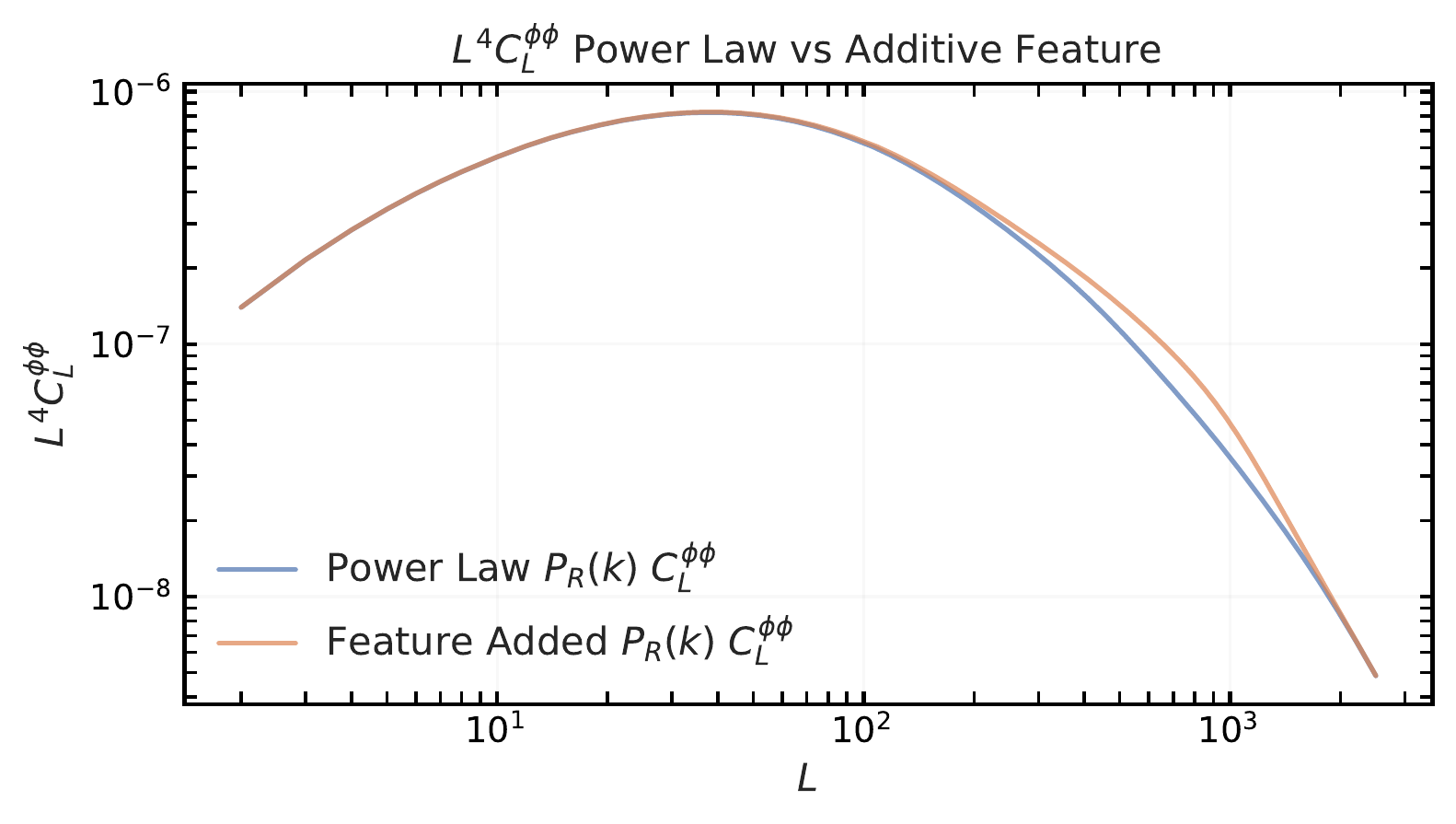}
  \caption{}
  \label{fig:pr4_feattest_2}
\end{subfigure}

\begin{subfigure}{0.5\textwidth}
  \includegraphics[width=\linewidth]{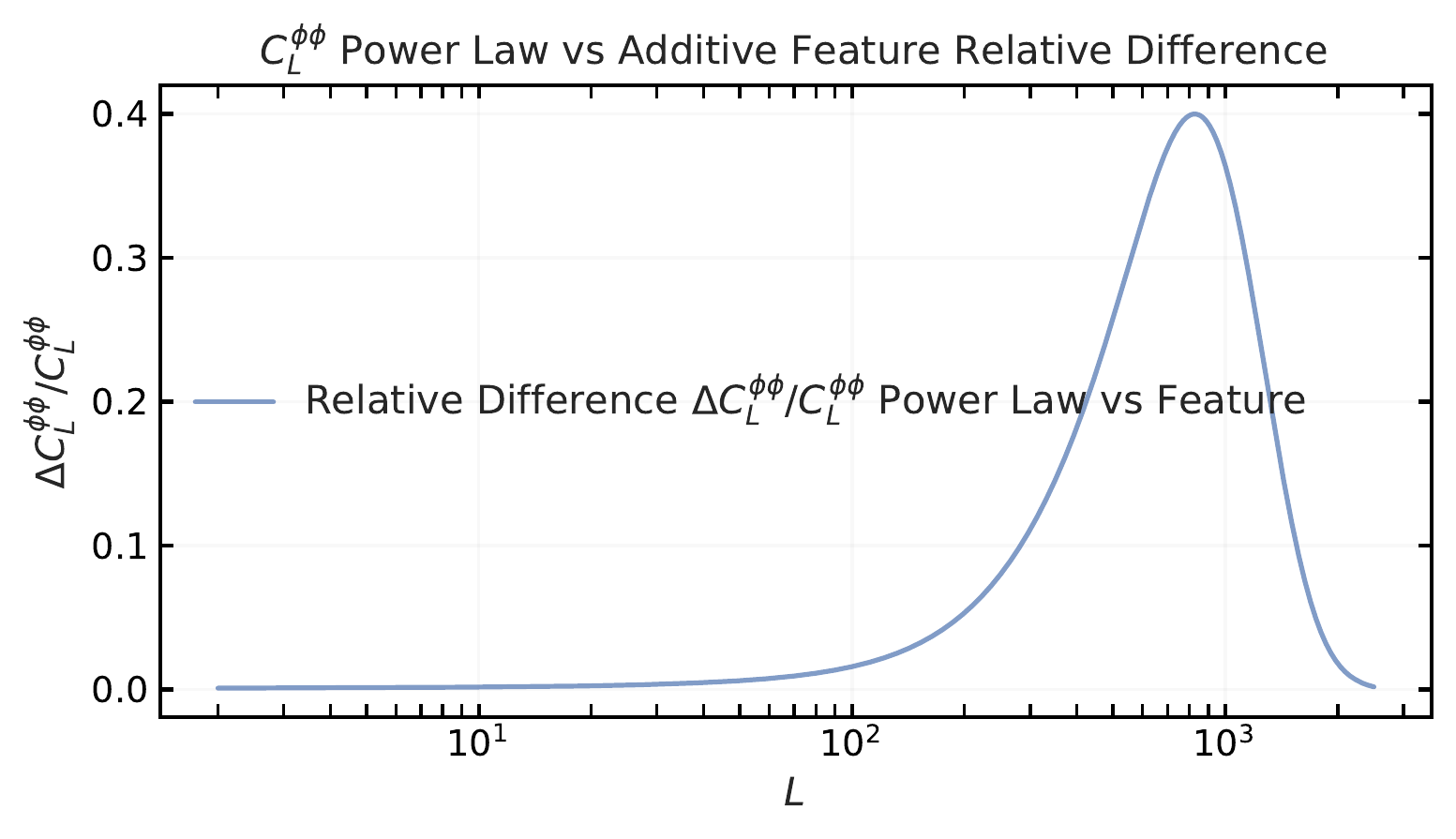}
  \caption{}
  \label{fig:pr4_feattest_3}
\end{subfigure}\hfil 
\begin{subfigure}{0.5\textwidth}
  \includegraphics[width=\linewidth]{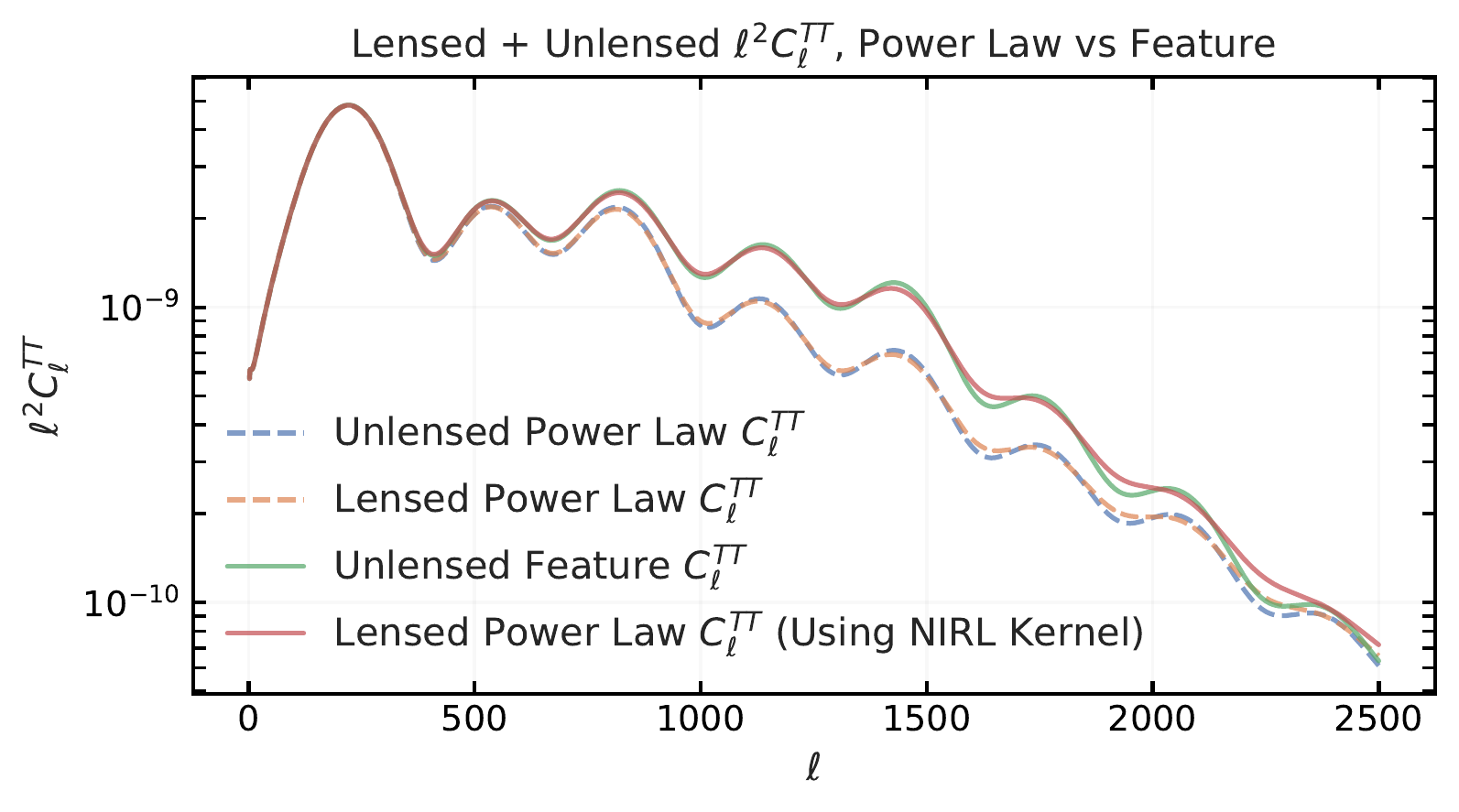}
  \caption{}
  \label{fig:pr4_feattest_4}
\end{subfigure}

\begin{subfigure}{0.5\textwidth}
  \includegraphics[width=\linewidth]{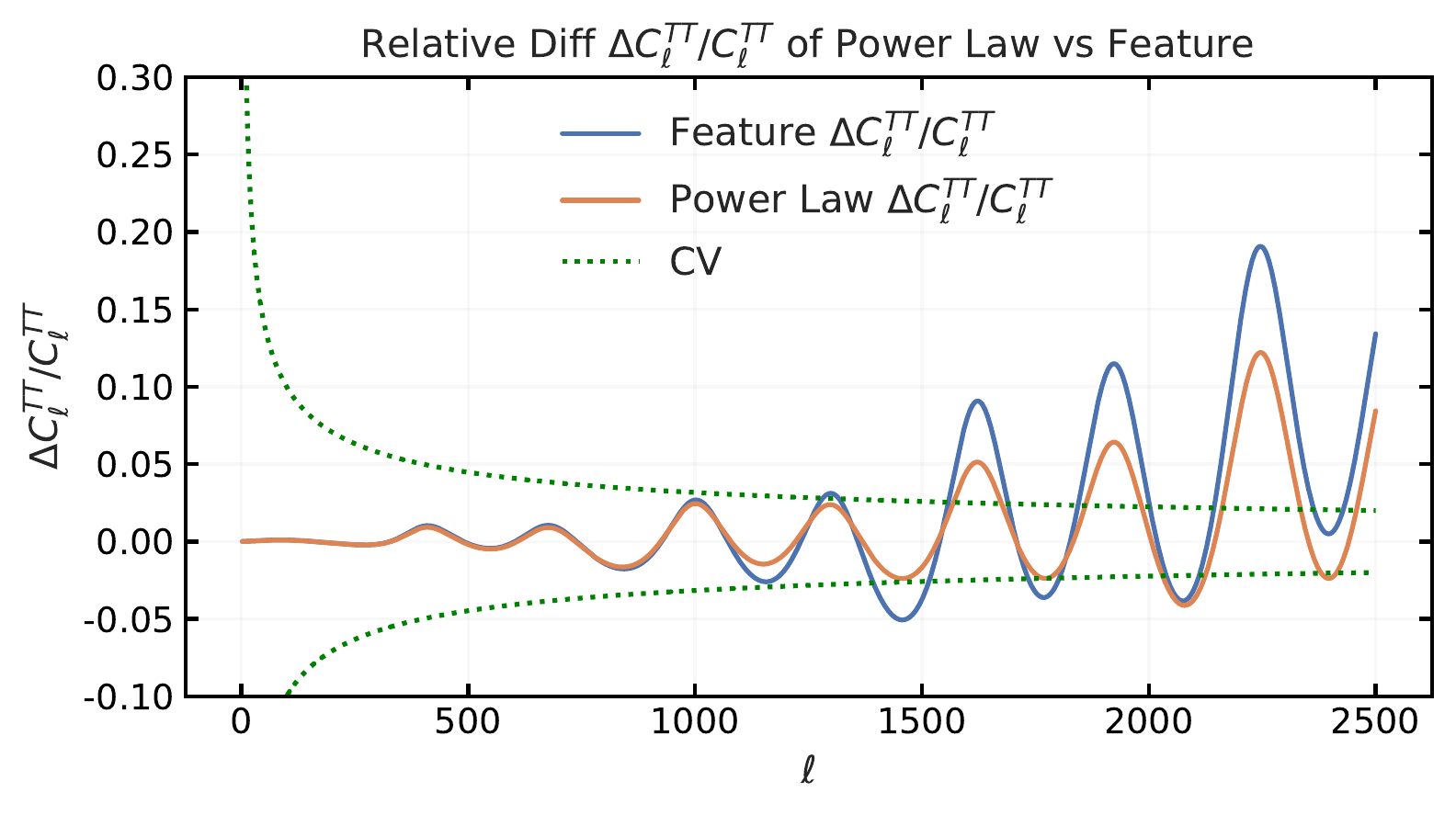}
  \caption{}
  \label{fig:pr4_feattest_5}
\end{subfigure}\hfil 
\begin{subfigure}{0.5\textwidth}
  \includegraphics[width=\linewidth]{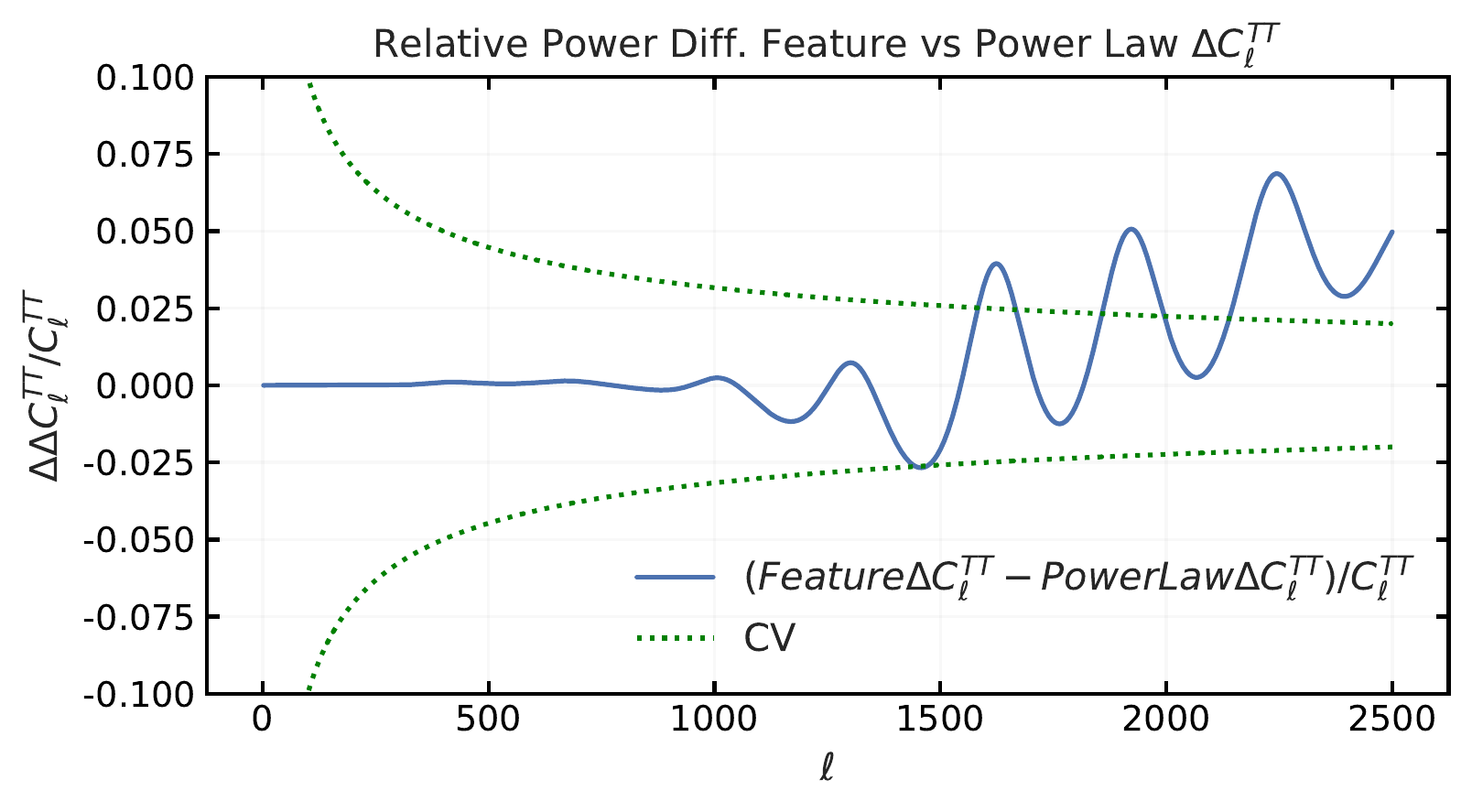}
  \caption{}
  \label{fig:pr4_feattest_6}
\end{subfigure}
\caption{These set of figures present the feature based simulated data we will use for NIRL validation. Figure \ref{fig:pr4_feattest_1} shows the feature based $P_R(k)$ in blue and power law reference in orange. Figure \ref{fig:pr4_feattest_2} shows the corresponding ${C}_L^{\phi\phi}$ with the feature based simulation in orange and power law simulation blue. Figure \ref{fig:pr4_feattest_3} shows the relative power difference due to the two ${C}_L^{\phi\phi}$. Figure \ref{fig:pr4_feattest_4} shows the lensed and unlensed ${C}_{\ell}^{TT}$ due to the feature (solid lines) versus the power law (dashed lines) based $P_R(k)$. Figure \ref{fig:pr4_feattest_5} shows the lensing contribution $\Delta{C}_{\ell}^{TT}$ due to the feature (blue line) and power law (orange line) $P_R(k)$. Figure \ref{fig:pr4_feattest_6} shows the relative power difference between the two $\Delta{C}_{\ell}^{TT}$ in blue, with respect to the cosmic variance limit in dashed green. }
\label{fig:pr4_feattest}
\end{figure}

We first present the simulated data and the underlying conditions used to generate it. The relevant plots are given in figure \ref{fig:pr4_feattest}.  We begin with a feature based $P_R(k)$, which consists of a power law with a Gaussian bump added at $k = 0.11$. This is depicted in \ref{fig:pr4_feattest_1}, with reference to the power law $P_R(k)$. In plot \ref{fig:pr4_feattest_2} and \ref{fig:pr4_feattest_3} the consequent ${C}_L^{\phi\phi}$ is depicted, from the feature versus power law simulation, as well as the relative difference between the two. As it visible the feature adds an excess power contribution around an $L\approx1000$. While the relative excess is large, at upto $40\%$, the net contribution to the lensing kernel in equation \ref{eqn:numer_lenscorr}, varies across ${\ell}$, with high ${\ell}$ receiving a strong coupling between the $l_1$ and $l_2$ due to the behaviour of the 3j symbol. The change in ${C}_{\ell}^{TT}$, both lensed and unlensed is depicted in \ref{fig:pr4_feattest_4}. In this plot it is clearly visible how there is an overall increase in power in the ${C}_{\ell}^{TT}$. But the more important factor is the weak lensing contribution, and the relative power difference between the lensed and unlensed ${C}_{\ell}^{TT}$, is given in \ref{fig:pr4_feattest_5}, with respect to the cosmic variance. It is clearly visible here that the feature causes a significant increase in the lensing contribution. The excess lensing contribution is also plotted in \ref{fig:pr4_feattest_6}, to show how the term in \ref{eqn:numer_lenscorr} changes when dealing with deviations from power law in the $P_R(k)$. This plot is key to appreciate the necessity for a high precision $P_R(k)$ estimator that makes no prior assumptions, as the power difference due to features can very well be above cosmic variance, as in the plot, and therefore observationally significant.

We now lay out the steps to be performed in order to quantify the results of the NIRL vs the template-based delensing approach, when considering a general underlying $P_R(k)$ origin for the temperature spectra, without any prior knowledge of it being a pure power law or otherwise. The key tests are listed in table \ref{table:feat_tests}.

\begin{table}
{\begin{center}
\begin{tabular}{ l r r l }
  \toprule
  Bump Feature Test Steps  \\
  \midrule
  1. IRL Reconstruction from Unlensed CAMB $C_{\ell}^{TT} \textbf{(Feature)}$ \\
  2. IRL Reconstruction from Lensed CAMB $\widetilde{C}_{\ell}^{TT} \textbf{(Feature)}$ \\
  3. IRL Reconstruction from $\widetilde{C}_{\ell}^{TT}\textbf{(Feature)} - \Delta C_{\ell}^{TT}\textbf{(Power Law)}$ \\
  4. NIRL Reconstruction from Lensed CAMB $\widetilde{C}_{\ell}^{TT} \textbf{(Feature)}$ \\
  \bottomrule
\end{tabular}
\end{center}}
\caption{This table lists the tests undertaken to validate the NIRL estimator and its convergence based on $P_R(k)$ incorporating a feature.}
\label{table:feat_tests}
\end{table}

\begin{figure}
    \centering 
\begin{subfigure}{0.47\textwidth}
  \includegraphics[width=\linewidth]{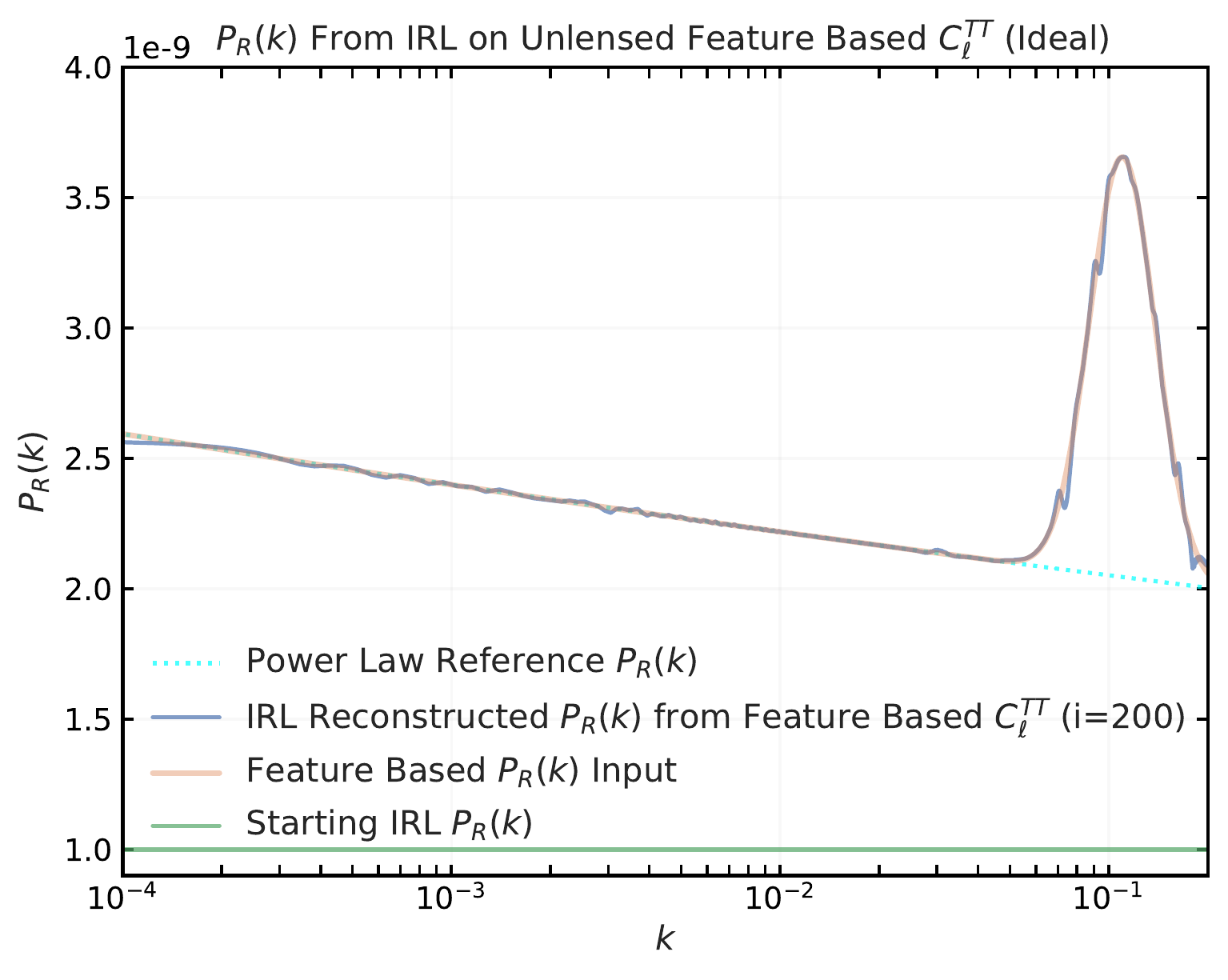}
  \caption{}
  \label{fig:pr4_feattest_reslt200_1}
\end{subfigure}\hfil 
\begin{subfigure}{0.47\textwidth}
  \includegraphics[width=\linewidth]{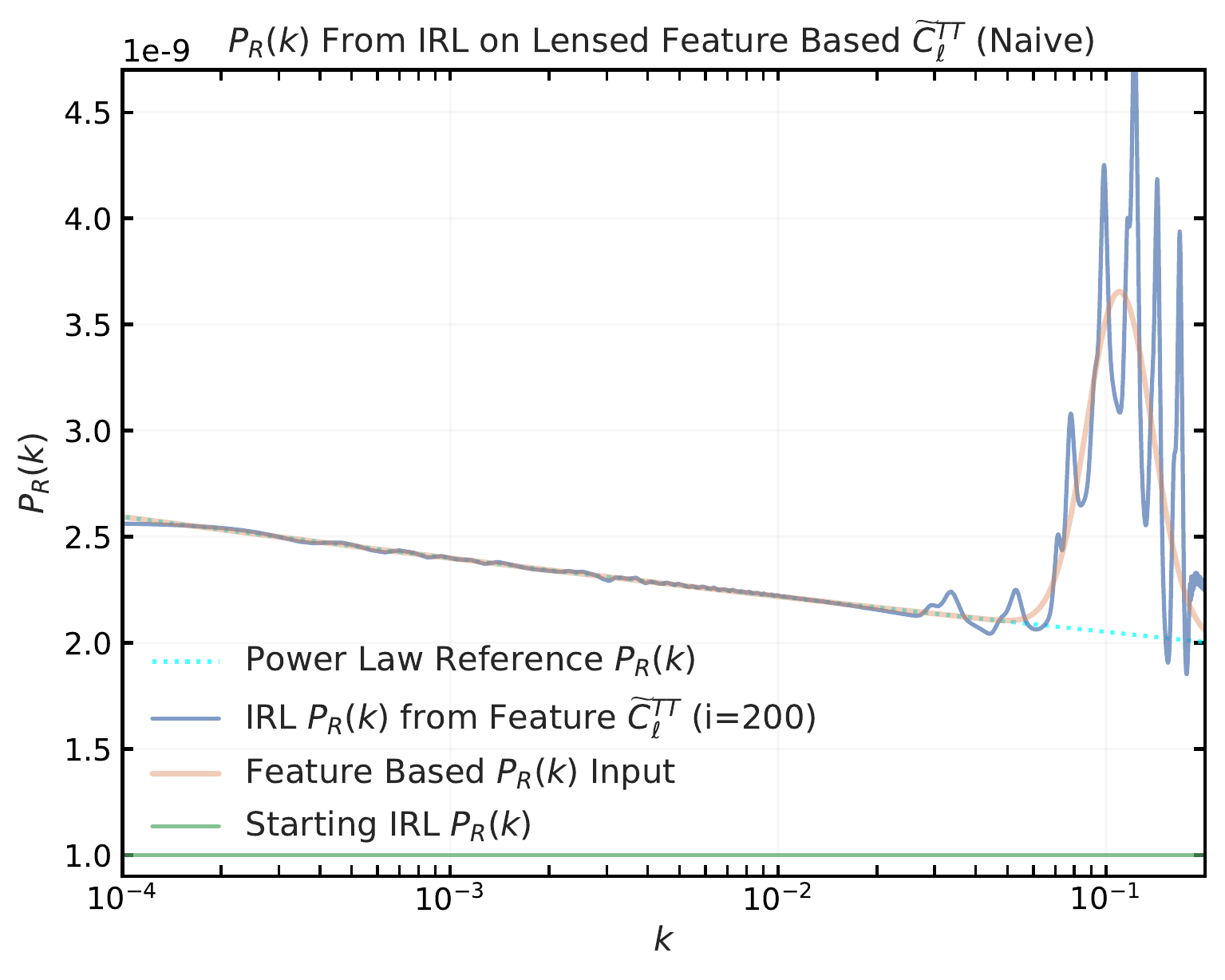}
  \caption{}
  \label{fig:pr4_feattest_reslt200_2}
\end{subfigure}

\begin{subfigure}{0.47\textwidth}
  \includegraphics[width=\linewidth]{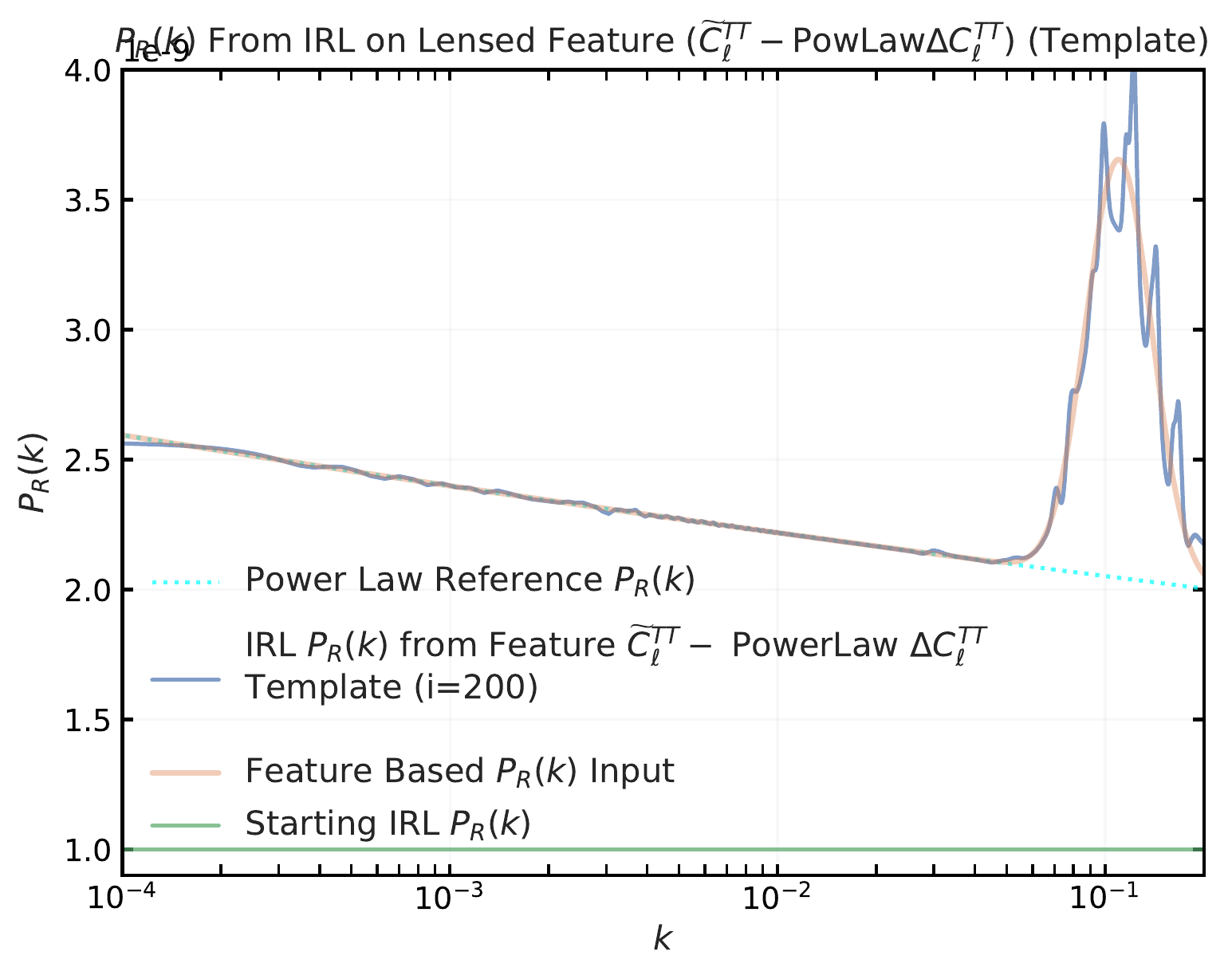}
  \caption{}
  \label{fig:pr4_feattest_reslt200_3}
\end{subfigure}\hfil 
\begin{subfigure}{0.47\textwidth}
  \includegraphics[width=\linewidth]{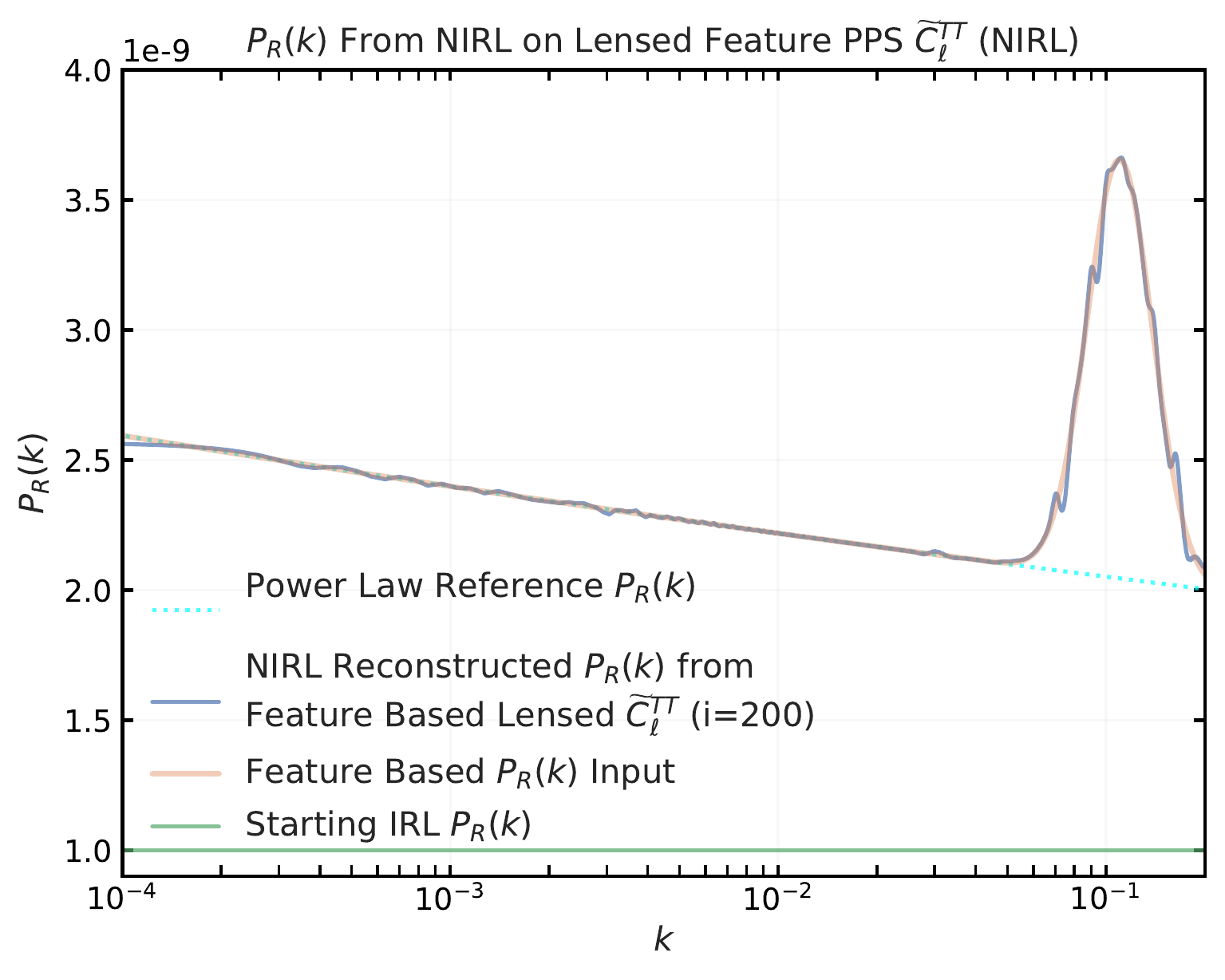}
  \caption{}
  \label{fig:pr4_feattest_reslt200_4}
\end{subfigure}
\caption{ The 4 figures show the reconstructed $P_R(k)$ from the feature based data, using 200 RL reconstruction iterations. The reference power law is given in cyan dotted lines. The reconstructed $P_R(k)$ are given in blue lines. The initial guess for the iterative estimator are given in green lines. The input feature $P_R(k)$ is given in orange lines.}
\label{fig:pr4_feattest_reslt200}
\end{figure}

\begin{figure}
    \centering 
\begin{subfigure}{0.47\textwidth}
  \includegraphics[width=\linewidth]{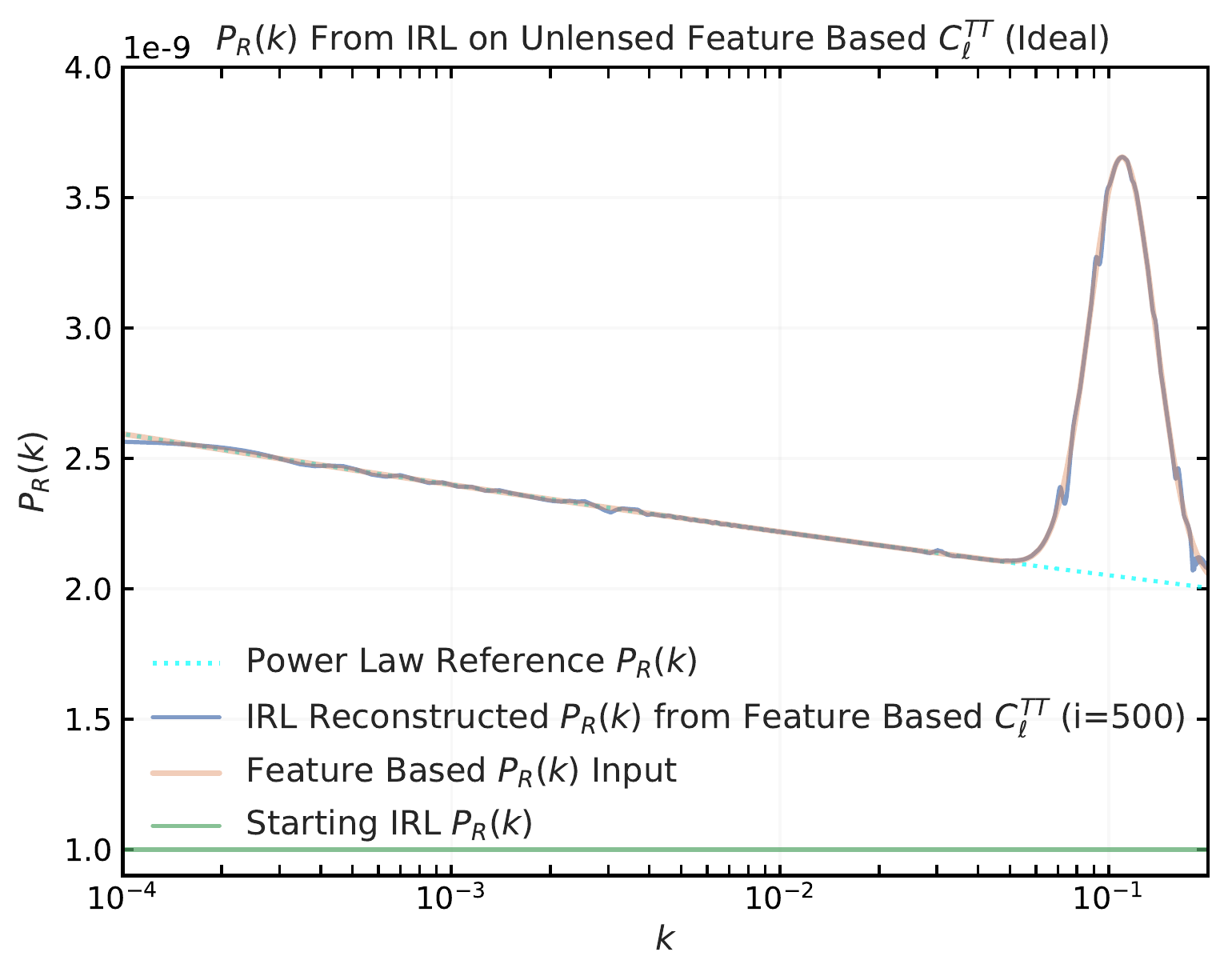}
  \caption{}
  \label{fig:pr4_feattest_reslt500_1}
\end{subfigure}\hfil 
\begin{subfigure}{0.47\textwidth}
  \includegraphics[width=\linewidth]{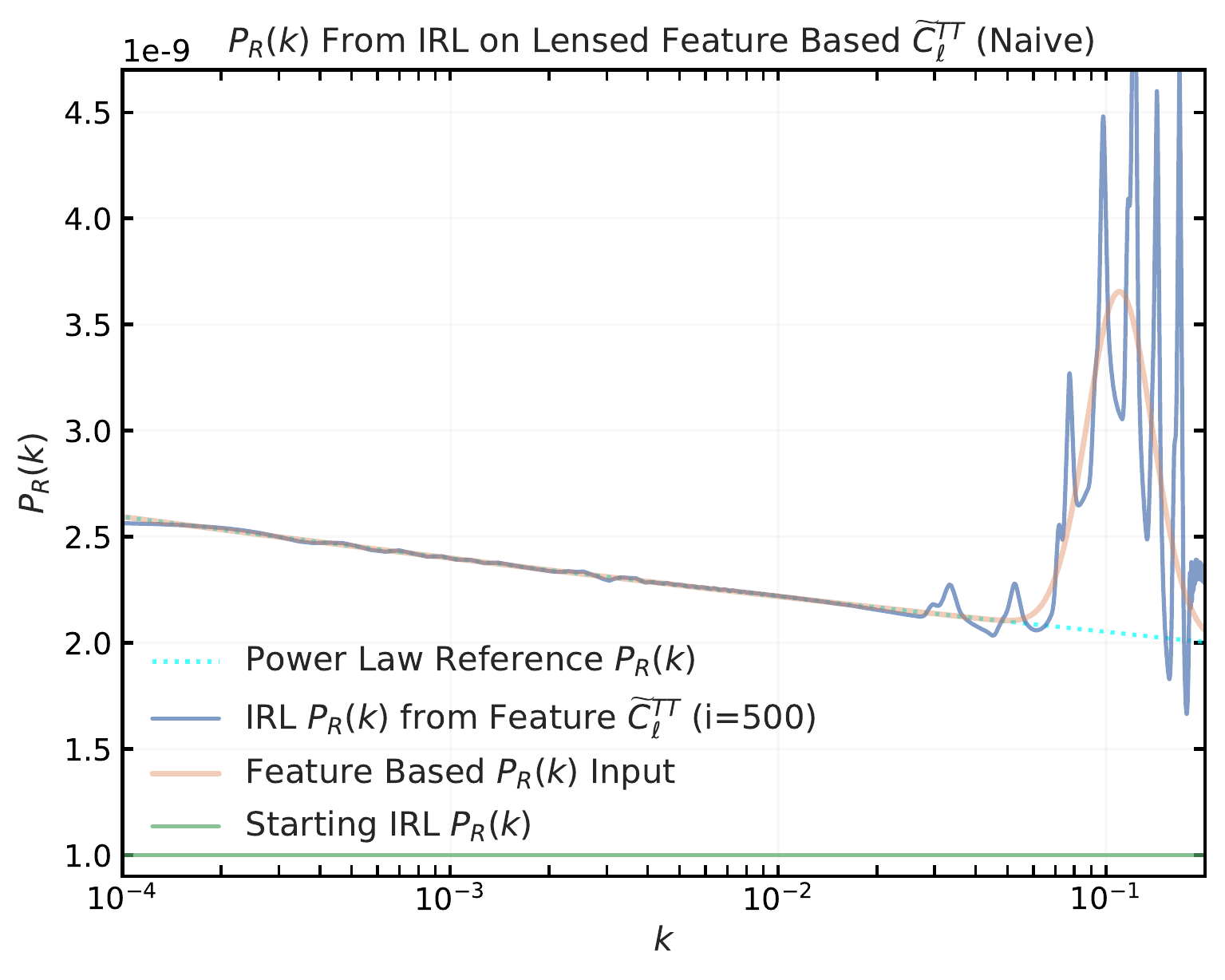}
  \caption{}
  \label{fig:pr4_feattest_reslt500_2}
\end{subfigure}

\begin{subfigure}{0.47\textwidth}
  \includegraphics[width=\linewidth]{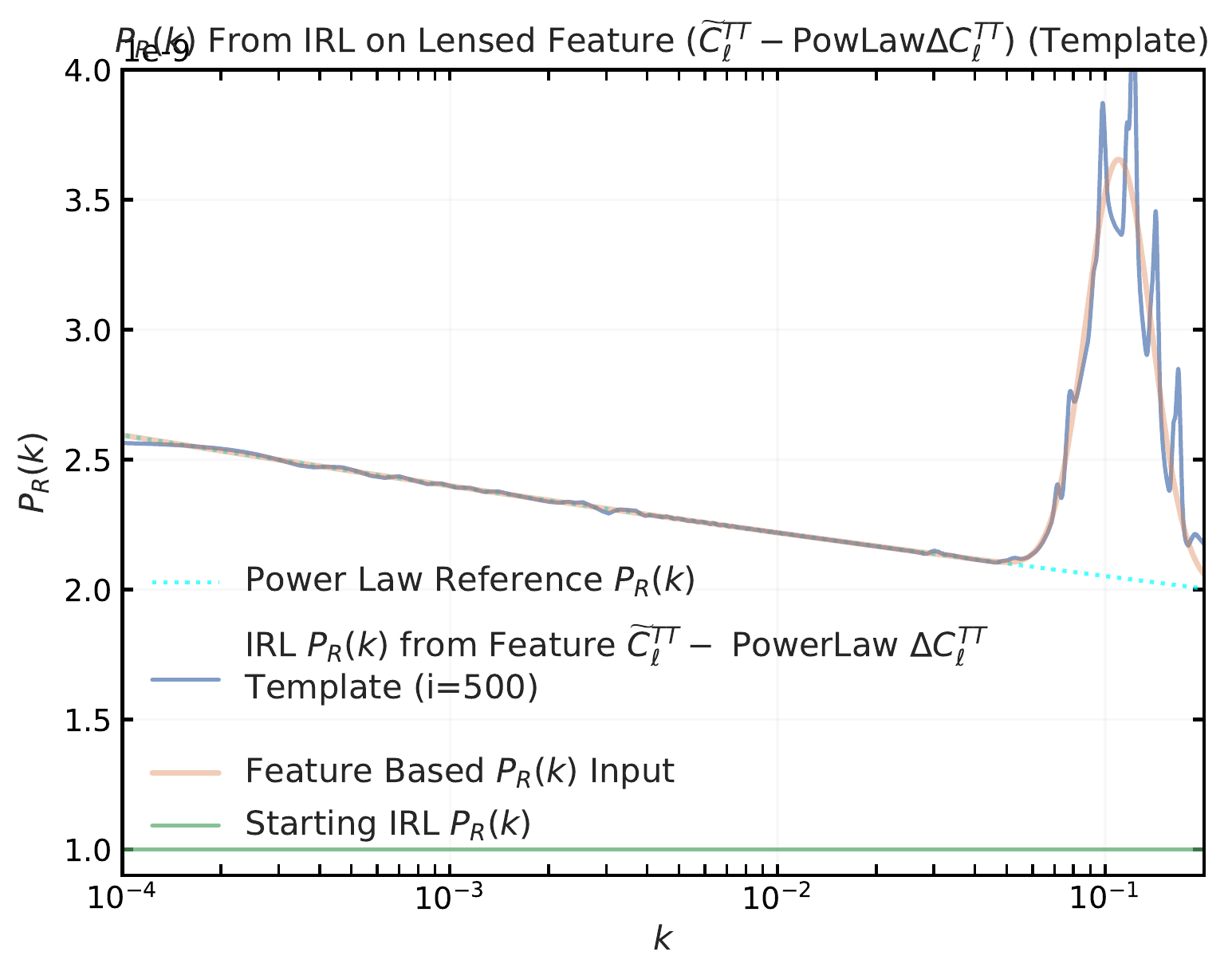}
  \caption{}
  \label{fig:pr4_feattest_reslt500_3}
\end{subfigure}\hfil 
\begin{subfigure}{0.47\textwidth}
  \includegraphics[width=\linewidth]{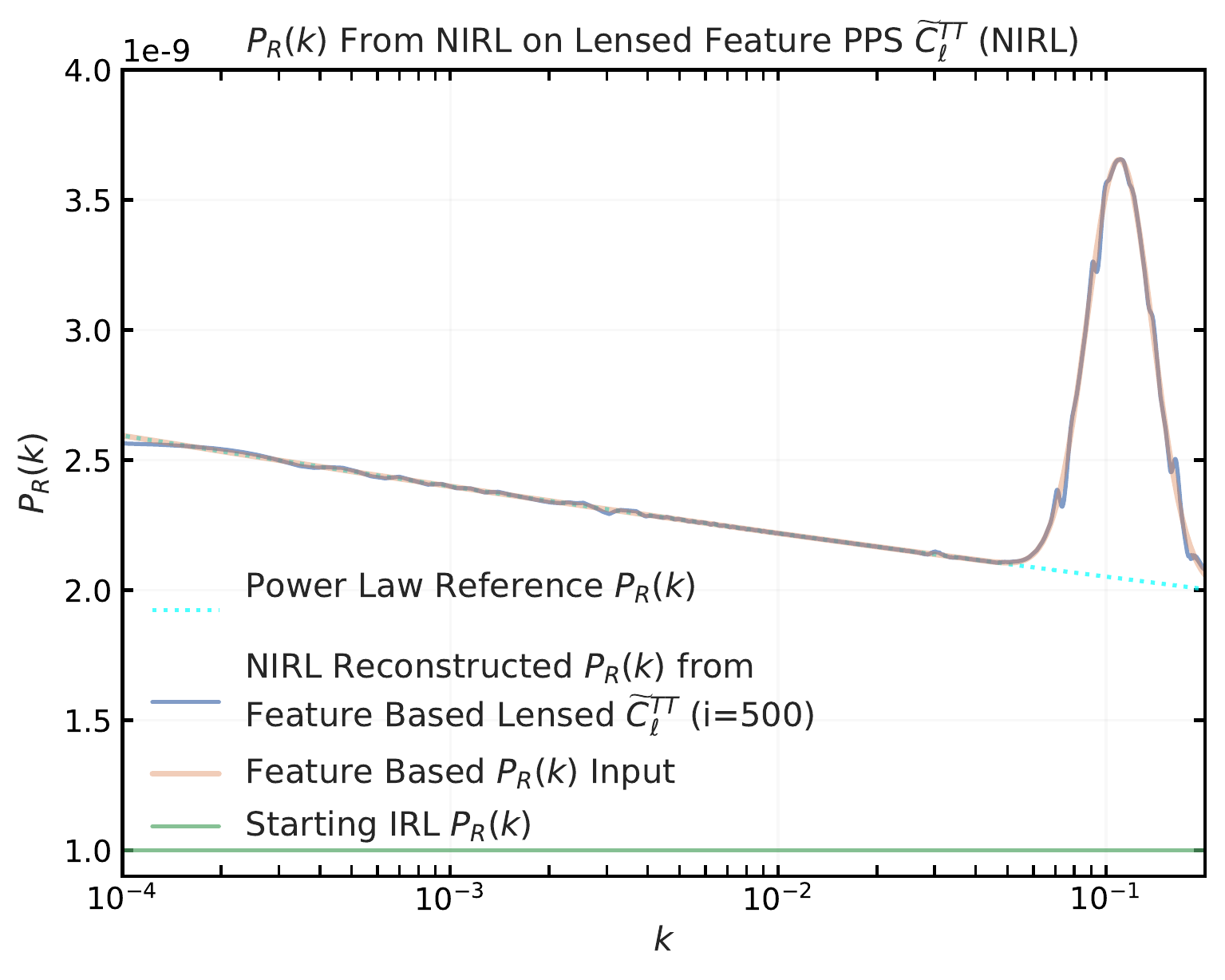}
  \caption{}
  \label{fig:pr4_feattest_reslt500_4}
\end{subfigure}
\caption{ The 4 figures show the reconstructed $P_R(k)$ from the feature based data, using 500 RL reconstruction iterations. The reference power law is given in cyan dotted lines. The reconstructed $P_R(k)$ are given in blue lines. The initial guess for the iterative estimator are given in green lines. The input feature $P_R(k)$ is given in orange lines.}
\label{fig:pr4_feattest_reslt500}
\end{figure}

\begin{figure}
\begin{subfigure}{1\textwidth}
\includegraphics[width=1.00\linewidth]{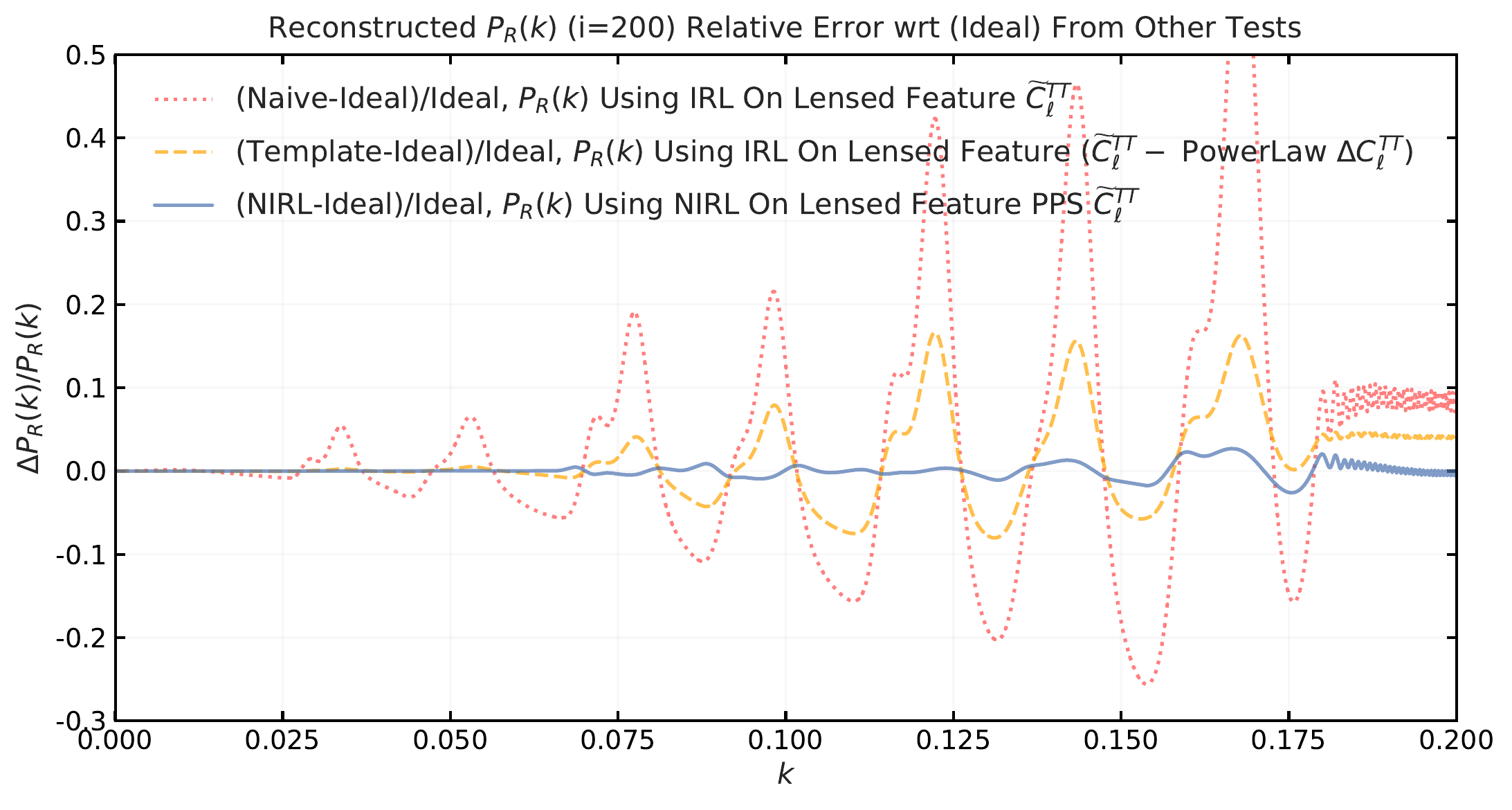}
\end{subfigure}
\caption{This figure shows the relative error of the 3 reconstructions of the $P_R(k)$ displayed in figure \ref{fig:pr4_feattest_reslt200_2},\ref{fig:pr4_feattest_reslt200_3}, and \ref{fig:pr4_feattest_reslt200_4} respectively, relative to the first reconstruction \ref{fig:pr4_feattest_reslt200_1} which is held as the ideal reference reconstruction. These are based on 200 iterations of the RL algorithm.}
\label{fig:pr4_feattest_reslt_relerr200}
\end{figure}

\begin{figure}
\begin{subfigure}{1\textwidth}
\includegraphics[width=1.00\linewidth]{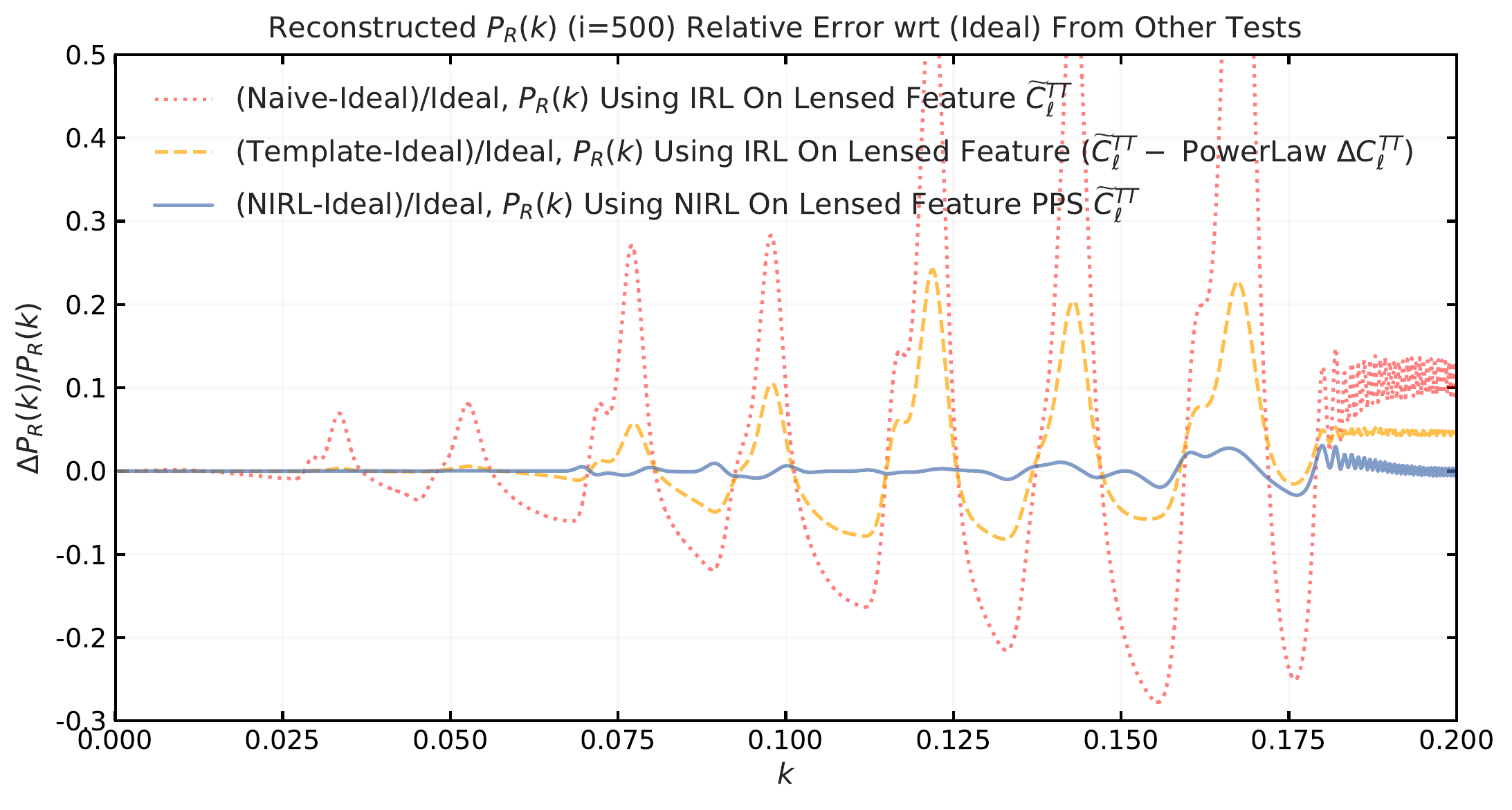}
\end{subfigure}
\caption{This figure shows the relative error of the 3 reconstructions of the $P_R(k)$ displayed in figure \ref{fig:pr4_feattest_reslt500_2},\ref{fig:pr4_feattest_reslt500_3}, and \ref{fig:pr4_feattest_reslt500_4} respectively, relative to the first reconstruction \ref{fig:pr4_feattest_reslt500_1} which is held as the ideal reference reconstruction. These are based on 500 iterations of the RL algorithm.}
\label{fig:pr4_feattest_reslt_relerr500}
\end{figure}

The two key outcomes we expect to observe are the NIRL estimators ability to approach the ideal exact reconstruction of $P_R(k)$ from the lensed data. the second, equally important aspect will be to see its relative accuracy with respect to a power law template based approach. These two results would decide the validity and usefulness of the NIRL reconstruction algorithm, which is a modification of the IRL estimator, in CMB analysis. The results of the tests are depicted in figure \ref{fig:pr4_feattest_reslt200}. The first plot \ref{fig:pr4_feattest_reslt200_1} is simply the $P_R(k)$ reconstructed from the unlensed feature origin unlensed ${C}_{\ell}^{TT}$, using the RL estimator. Since this is the ideal clean data, the reconstruction is the ideal reference for comparison for the subsequent ones. We can see the feature is well recovered, within the limit of 200 iterations for the RL estimator. There are some small kinks in the overall shape, but it is expected that with increasing number of RL iterations they should soften out, as this is the trend observed during increasing iterations from 1 to 200. The next plot \ref{fig:pr4_feattest_reslt200_2} shows the 'naive' reconstruction of the $P_R(k)$ from the feature origin Lensed $\widetilde{C}_{\ell}^{TT}$, using RL estimator, and clearly shows the reconstruction contains the lensing origin spurious features superimposed on the injected Gaussian feature. In the next plot in \ref{fig:pr4_feattest_reslt200_3} we plot the $P_R(k)$ reconstructed from the $\widetilde{C}_{\ell}^{TT}\textbf{(Feature)} - \Delta C_{\ell}^{TT}\textbf{(Power Law)}$, using the RL estimator. Here the reconstruction displays clearly that even though it is an improvement over the 'naive' lensing based RL reconstruction, it is still suboptimal as the Feature origin lensing contributions are still missed and show up as artefacts superimposed on the recovered Feature itself. In the final plot in figure \ref{fig:pr4_feattest_reslt200_4} we reconstruct the $P_R(k)$ using the NIRL method directly from the feature origin lensed $\widetilde{C}_{\ell}^{TT}$. In this plot, the improvement is clearly visible. The reconstruction recovers the $P_R(k)$ feature faithfully and is much cleaner, close to the exact reconstruction. The plot in figure \ref{fig:pr4_feattest_reslt_relerr200} plots the relative error between the last 3 plots in the previous figure, with respect to the first ideal reconstruction. We can clearly see the improvement in the reconstruction accuracy, with the NIRL version more or less exactly approaching the ideal reconstruction. 
We also provide the same set of plots in figure \ref{fig:pr4_feattest_reslt500}, but this time for 500 RL iterations instead. As mentioned earlier, the small kinks in the ideal and NIRL reconstruction get damped further, while the spurious features in the non optimized reconstructions get amplified further. This lends some insight that for any underlying PPS with a higher degree of features, a larger number of RL iterations are required for convergence. This iteration limit may be set for some convergence criteria, such as a 1\% relative difference between successive iterations. The improvement due to higher iterations is also visible in the relative error plot with respect to the ideal reconstruction, in figure \ref{fig:pr4_feattest_reslt_relerr500}. We note that while the NIRL reconstruction remains nearly the same with respect to the ideal reconstruction, the non optimized reconstruction move further away from the ideal one, showing that more of the unoptimized power is being fitted to the reconstructed $P_R(k)$. Hence ideally, in future studies we should aim for higher number of RL iterations, while looking for ways to reduce overfitting of noise. The summary of the test results are given in table \ref{table:pr4_feattest_reslt_summ}

\begin{table}
{\begin{center}
\begin{tabular}{ l r r l }
  \toprule
  Data & \multicolumn{2}{c}{Algorithm}{Result} \\
  \midrule
  Unlensed ${C}_{\ell}^{TT}\textbf{(Feature)}$ & IRL & Exact \\
  Lensed $\widetilde{C}_{\ell}^{TT}\textbf{(Feature)}$ & IRL & Poor \\
  Lensed $\widetilde{C}_{\ell}^{TT}\textbf{(Feature)}-\Delta C_{\ell}^{TT}\textbf{(Power Law)}$ & IRL & Average \\
  Lensed $\widetilde{C}_{\ell}^{TT}\textbf{(Feature)}$ & NIRL & Good \\
  \bottomrule
\end{tabular}
\end{center}}
\caption{This table lists the last 4 Reconstruction tests from \ref{table:feat_tests} with their reconstruction quality.}
\label{table:pr4_feattest_reslt_summ}
\end{table}

\subsection{Results iii): Wavepacket Reconstruction with Delensing}
\label{sub_sec:wavepk_NIRL_delen}

We have seen that the NIRL performs well for a synthetic bump-like feature introduced in the $P_R(k)$ in the previous section. As the feature contribution of the excess power in the power spectrum is above the cosmic variance (CV) threshold, as shown in figure \ref{fig:pr4_feattest}, we can expect that for any observation data, at least as precise as the CV limit, will benefit from the non-linear correcting iterative Richardson-Lucy estimator, by providing a more accurate reconstruction of $P_R(k)$, validated by high precision.
However it is clear from current observations that the level of deviations caused in the $C_{\ell}^{TT}$ given in figure \ref{fig:pr4_feattest_4}, are immediately ruled out, and our synthetic test feature is unphysical and only a demonstration. Therefore it may make sense to test our method on a more physically or at least observationally motivated, but synthetic feature based $P_R(k)$, and to see how the NIRL estimator performs. We expect that for more conservative features, which impact the $C_{\ell}^{TT}$ below current observational precision, the NIRL reconstruction will improve the accuracy somewhat, though the precision would remain the same. 
Previous work on the topic of power law deviations based on data, have been carried out earlier in \cite{Polarski:1992dq,Hazra:2018opk,Hazra:2014goa,Keeley:2020rmo}.

\begin{figure}
    \centering 
\begin{subfigure}{0.5\textwidth}
  \includegraphics[width=\linewidth]{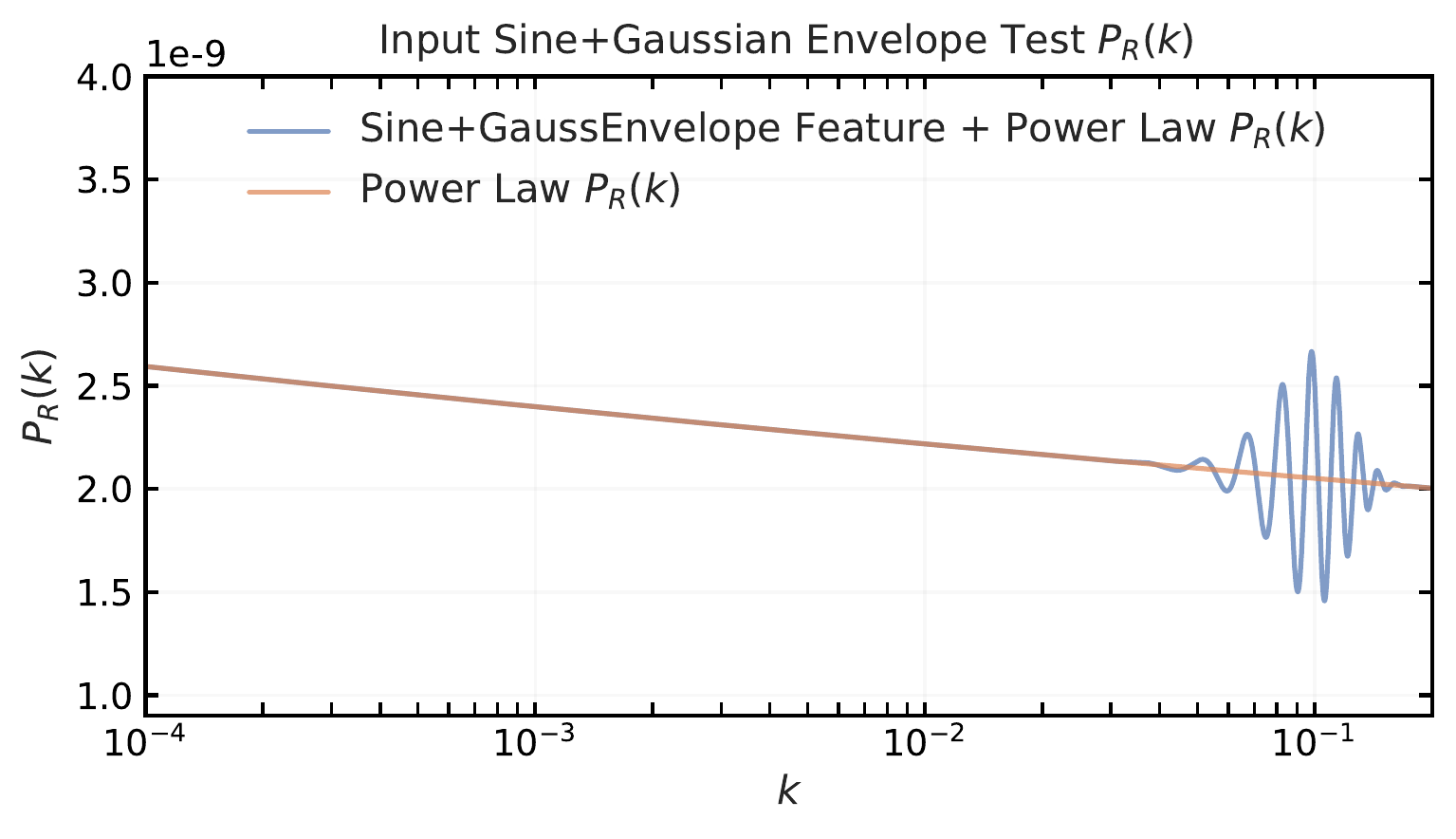}
  \caption{}
  \label{fig:pr4_feattestwv_1}
\end{subfigure}\hfil 
\begin{subfigure}{0.5\textwidth}
  \includegraphics[width=\linewidth]{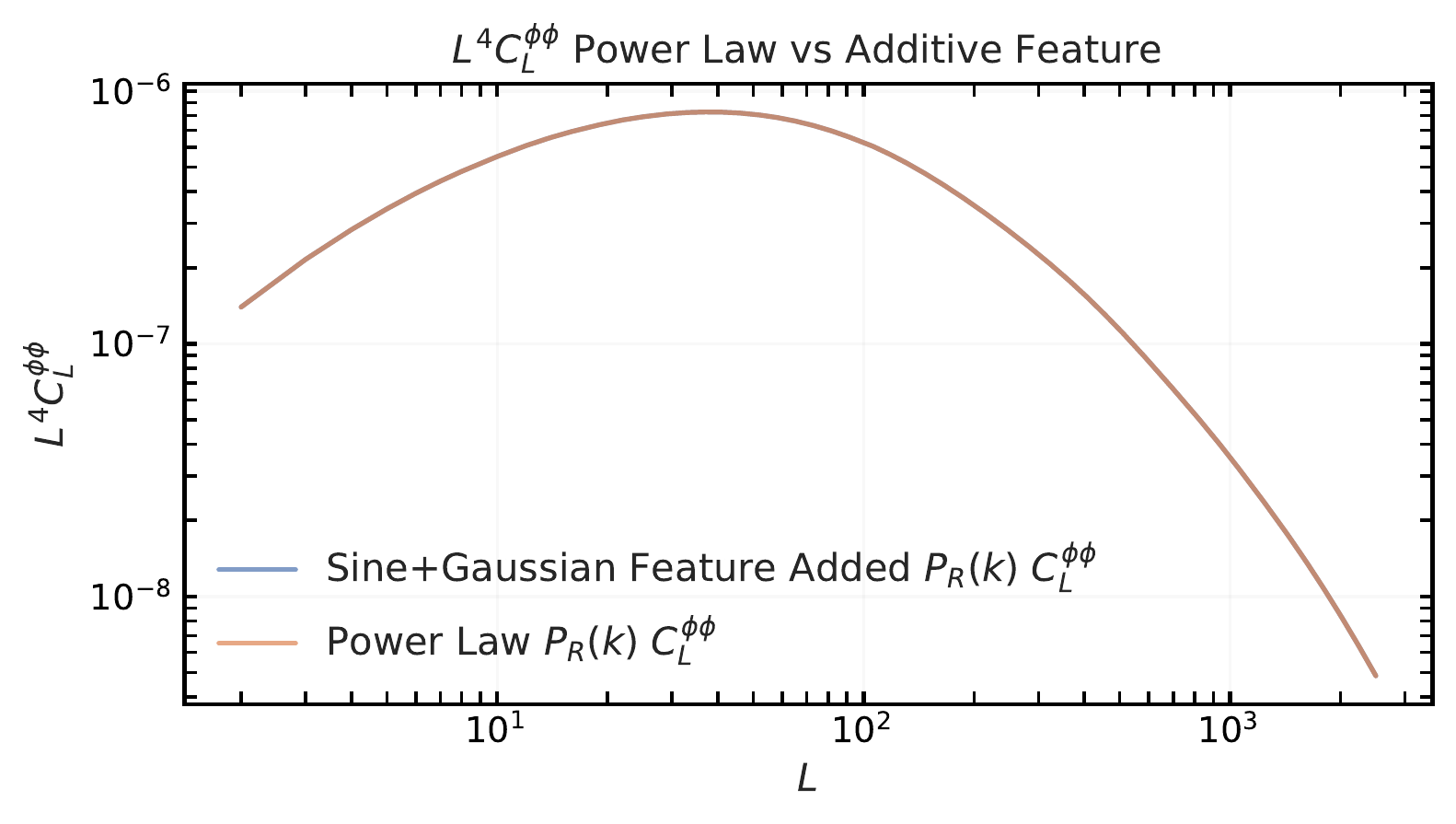}
  \caption{}
  \label{fig:pr4_feattestwv_2}
\end{subfigure}

\begin{subfigure}{0.5\textwidth}
  \includegraphics[width=\linewidth]{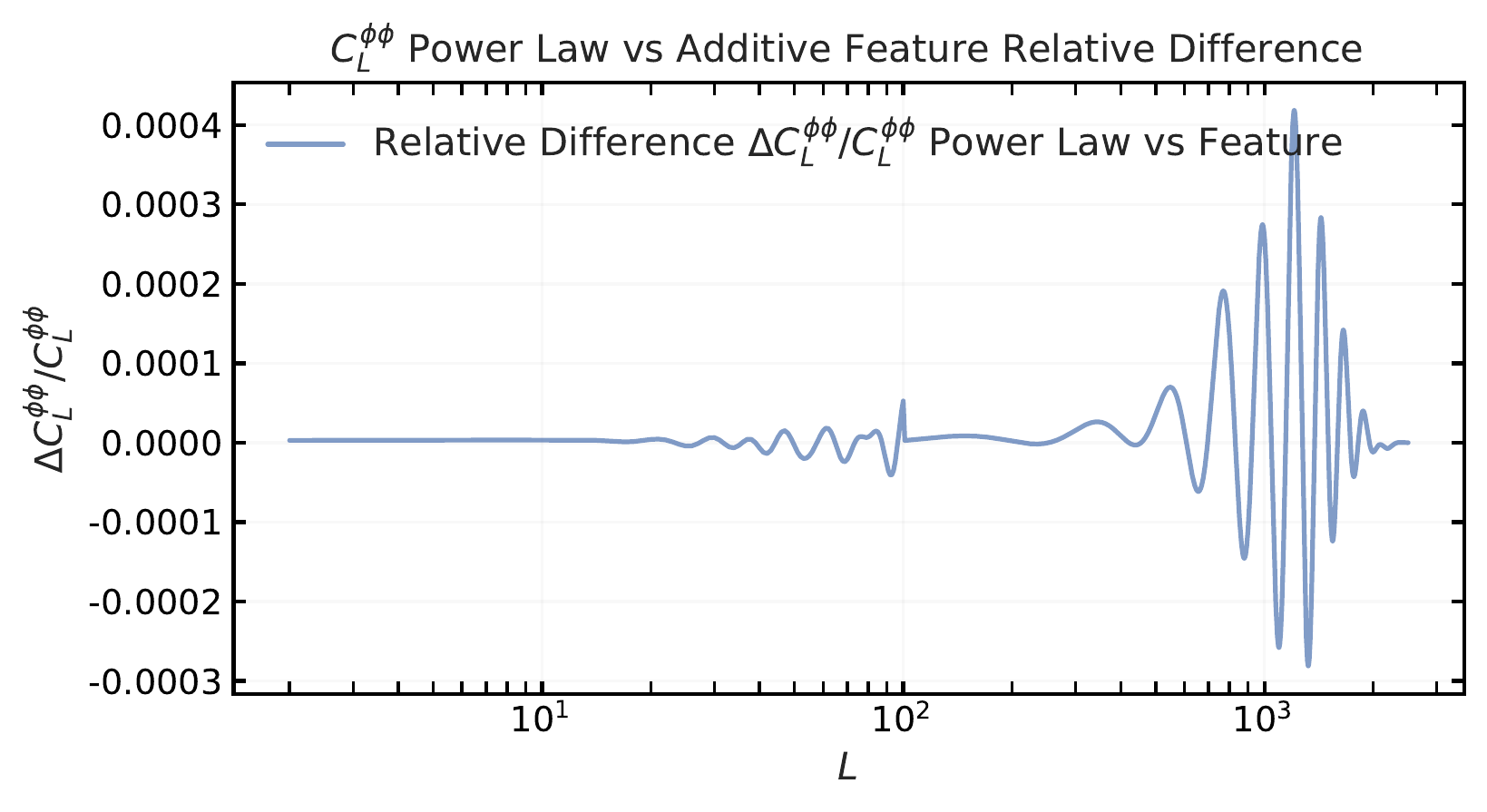}
  \caption{}
  \label{fig:pr4_feattestwv_3}
\end{subfigure}\hfil 
\begin{subfigure}{0.5\textwidth}
  \includegraphics[width=\linewidth]{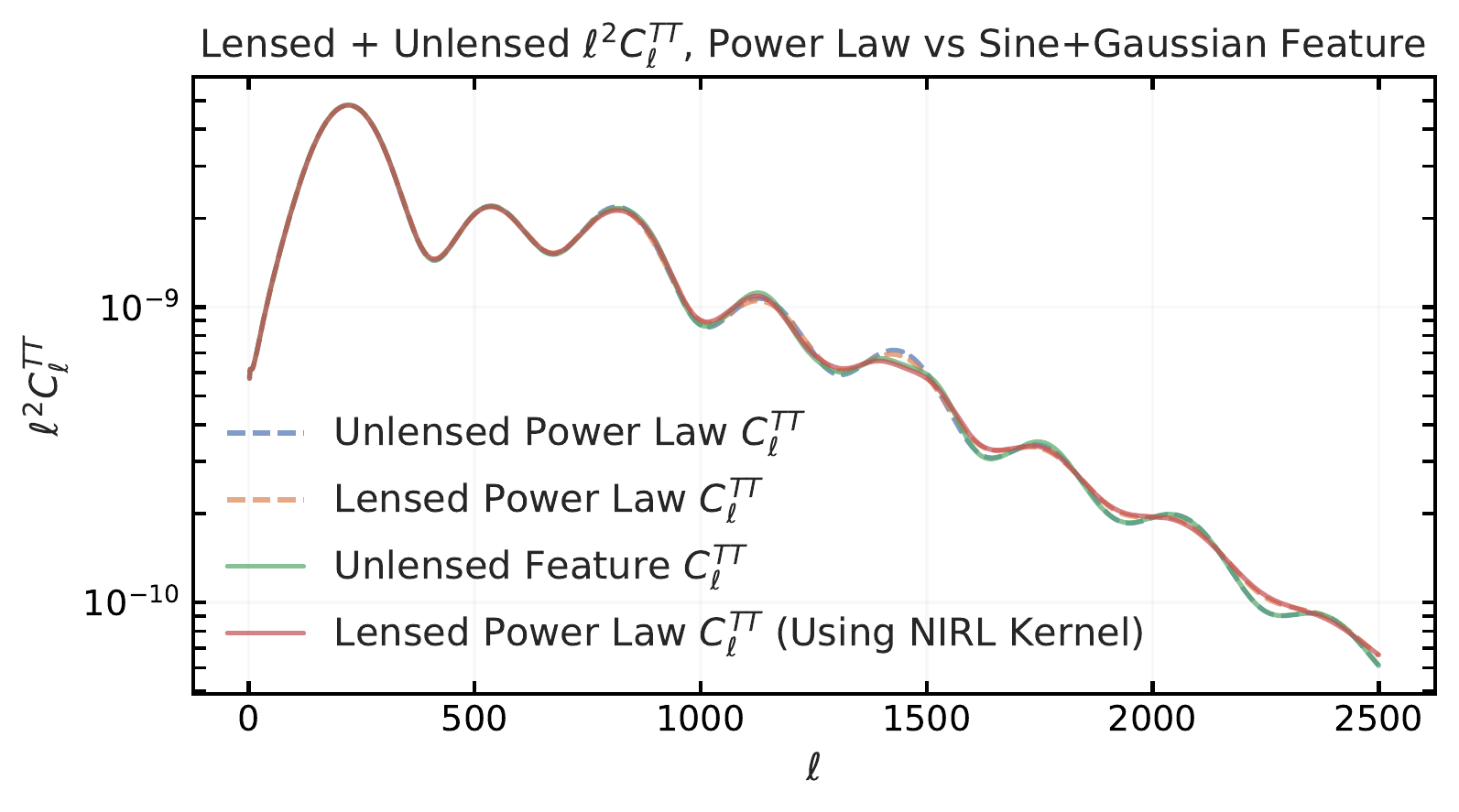}
  \caption{}
  \label{fig:pr4_feattestwv_4}
\end{subfigure}

\begin{subfigure}{0.5\textwidth}
  \includegraphics[width=\linewidth]{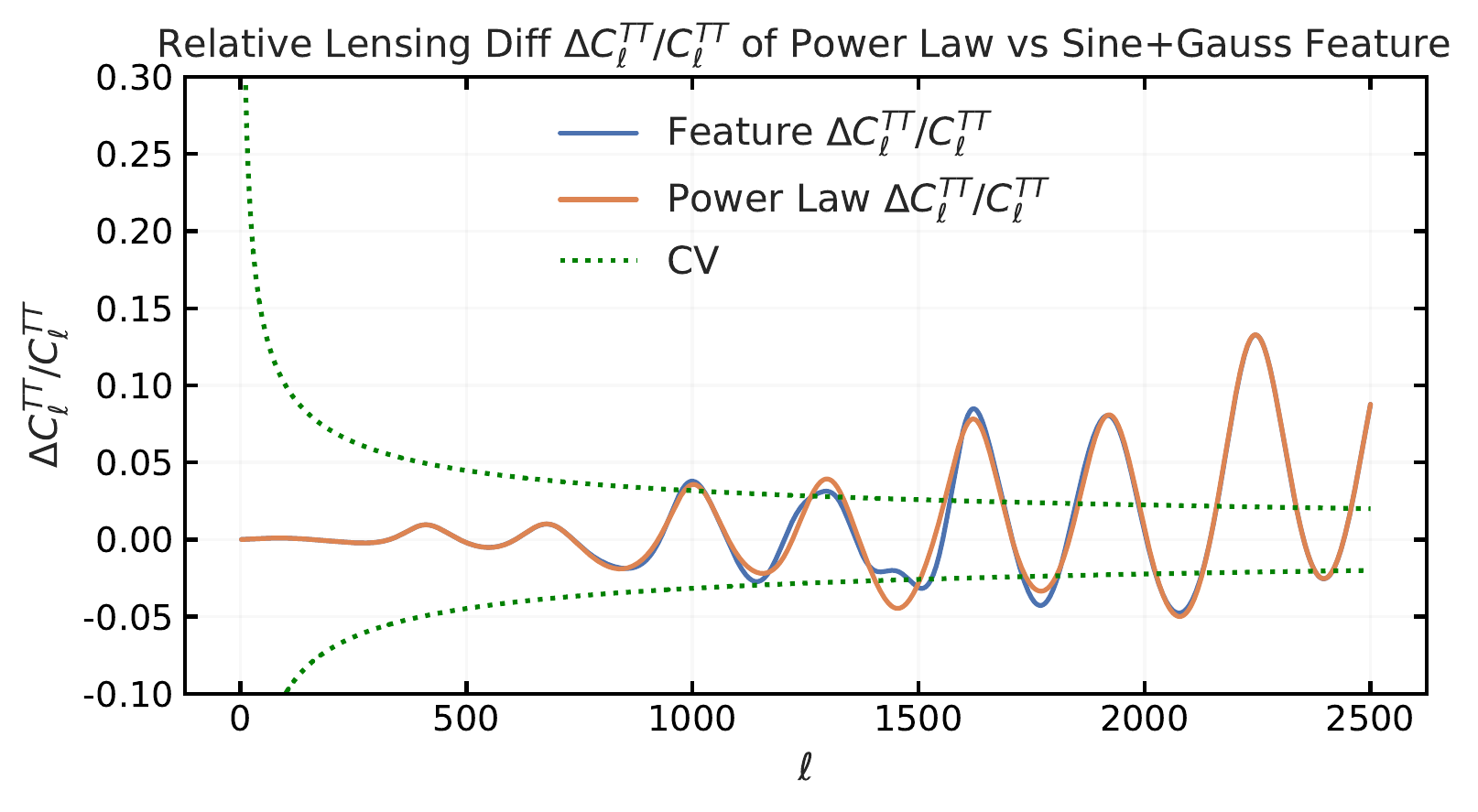}
  \caption{}
  \label{fig:pr4_feattestwv_5}
\end{subfigure}\hfil 
\begin{subfigure}{0.5\textwidth}
  \includegraphics[width=\linewidth]{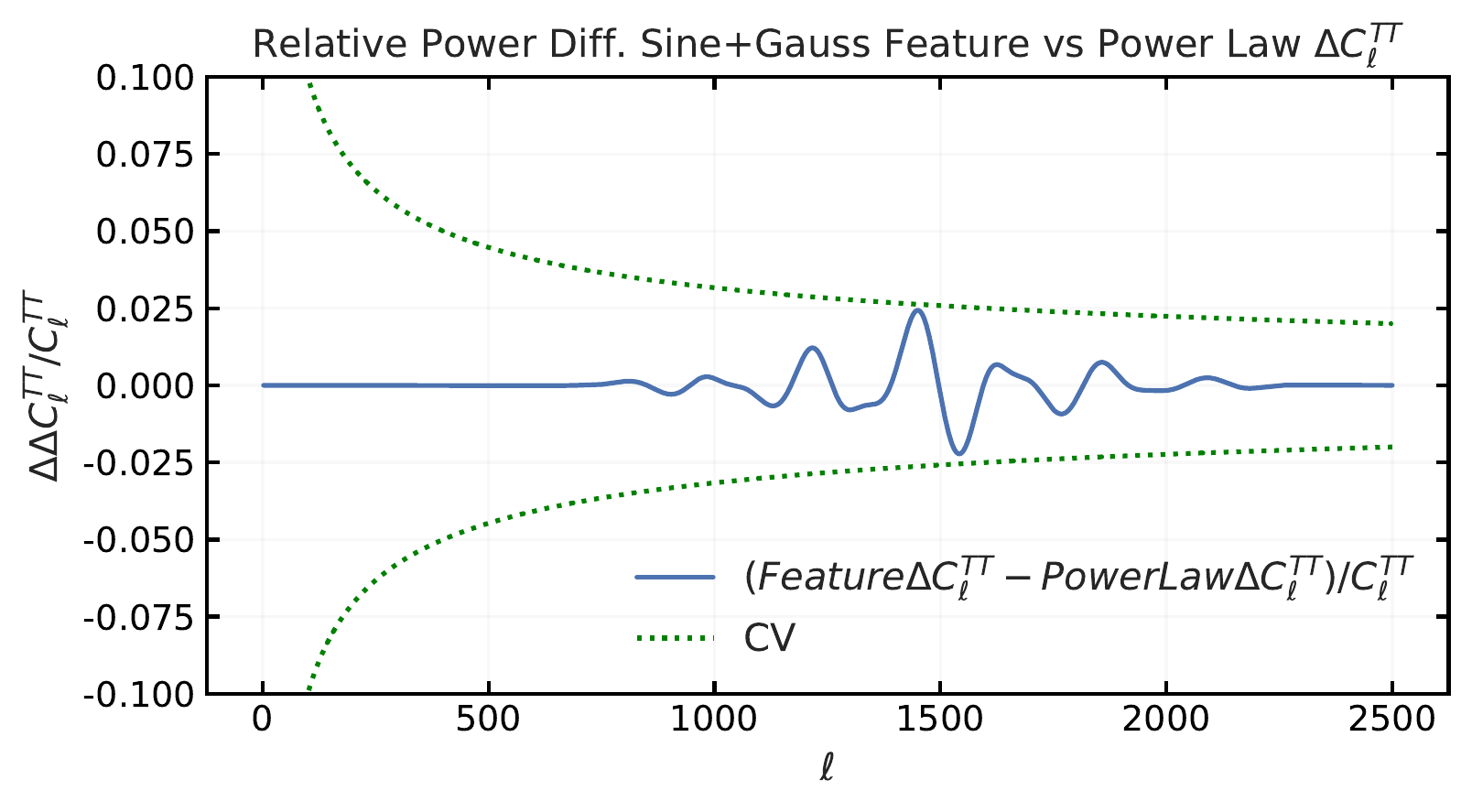}
  \caption{}
  \label{fig:pr4_feattestwv_6}
\end{subfigure}
\caption{These set of figures present the wavepacket feature added simulated data used for NIRL validation. Figure \ref{fig:pr4_feattestwv_1} shows the feature based $P_R(k)$ in blue and power law reference in orange. Figure \ref{fig:pr4_feattestwv_2} shows the corresponding ${C}_L^{\phi\phi}$ with the feature based simulation in orange and power law simulation blue. Figure \ref{fig:pr4_feattestwv_3} shows the relative power difference due to the two ${C}_L^{\phi\phi}$. Figure \ref{fig:pr4_feattestwv_4} shows the lensed and unlensed ${C}_{\ell}^{TT}$ due to the feature (solid lines) vs the power law (dashed lines) based $P_R(k)$. Figure \ref{fig:pr4_feattestwv_5} shows the lensing contribution $\Delta{C}_{\ell}^{TT}$ due to the feature (blue line) and power law (orange line) $P_R(k)$. Figure \ref{fig:pr4_feattestwv_6} shows the relative power difference between the two $\Delta{C}_{\ell}^{TT}$ in blue, with respect to the cosmic variance limit in dashed green. }
\label{fig:pr4_feattestwv}
\end{figure}

We therefore synthetically add a 'wavepacket' like feature added on the power law $P_R(k)$, which consists of a sine wave with a Gaussian envelope, centered at $k = 0.1$. We plot this in figure \ref{fig:pr4_feattestwv_1}, along with the reference power law. In figure \ref{fig:pr4_feattestwv_2} and \ref{fig:pr4_feattestwv_3} the consequent ${C}_L^{\phi\phi}$ is depicted, from the feature vs power law simulation, as well as the relative difference between the two. As can be seen in this plot, the relative difference in the weak lensing power spectrum is negligible, due to the rapidly oscillating behaviour of the feature. Due to the smoothness of the lensing kernel, this results in negligible power transfer to ${C}_L^{\phi\phi}$. The change in ${C}_{\ell}^{TT}$, both lensed and unlensed is depicted in \ref{fig:pr4_feattestwv_4}. We can see that the change in the powerlaw versus the wavepacket feature based $P_R(k)$ is not drastic and produces only slight changes, with the most significant effect being around $L \approx 1500$. The lensing modulation of the $C_{\ell}^{TT}$ between the two different $P_R(k)$ is given in figure \ref{fig:pr4_feattestwv_5} and the fractional difference between the two lensing modulations are given in figure \ref{fig:pr4_feattestwv_6}. The last two plots show the effect of the new wavepacket feature is much less drastic and could conceivably be lurking just on the boundary of the CV precision. Overall, such features, both motivated by observation and theory in previous literature, provide a good hypothetical feature to test NIRL with.

We again lay out the steps to be performed in order to quantify the results of the NIRL vs the template-based delensing approach, when considering a general underlying $P_R(k)$ origin for the temperature spectra, without any prior knowledge of it being a pure power law or otherwise. The key tests are listed in table \ref{table:feat_testswv}.

\begin{table}
{\begin{center}
\begin{tabular}{ l r r l }
  \toprule
  Wavepacket Feature Test Steps  \\
  \midrule
  1. IRL Reconstruction from Unlensed CAMB $C_{\ell}^{TT} \textbf{(Sine Feature)}$ \\
  2. IRL Reconstruction from Lensed CAMB $\widetilde{C}_{\ell}^{TT} \textbf{(Sine Feature)}$ \\
  3. IRL Reconstruction from $\widetilde{C}_{\ell}^{TT}\textbf{(Sine Feature)} - \Delta C_{\ell}^{TT}\textbf{(Power Law)}$ \\
  4. NIRL Reconstruction from Lensed CAMB $\widetilde{C}_{\ell}^{TT} \textbf{(Sine Feature)}$ \\
  \bottomrule
\end{tabular}
\end{center}}
\caption{This table lists the tests undertaken to validate the NIRL estimator and its convergence based on $P_R(k)$ incorporating a feature.}
\label{table:feat_testswv}
\end{table}

\begin{figure}
    \centering 
\begin{subfigure}{0.47\textwidth}
  \includegraphics[width=\linewidth]{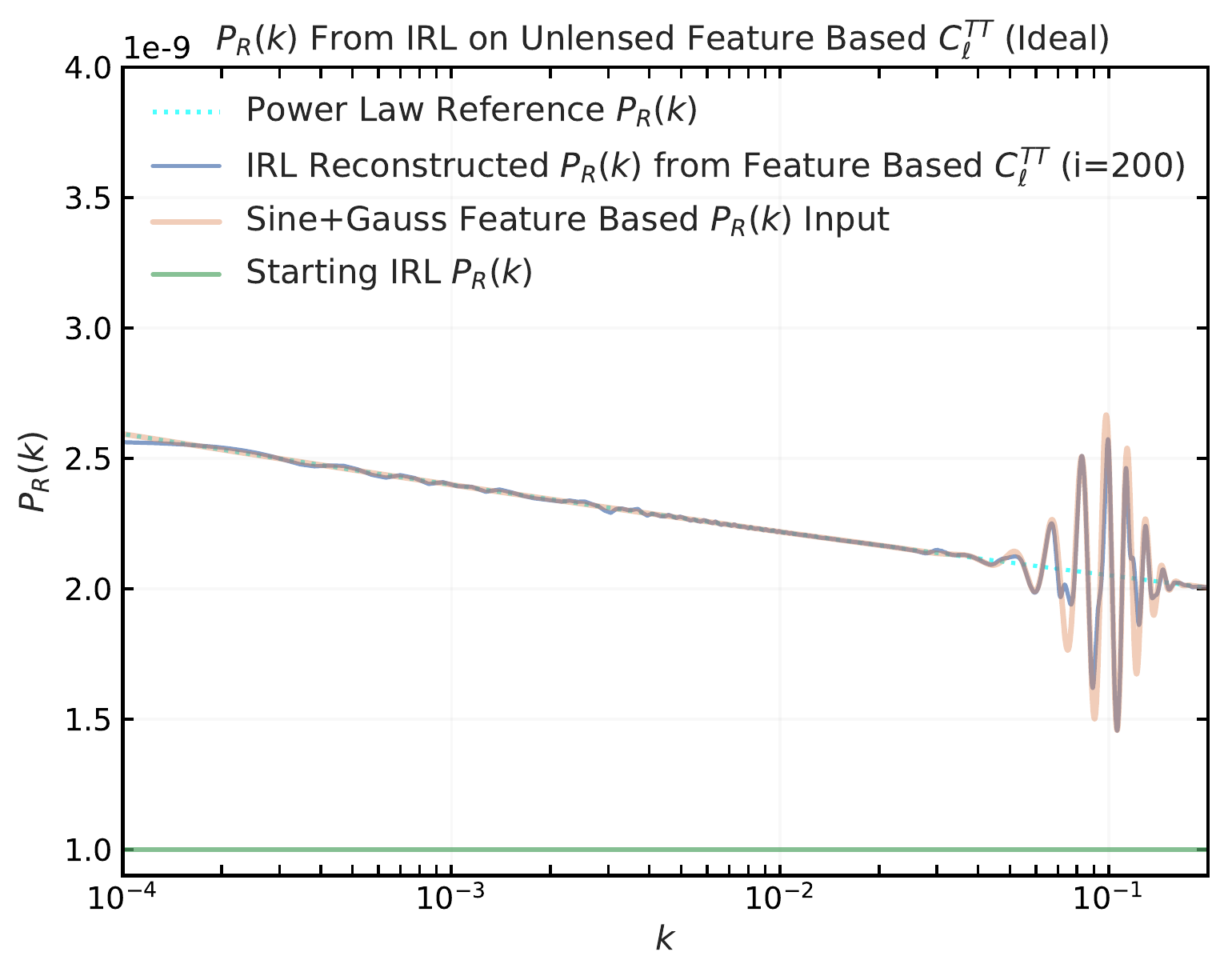}
  \caption{}
  \label{fig:pr4_feattestwv_reslt200_1}
\end{subfigure}\hfil 
\begin{subfigure}{0.47\textwidth}
  \includegraphics[width=\linewidth]{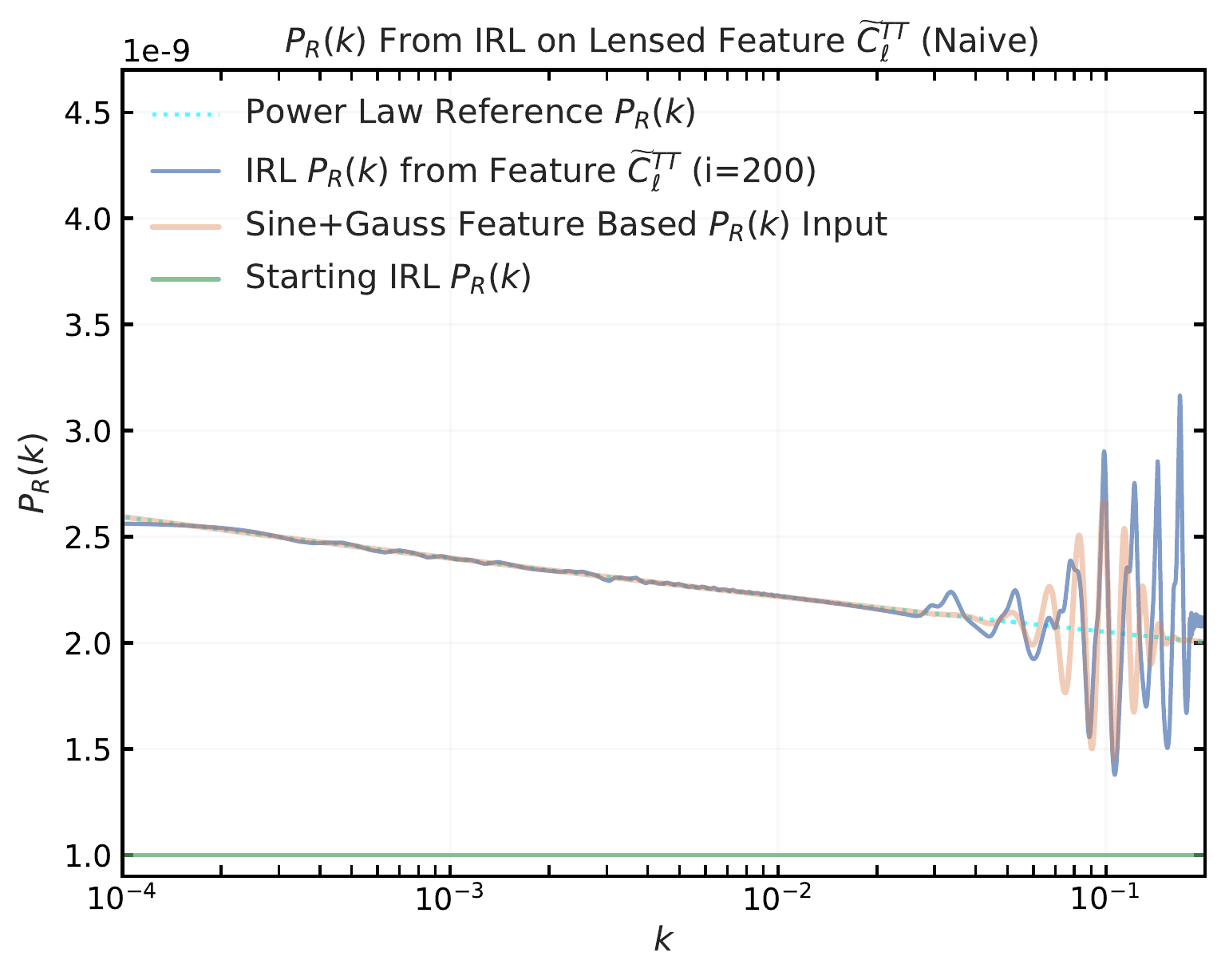}
  \caption{}
  \label{fig:pr4_feattestwv_reslt200_2}
\end{subfigure}

\begin{subfigure}{0.47\textwidth}
  \includegraphics[width=\linewidth]{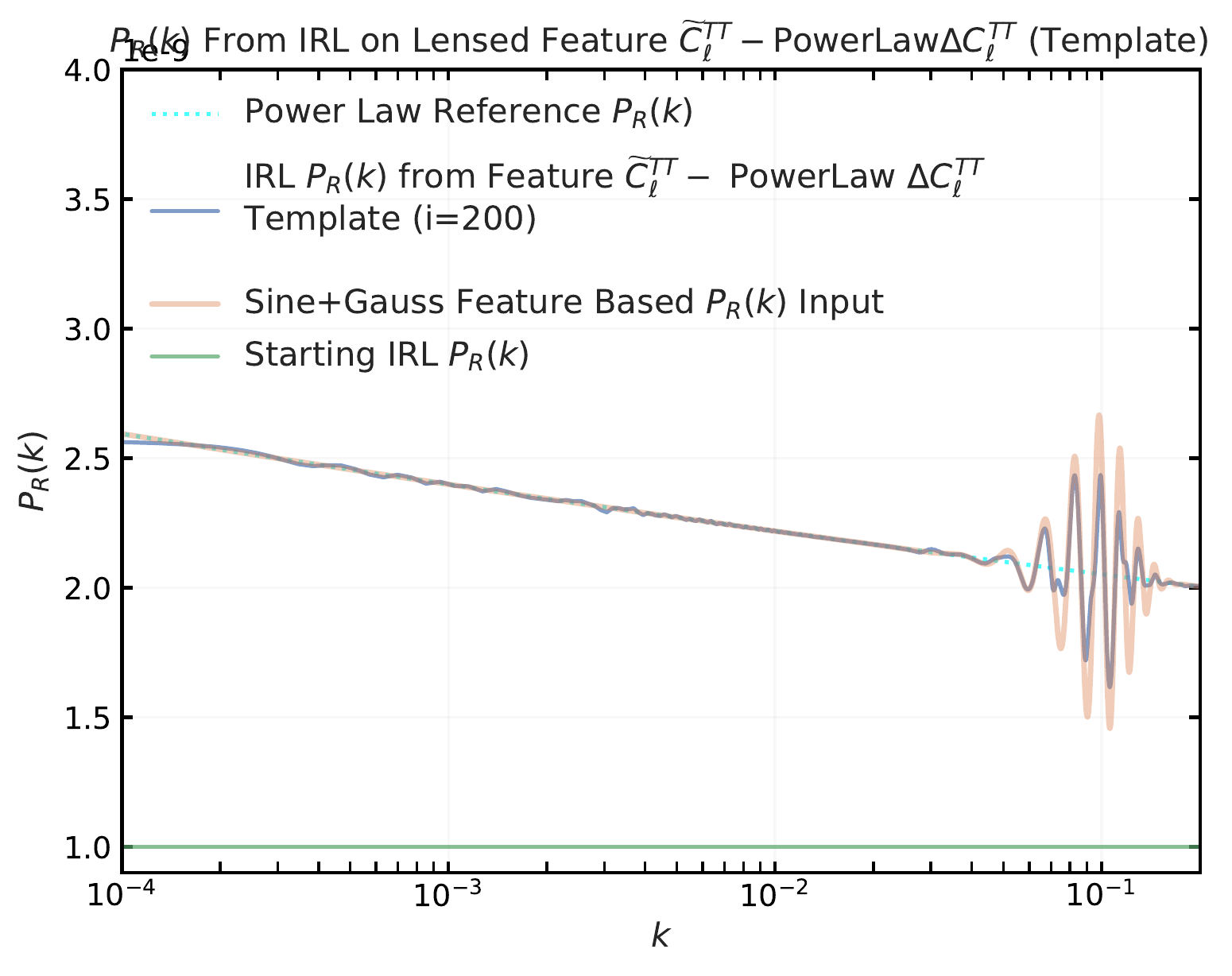}
  \caption{}
  \label{fig:pr4_feattestwv_reslt200_3}
\end{subfigure}\hfil 
\begin{subfigure}{0.47\textwidth}
  \includegraphics[width=\linewidth]{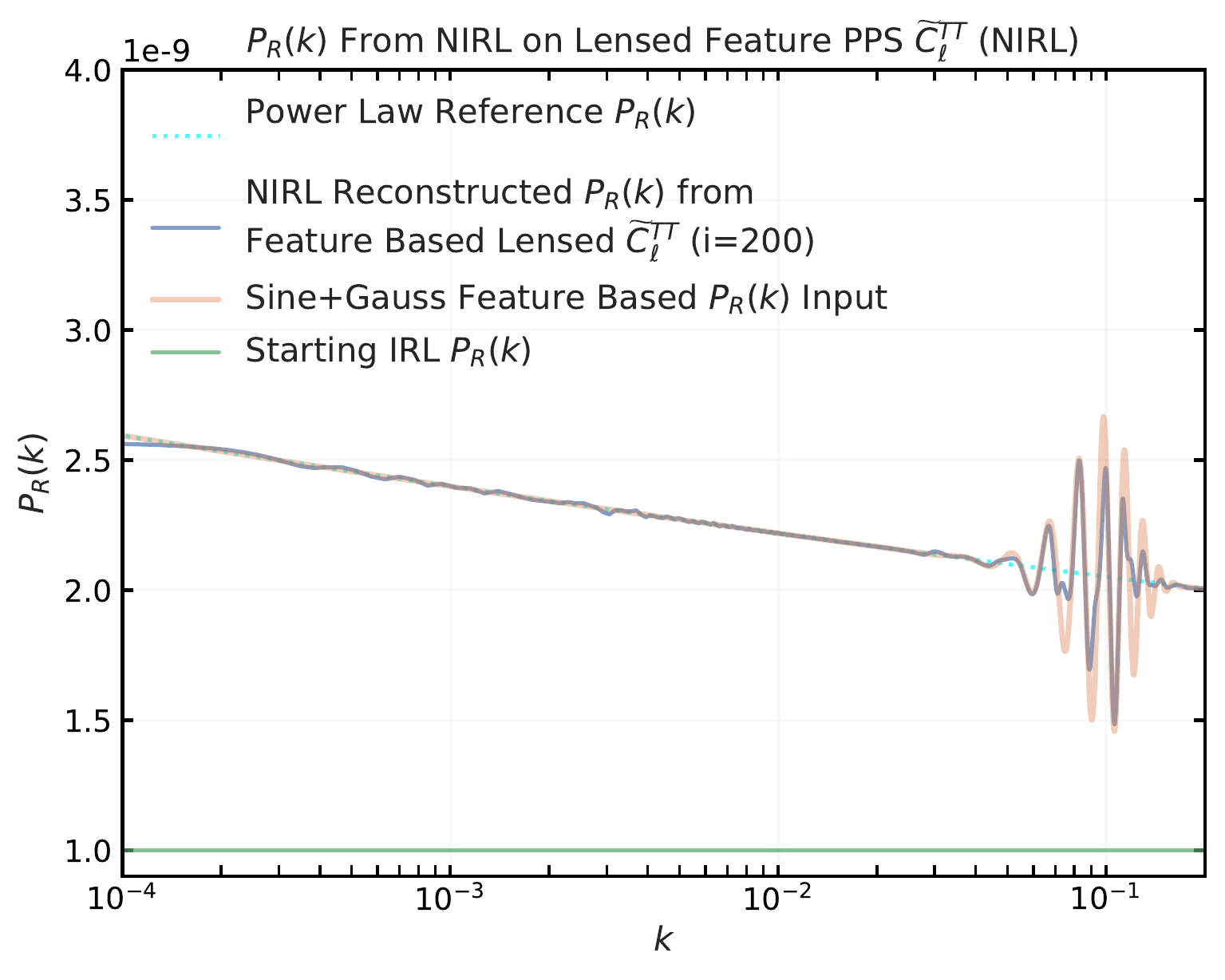}
  \caption{}
  \label{fig:pr4_feattestwv_reslt200_4}
\end{subfigure}
\caption{ The 4 figures show the reconstructed $P_R(k)$ from the feature based data, using 200 RL reconstruction iterations. The reference power law is given in cyan dotted lines. The reconstructed $P_R(k)$ are given in blue lines. The initial guess for the iterative estimator are given in green lines. The input feature $P_R(k)$ is given in orange lines.}
\label{fig:pr4_feattestwv_reslt200}
\end{figure}

\begin{figure}
    \centering 
\begin{subfigure}{0.47\textwidth}
  \includegraphics[width=\linewidth]{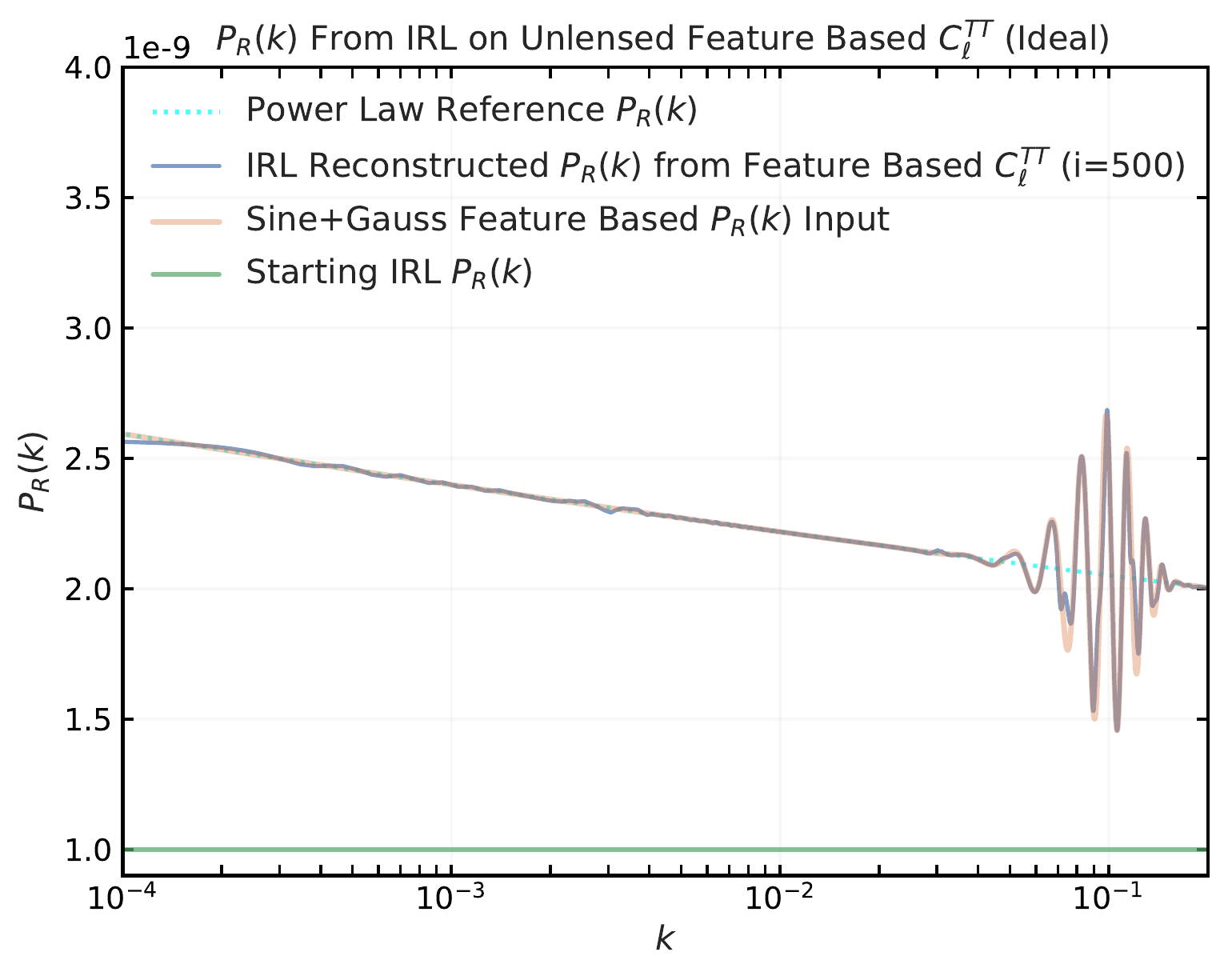}
  \caption{}
  \label{fig:pr4_feattestwv_reslt500_1}
\end{subfigure}\hfil 
\begin{subfigure}{0.47\textwidth}
  \includegraphics[width=\linewidth]{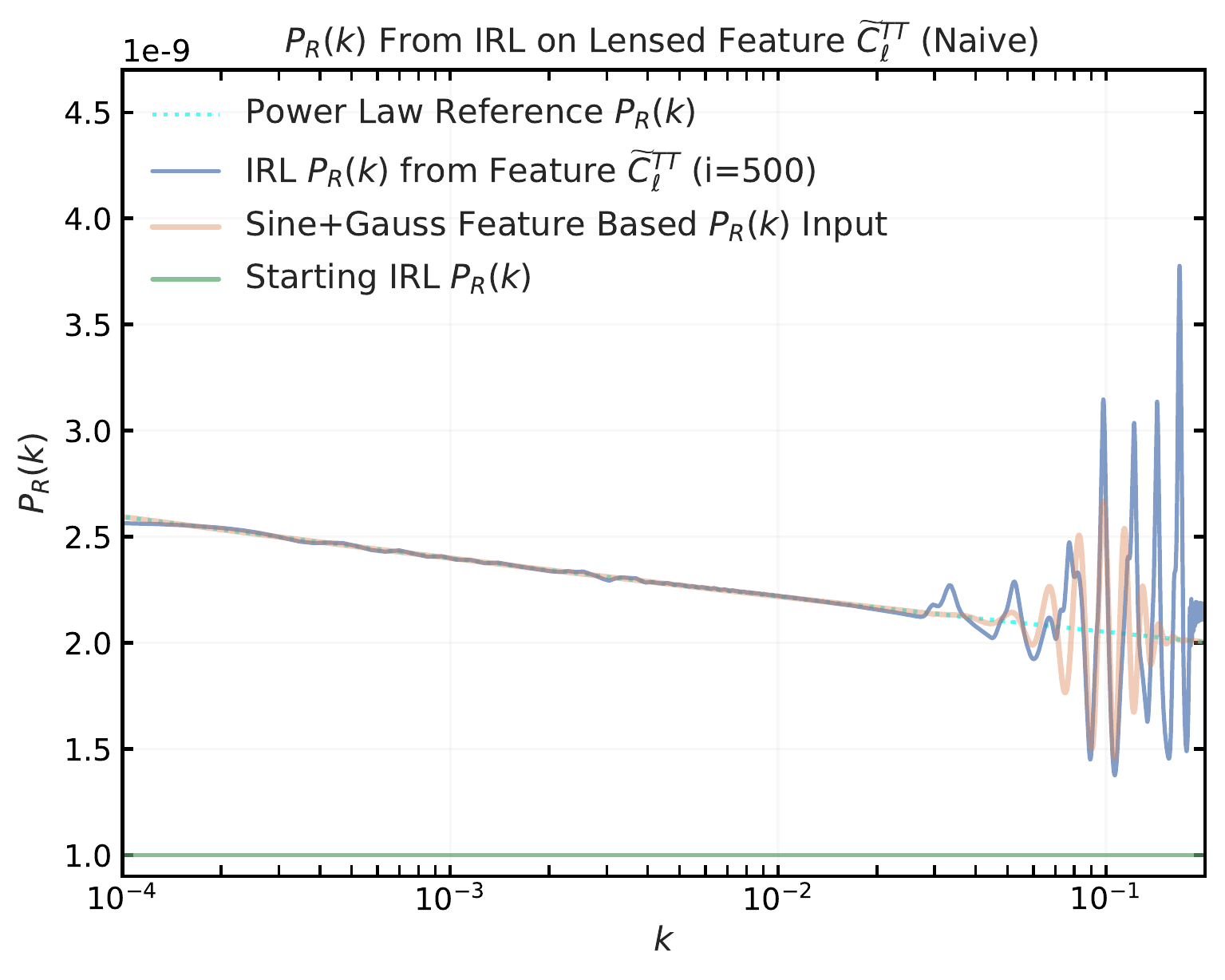}
  \caption{}
  \label{fig:pr4_feattestwv_reslt500_2}
\end{subfigure}

\begin{subfigure}{0.47\textwidth}
  \includegraphics[width=\linewidth]{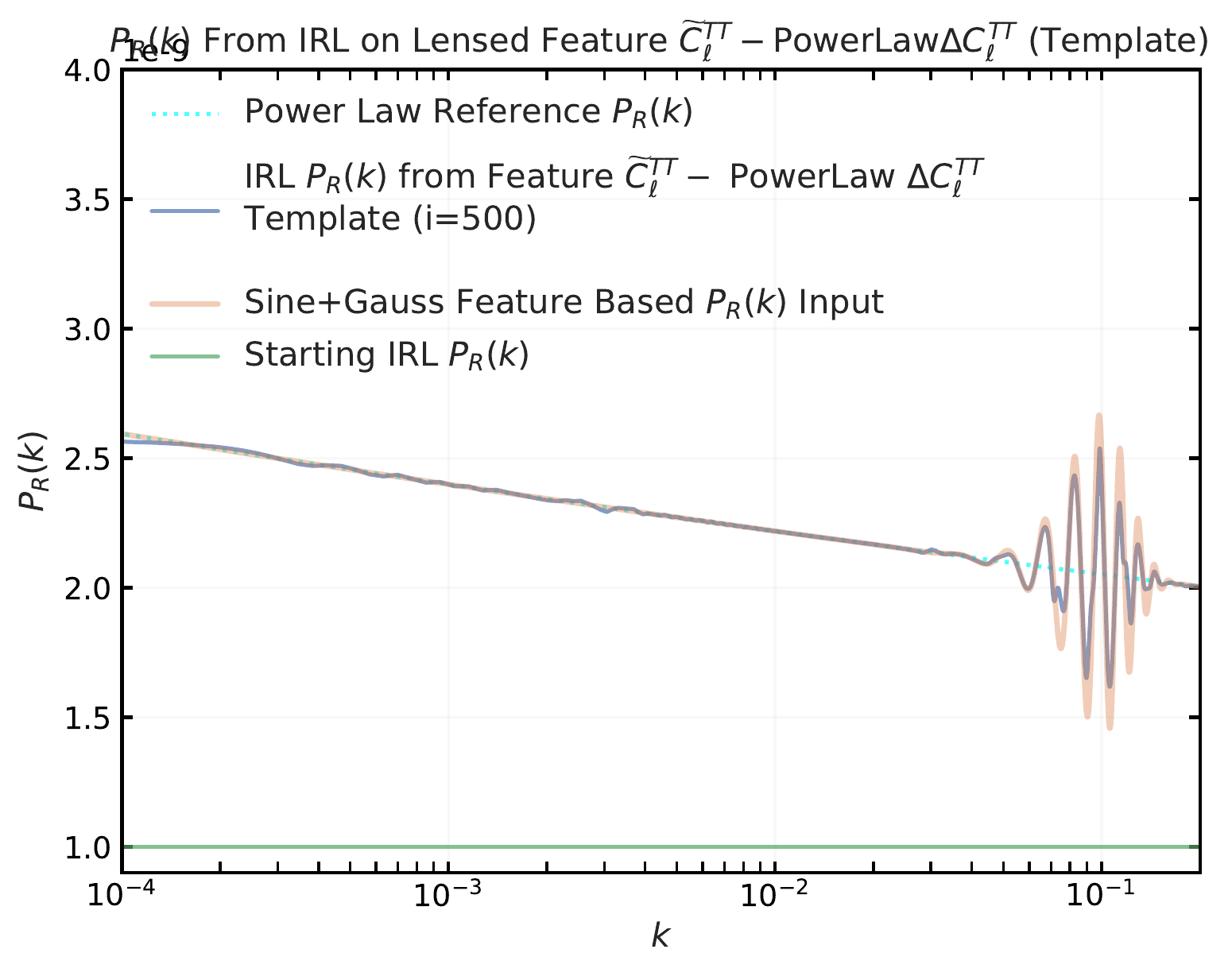}
  \caption{}
  \label{fig:pr4_feattestwv_reslt500_3}
\end{subfigure}\hfil 
\begin{subfigure}{0.47\textwidth}
  \includegraphics[width=\linewidth]{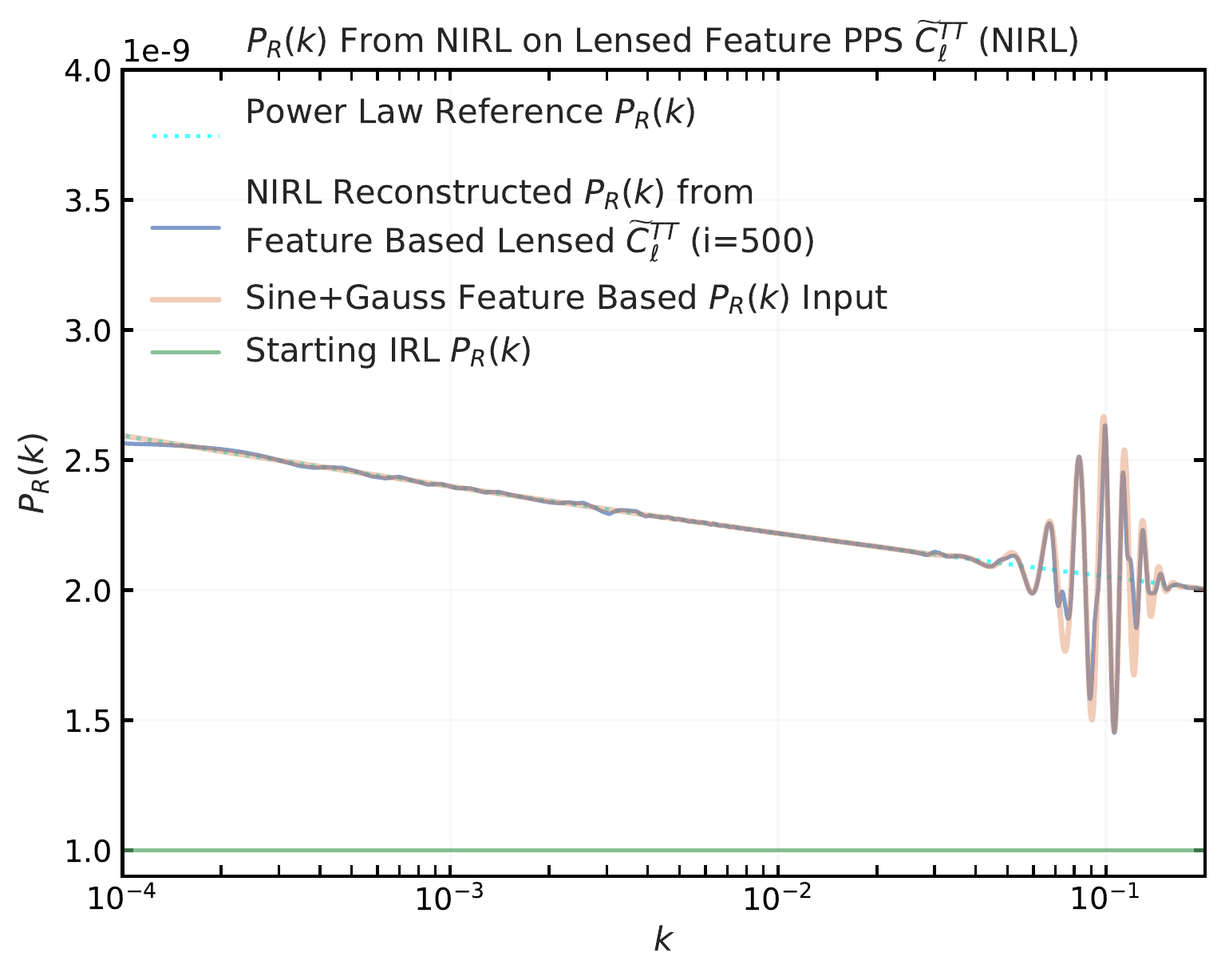}
  \caption{}
  \label{fig:pr4_feattestwv_reslt500_4}
\end{subfigure}
\caption{ The 4 figures show the reconstructed $P_R(k)$ from the feature based data, using 500 RL reconstruction iterations. The reference power law is given in cyan dotted lines. The reconstructed $P_R(k)$ are given in blue lines. The initial guess for the iterative estimator are given in green lines. The input feature $P_R(k)$ is given in orange lines.}
\label{fig:pr4_feattestwv_reslt500}
\end{figure}

\begin{figure}
\begin{subfigure}{1\textwidth}
\includegraphics[width=1.00\linewidth]{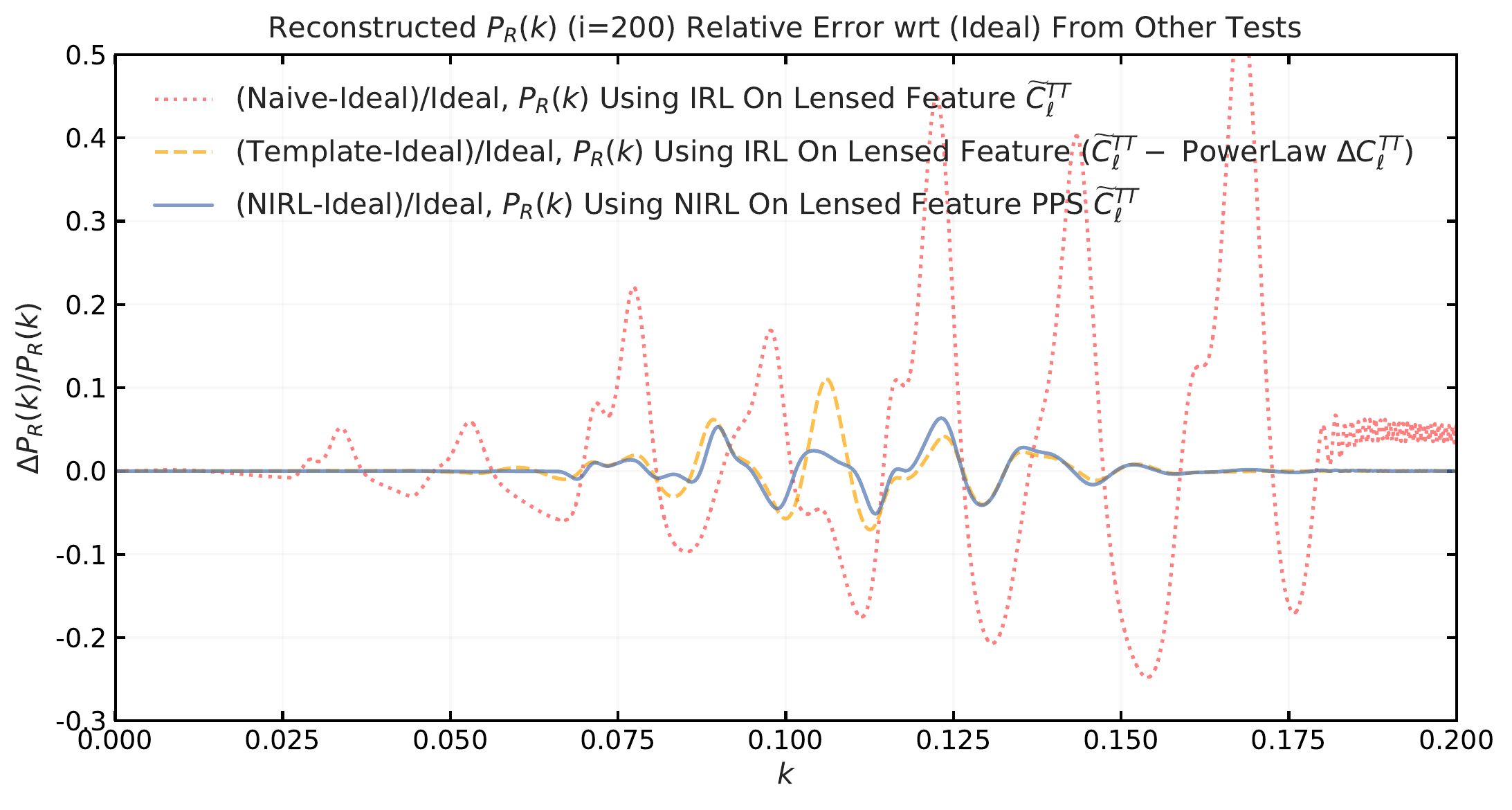}
\end{subfigure}
\caption{This figure shows the relative error of the 3 reconstructions of the $P_R(k)$ displayed in figure \ref{fig:pr4_feattestwv_reslt200_2},\ref{fig:pr4_feattestwv_reslt200_3}, and \ref{fig:pr4_feattestwv_reslt200_4} respectively, relative to the first reconstruction \ref{fig:pr4_feattestwv_reslt200_1} which is held as the ideal reference reconstruction. These are based on 200 iterations of the RL algorithm.}
\label{fig:pr4_feattestwv_reslt_relerr200}
\end{figure}

\begin{figure}
\begin{subfigure}{1\textwidth}
\includegraphics[width=1.00\linewidth]{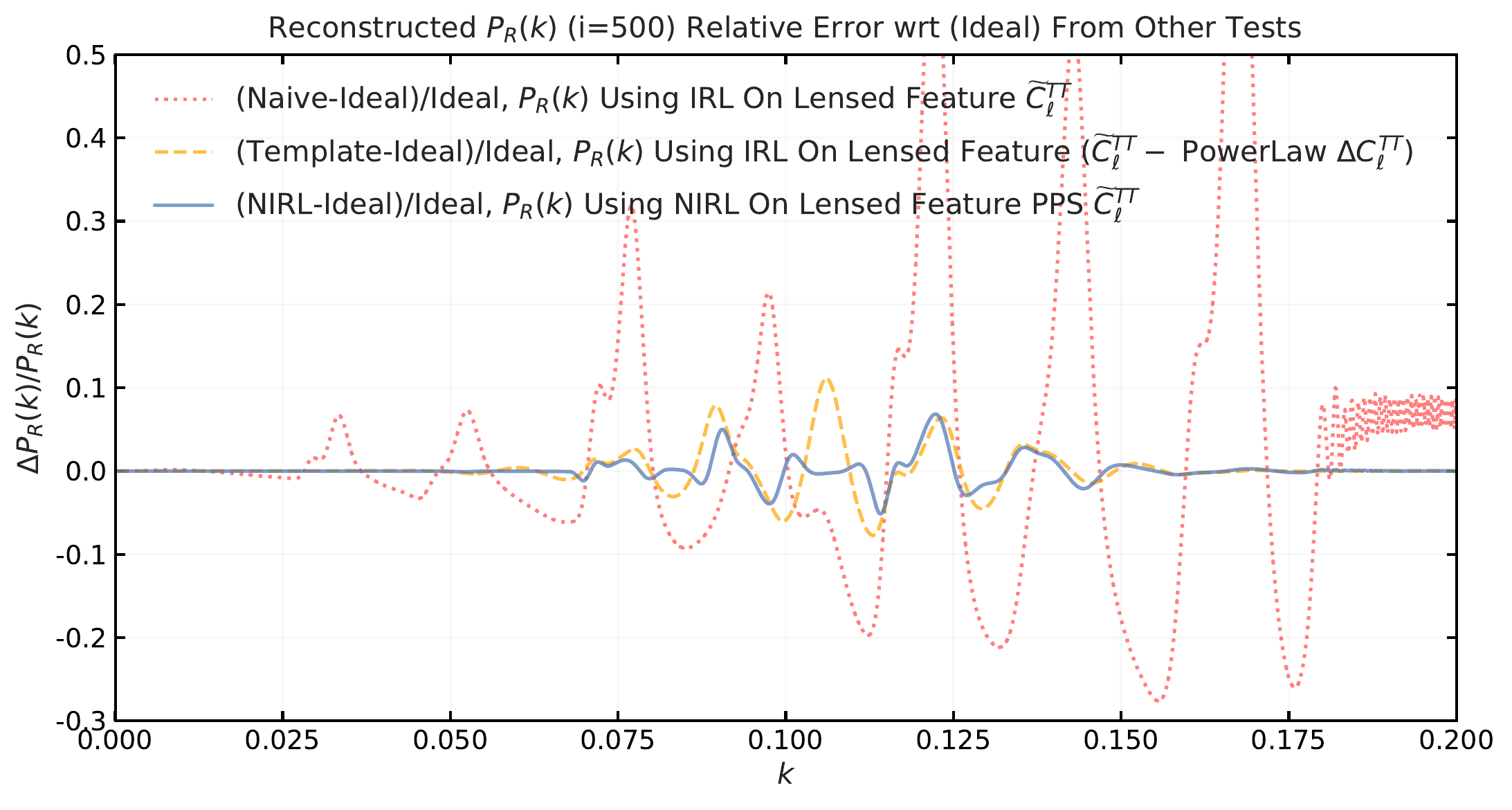}
\end{subfigure}
\caption{This figure shows the relative error of the 3 reconstructions of the $P_R(k)$ displayed in figure \ref{fig:pr4_feattestwv_reslt500_2},\ref{fig:pr4_feattestwv_reslt500_3}, and \ref{fig:pr4_feattestwv_reslt500_4} respectively, relative to the first reconstruction \ref{fig:pr4_feattestwv_reslt500_1} which is held as the ideal reference reconstruction. These are based on 500 iterations of the RL algorithm.}
\label{fig:pr4_feattestwv_reslt_relerr500}
\end{figure}

We lay out the results of the Non-Linear Iterative Richardson-Lucy reconstruction applied on this data in figures \ref{fig:pr4_feattestwv_reslt200} and \ref{fig:pr4_feattestwv_reslt500}. Like before we want to first obtain the ideal reconstruction of the sine plus Gaussian envelope wavepacket $P_R(k)$ from the unlensed ${C}_{\ell}^{TT}$, and compare the equivalent reconstructions of the $P_R(k)$ from the IRL estimator on lensed ${C}_{\ell}^{TT}$, IRL estimator on the lensed ${C}_{\ell}^{TT}$ minus a power law template factor, and the NIRL estimator on the lensed ${C}_{\ell}^{TT}$. We perform the ideal IRL  reconstruction of $P_R(k)$ on the unlensed ${C}_{\ell}^{TT}$ at 200 and 500 RL iterations. This is depicted in figures \ref{fig:pr4_feattestwv_reslt200_1} and \ref{fig:pr4_feattestwv_reslt500_1}. We can see that in the ideal case, it takes a larger number of iterations at 500 to converge properly to the input wavepacket based $P_R(k)$. This suggests that for a higher level of complexity in the free-form $P_R(k)$, for a given number of degree of freedoms, the number of iterations required to converge are higher. This is in the situation where the error weighting term in the NIRL and IRL algorithm is absent, such as in equation \ref{eq:IRL_eqns_nonlin_notanerr}.
We will look at the consequences of a large number of iterations and error weighting in the next section. We move on to $P_R(k)$ reconstruction using the 'Naive' IRL estimator on the lensed ${C}_{\ell}^{TT}$ data for 200 and 500 iterations and obtain the plots in figures \ref{fig:pr4_feattestwv_reslt200_2} and \ref{fig:pr4_feattestwv_reslt500_2}. We can clearly observe the fitting of the unaccounted for lensing power and the increasing iterations only accentuates the reconstructed spurious deviations. Next we perform the 'Template' based approach where we use the IRL algorithm on the lensed ${C}_{\ell}^{TT}$ but with a theoretical power law generated lensing template subtracted from the data. As we can see in figures \ref{fig:pr4_feattestwv_reslt200_3} and \ref{fig:pr4_feattestwv_reslt500_3}, the increasing number of iterations improves the fit somewhat and reduces the egregious levels of spurious features, but eventually converges to a sub-optimal fit and misses some fraction of the power in the original wavepacket $P_R(k)$. When we finally apply the NIRL estimator on the ${C}_{\ell}^{TT}$ data directly, we obtain the results in figures \ref{fig:pr4_feattestwv_reslt200_4} and \ref{fig:pr4_feattestwv_reslt500_4}, and observe that at 500 iterations, the reconstruction is better than the others and approaches the ideal one in figure \ref{fig:pr4_feattestwv_reslt500_1}. We further plot the relative percentage difference between the three reconstruction techniques with respect to the ideal reconstruction from the unlensed ${C}_{\ell}^{TT}$ in figures \ref{fig:pr4_feattestwv_reslt_relerr200} and \ref{fig:pr4_feattestwv_reslt_relerr500}. We can see that the NIRL estimator (blue line) has distinct improvements over the template (orange dashed line) and the naive (red dotted line) reconstructions, ranging from 2\% to 10\% reduction in the relative error. We also should mention that in the NIRL estimator correction given in equations \ref{eqn:numer_lenscorr}, the $P_{k_1}^{(i)}$  term that feeds back the iterative update into the kernel, is sparsely downsampled at high $k$ regions, via interpolation, from the $P_{k}^{(i)}$ in equation \ref{eq:IRL_eqns_nonlin_notanerr}. This is to reduce the computation time of the iterative correction. However if we observe the figure \ref{fig:pr4_feattestwv_reslt_relerr500}, we can clearly see that the NIRL falls short of increasing accuracy with respect to iterations, especially at the high $k$ regions. It may be a worthwhile exercise to perform the same analysis under the same $k$ sampling for the iterative corrector to check if the high $k$ region improves in the NIRL reconstruction. Due to correlation across the $k$ grid, a behaviour observed earlier in full covariance matrix analysis of RL reconstructions \cite{Chandra:2021ydm}, improvement at high $k$ may also affect reconstruction in other $k$ regions as well. However this is beyond the scope of this paper and we defer it for future studies, especially when working with actual observation data. The summary of the test results are given in table \ref{table:pr4_feattestwv_reslt_summ}

\begin{table}
{\begin{center}
\begin{tabular}{ l r r l }
  \toprule
  Data & \multicolumn{2}{c}{Algorithm}{Result} \\
  \midrule
  Unlensed ${C}_{\ell}^{TT}\textbf{(Wave Feature)}$ & IRL & Exact \\
  Lensed $\widetilde{C}_{\ell}^{TT}\textbf{(Wave Feature)}$ & IRL & Poor \\
  Lensed $\widetilde{C}_{\ell}^{TT}\textbf{(Wave Feature)}-\Delta C_{\ell}^{TT}\textbf{(Power Law)}$ & IRL & Average \\
  Lensed $\widetilde{C}_{\ell}^{TT}\textbf{(Wave Feature)}$ & NIRL & Good \\
  \bottomrule
\end{tabular}
\end{center}}
\caption{This table lists the last 4 Reconstruction tests from \ref{table:feat_testswv} with their reconstruction quality.}
\label{table:pr4_feattestwv_reslt_summ}
\end{table}

\subsection{Results iv): Bump Feature Reconstruction with Delensing and Realisation Noise}
\label{sub_sec:feat_NIRL_delen_rlznnoise}

\begin{figure}
    \centering 
\begin{subfigure}{1.0\textwidth}
  \includegraphics[width=\linewidth]{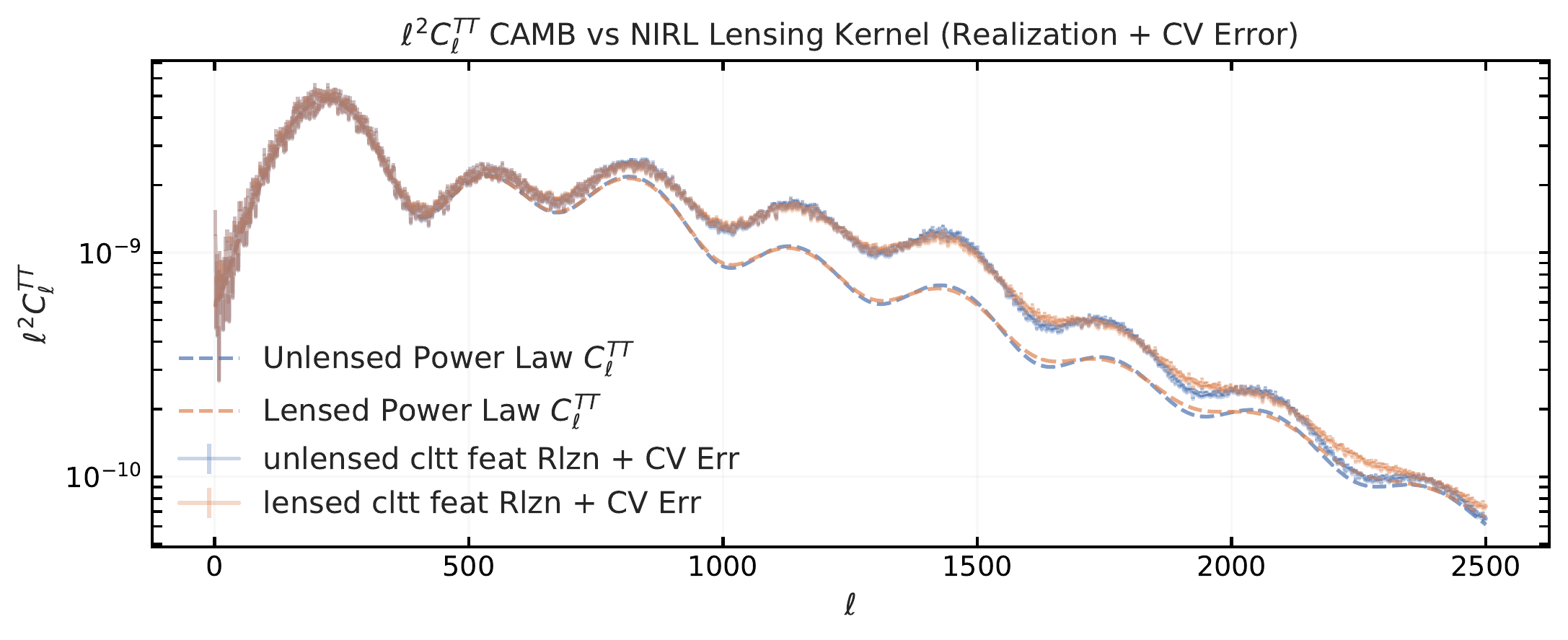}
  \caption{}
\end{subfigure}
\caption{ This plot shows the lensed and unlensed $C_{\ell}^{TT}$ using the Gaussian bump feature, in solid lines, generated as a realization based on cosmic variance error bars. It also shows the power law version, without any error bars or realization noise, in dashes lines.}
\label{fig:pr4_bumpfeat_rlzn}
\end{figure}

In this section we carry out one more proof/validation of concept test for the NIRL modified deconvolution algorithm. Referring to section \ref{sub_sec:feat_NIRL_delen}, we want to observe the effects of realization/observation noise in data and how it affects the $P_R(k)$ reconstruction. Since it is fact that the RL algorithm tends to fit noise in data as spurious features, especially for high number of RL iterations, it needs to studied whether NIRL reconstruction of $P_R(k)$, which requires a high number of iterations (as observed in the previous section on wavepacket reconstruction), are capable of production a $P_R(k)$ that can be processed to separate between noise induced features versus actual injected features present in the $P_R(k)$.
To this end, we reproduce the data used in \ref{fig:pr4_feattest_4}, but this time add a cosmic variance limited error bar and a Gaussian random realisation of the data using those error bars as variance and the simulated data as mean, and we plot this in figure \ref{fig:pr4_bumpfeat_rlzn}. The solid line with the error bars shows the bump feature induced power spectrum and the contribution from the realisation noise generated by cosmic variance limited error bars.

We again lay out the steps to be performed in order to quantify the results of the NIRL versus the template-based delensing approach. The key tests are listed in table \ref{table:feat_testsbmprlzn}.

\begin{table}
{\begin{center}
\begin{tabular}{ l r r l }
  \toprule
  Bump Feature (Realisation) Test Steps  \\
  \midrule
  1. IRL Reconstruction from Unlensed CAMB $C_{\ell}^{TT} \textbf{(Bump Feature Realisation)}$ \\
  2. IRL Reconstruction from Lensed CAMB $\widetilde{C}_{\ell}^{TT} \textbf{(Bump Feature Realisation)}$ \\
  3. IRL Reconstruction from $\widetilde{C}_{\ell}^{TT}\textbf{(Bump Feature Realisation)} - \Delta C_{\ell}^{TT}\textbf{(Power Law)}$ \\
  4. NIRL Reconstruction from Lensed CAMB $\widetilde{C}_{\ell}^{TT} \textbf{(Bump Feature Realisation)}$ \\
  \bottomrule
\end{tabular}
\end{center}}
\caption{This table lists the tests undertaken to validate the NIRL estimator and its convergence based on $P_R(k)$ incorporating a bump like feature with realisation noise and CV error bars.}
\label{table:feat_testsbmprlzn}
\end{table}

\begin{figure}
    \centering 
\begin{subfigure}{0.47\textwidth}
  \includegraphics[width=\linewidth]{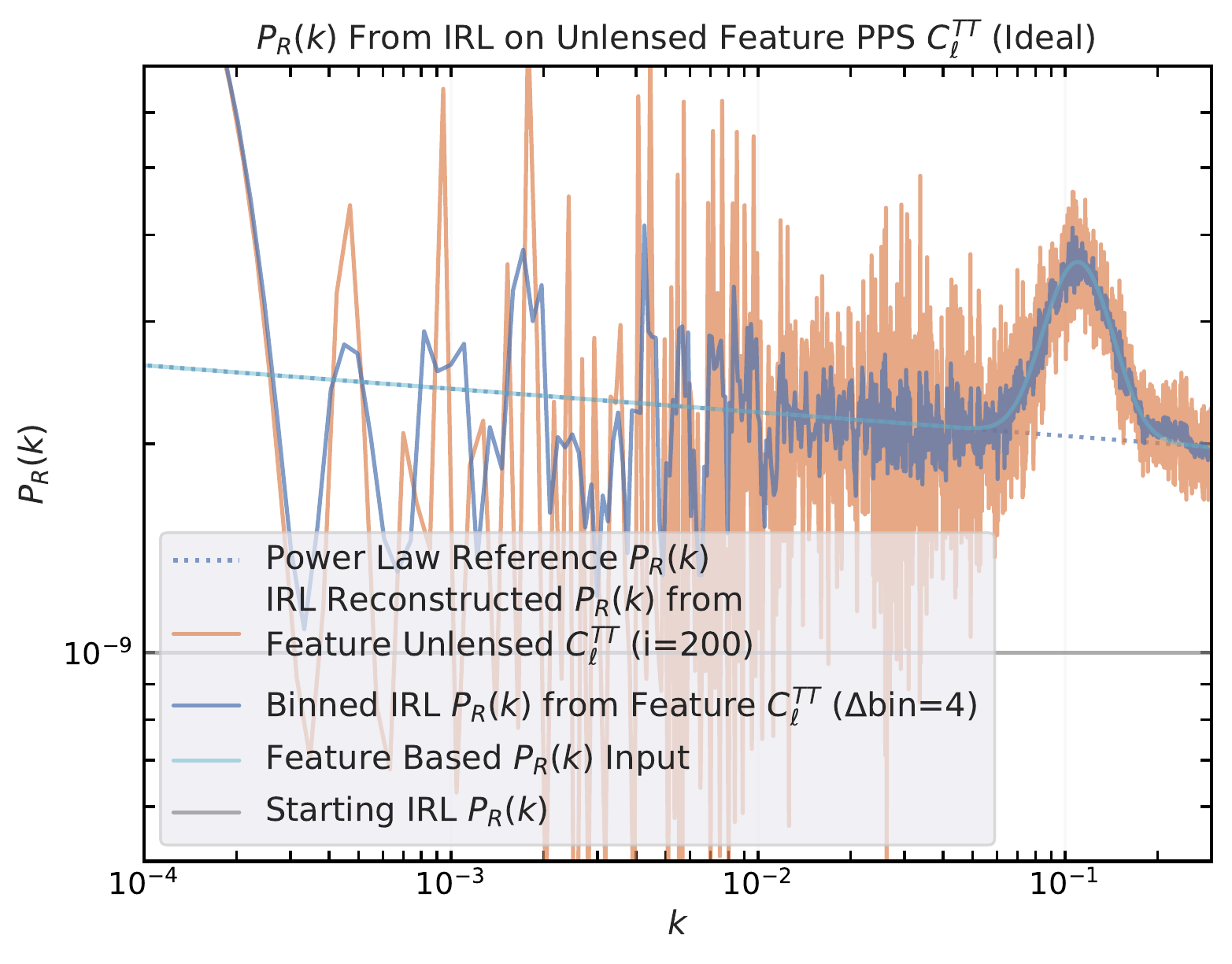}
  \caption{}
  \label{fig:pr4_feattestrlzn_reslt200_1}
\end{subfigure}\hfil 
\begin{subfigure}{0.47\textwidth}
  \includegraphics[width=\linewidth]{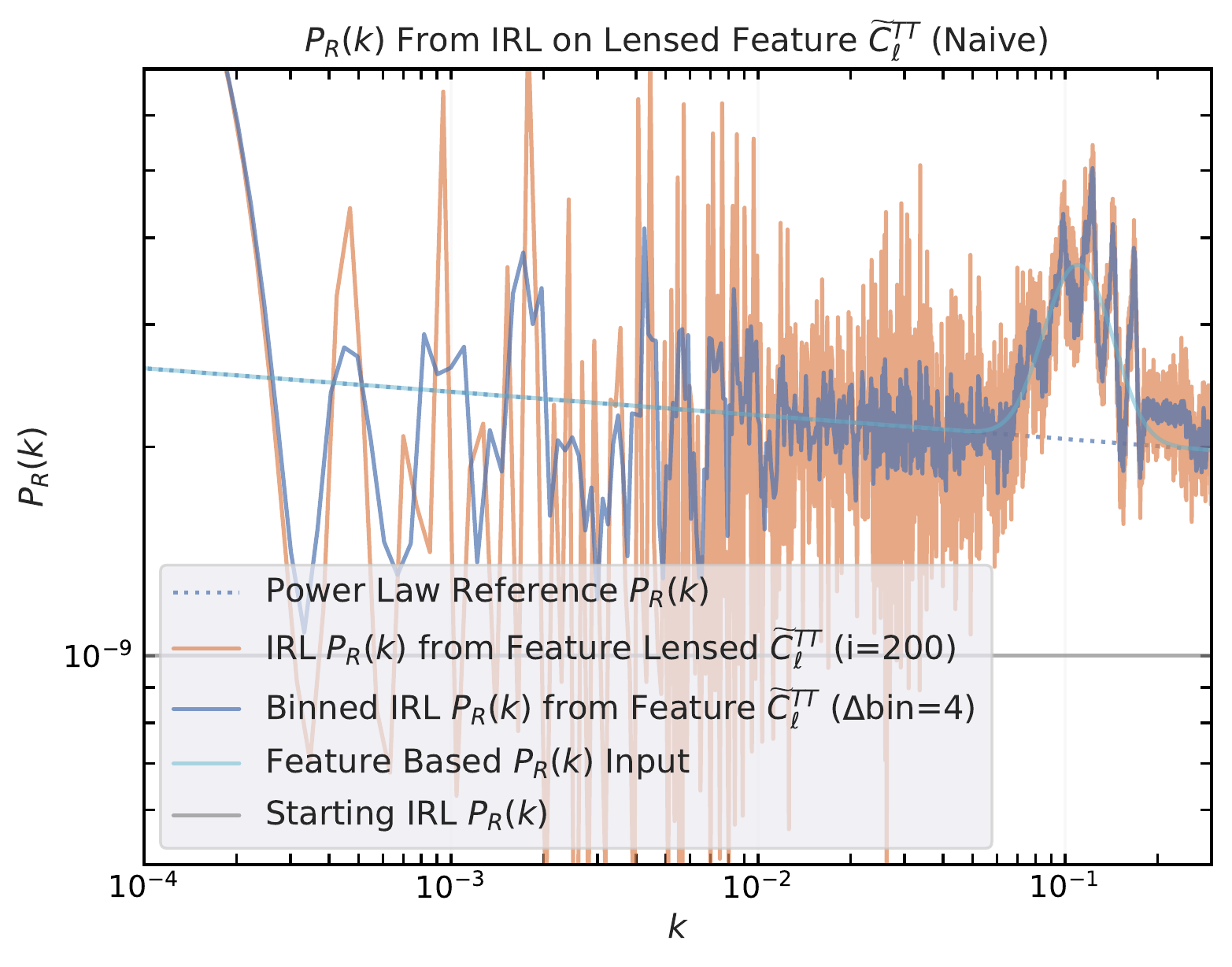}
  \caption{}
  \label{fig:pr4_feattestrlzn_reslt200_2}
\end{subfigure}

\begin{subfigure}{0.47\textwidth}
  \includegraphics[width=\linewidth]{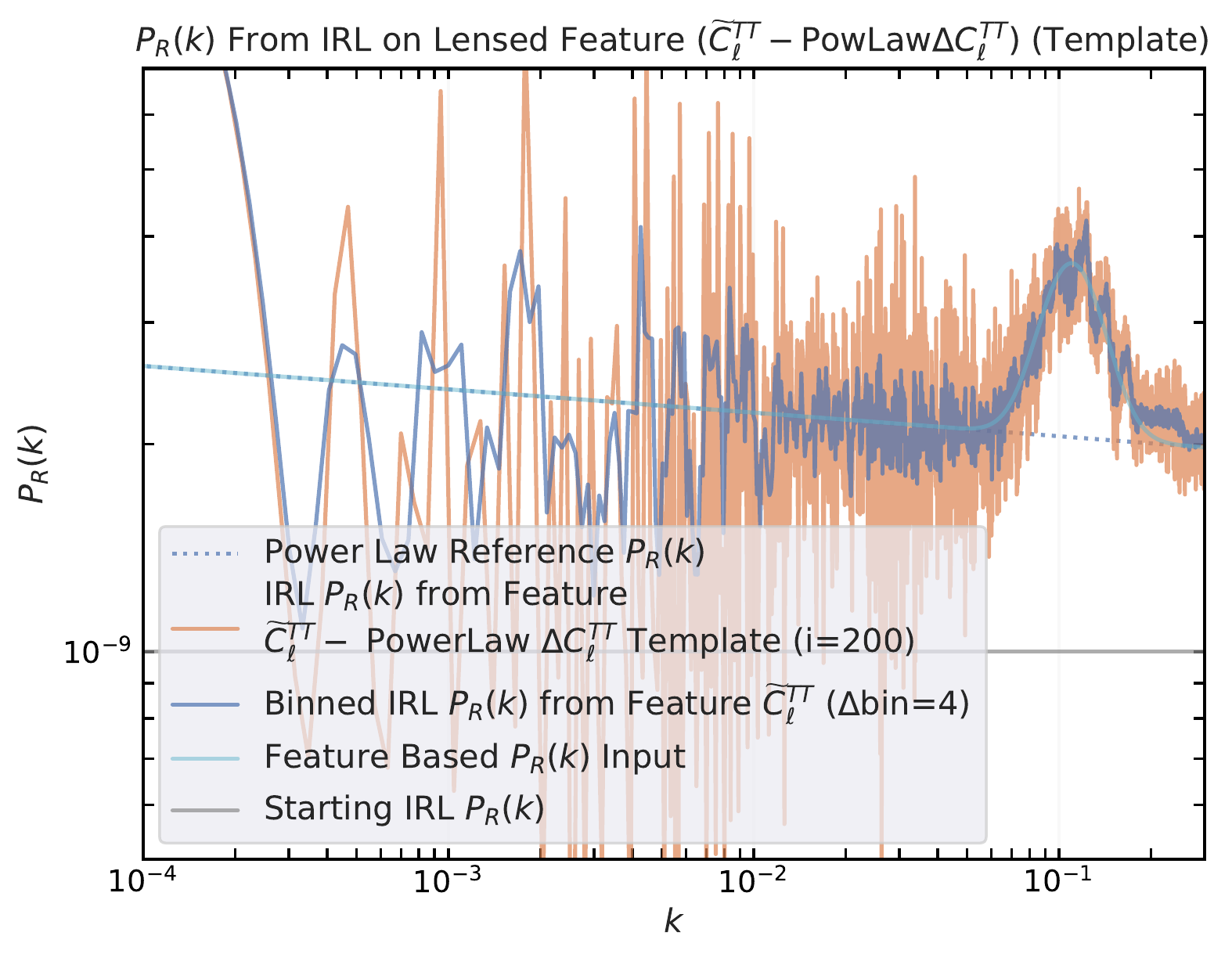}
  \caption{}
  \label{fig:pr4_feattestrlzn_reslt200_3}
\end{subfigure}\hfil 
\begin{subfigure}{0.47\textwidth}
  \includegraphics[width=\linewidth]{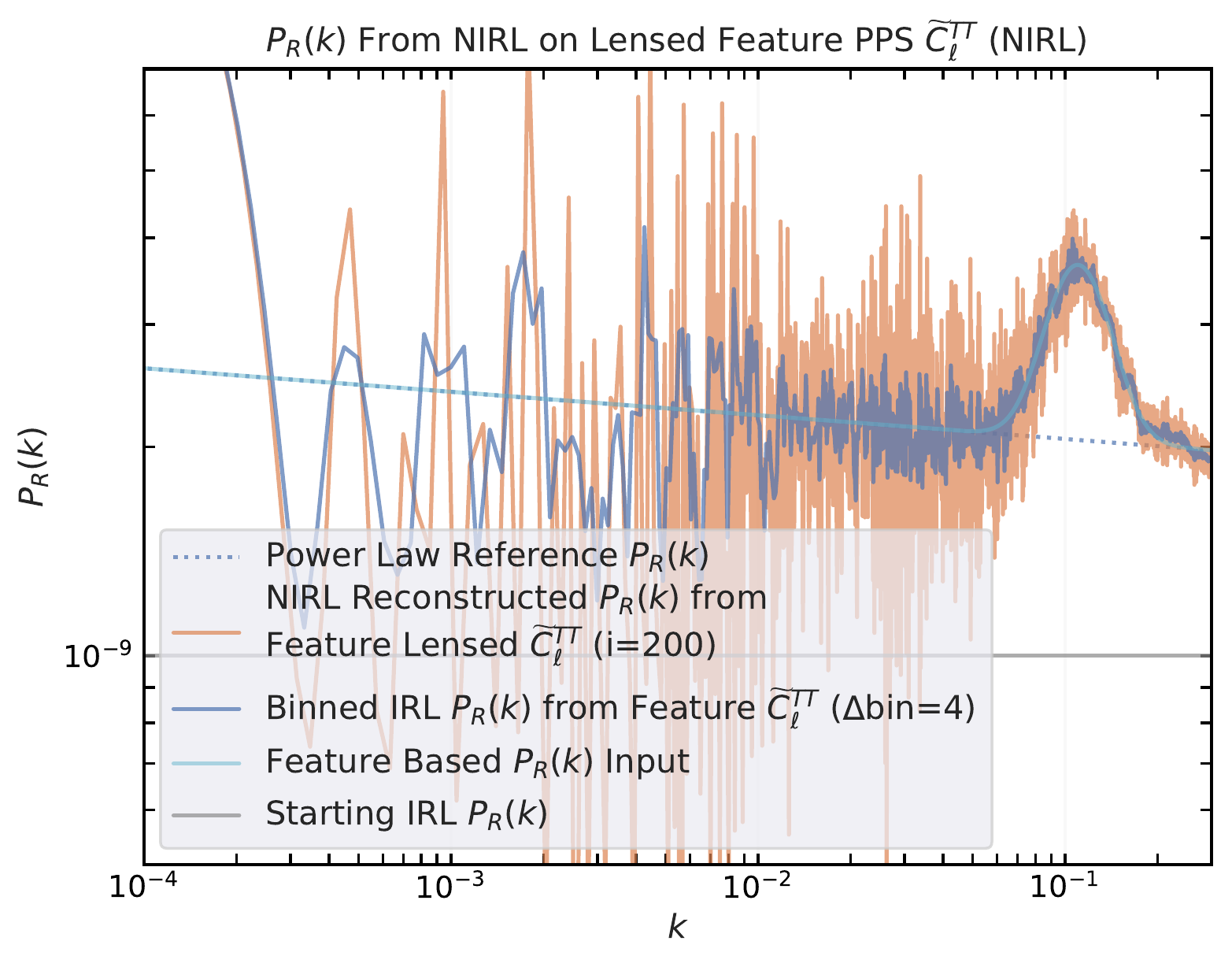}
  \caption{}
  \label{fig:pr4_feattestrlzn_reslt200_4}
\end{subfigure}
\caption{ The 4 figures show the reconstructed $P_R(k)$ from the feature based data, using 200 RL reconstruction iterations. The reference power law is given in cyan dotted lines. The reconstructed $P_R(k)$ are given in orange lines, with the binned reconstruction in blue lines. The input feature $P_R(k)$ is given in cyan solid lines.}
\label{fig:pr4_feattestrlzn_reslt200}
\end{figure}

\begin{figure}
    \centering 
\begin{subfigure}{0.47\textwidth}
  \includegraphics[width=\linewidth]{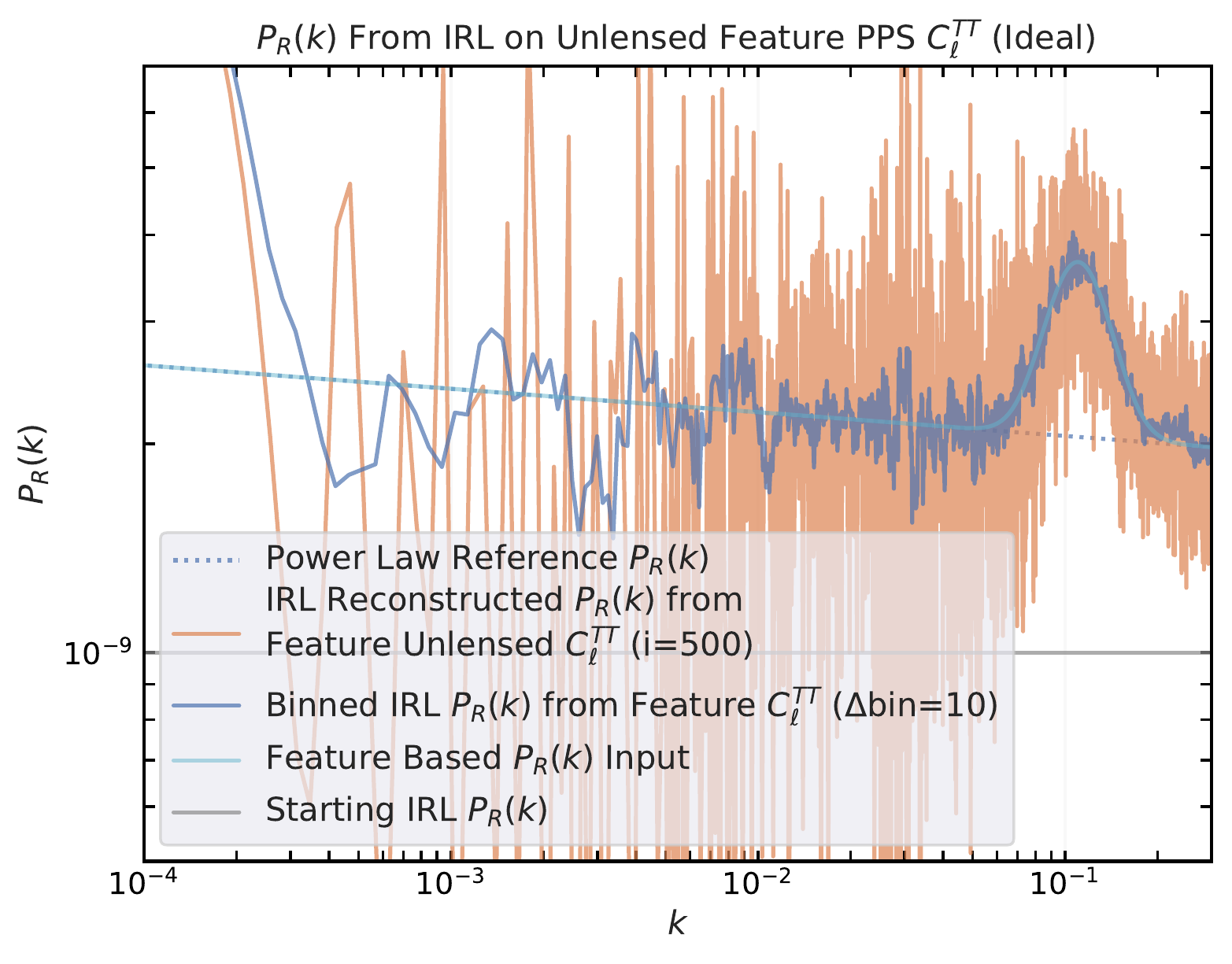}
  \caption{}
  \label{fig:pr4_feattestrlzn_reslt500_1}
\end{subfigure}\hfil 
\begin{subfigure}{0.47\textwidth}
  \includegraphics[width=\linewidth]{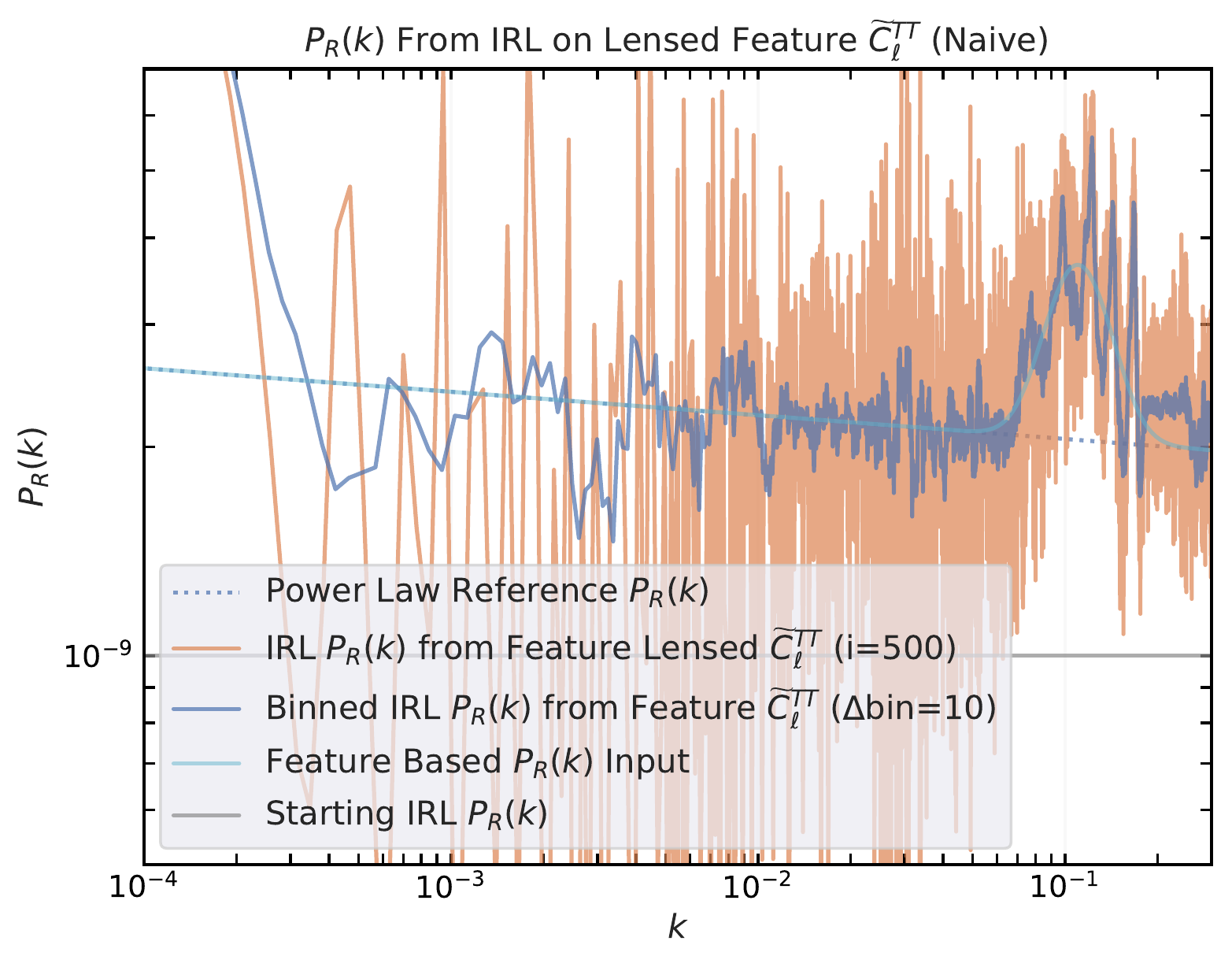}
  \caption{}
  \label{fig:pr4_feattestrlzn_reslt500_2}
\end{subfigure}

\begin{subfigure}{0.47\textwidth}
  \includegraphics[width=\linewidth]{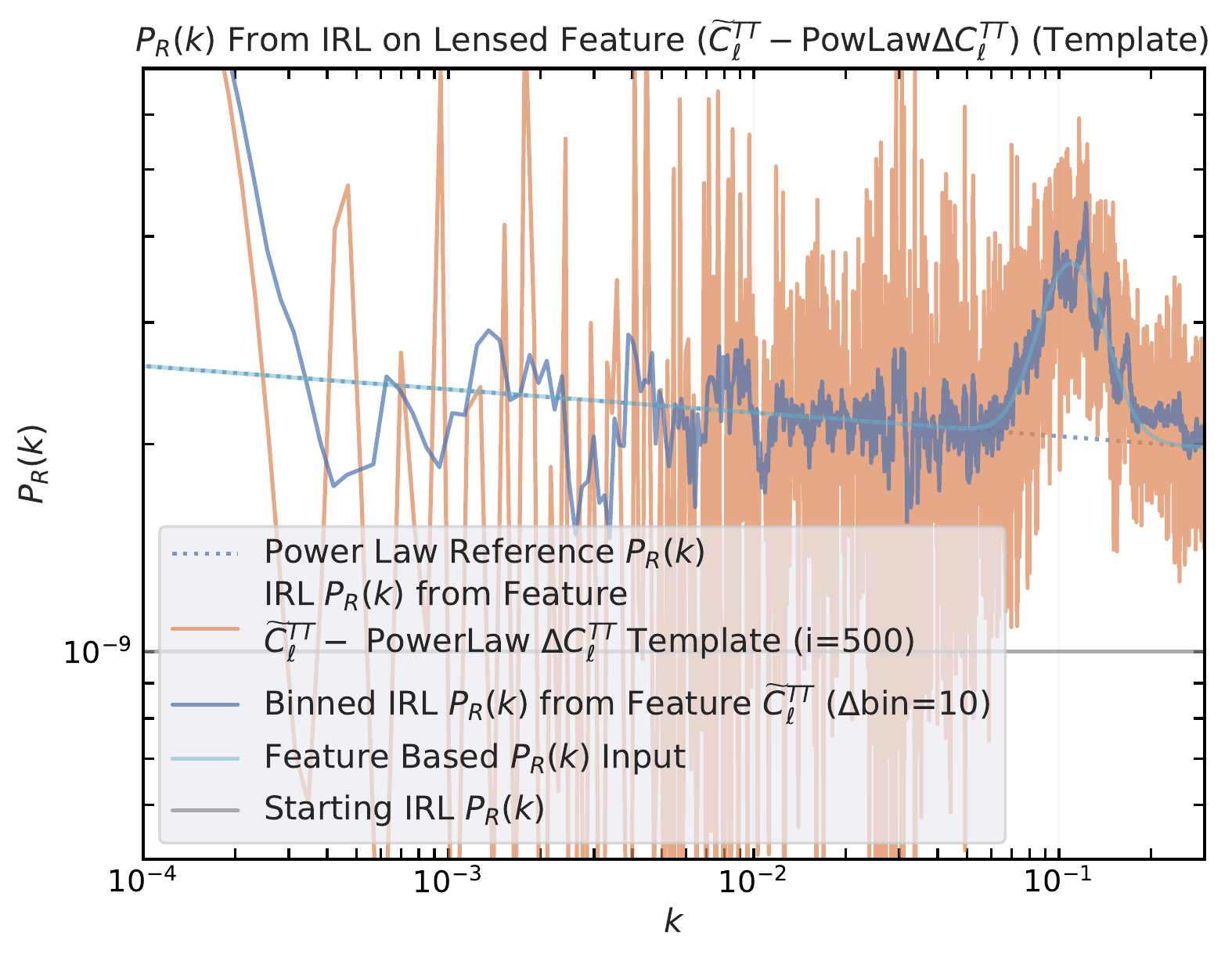}
  \caption{}
  \label{fig:pr4_feattestrlzn_reslt500_3}
\end{subfigure}\hfil 
\begin{subfigure}{0.47\textwidth}
  \includegraphics[width=\linewidth]{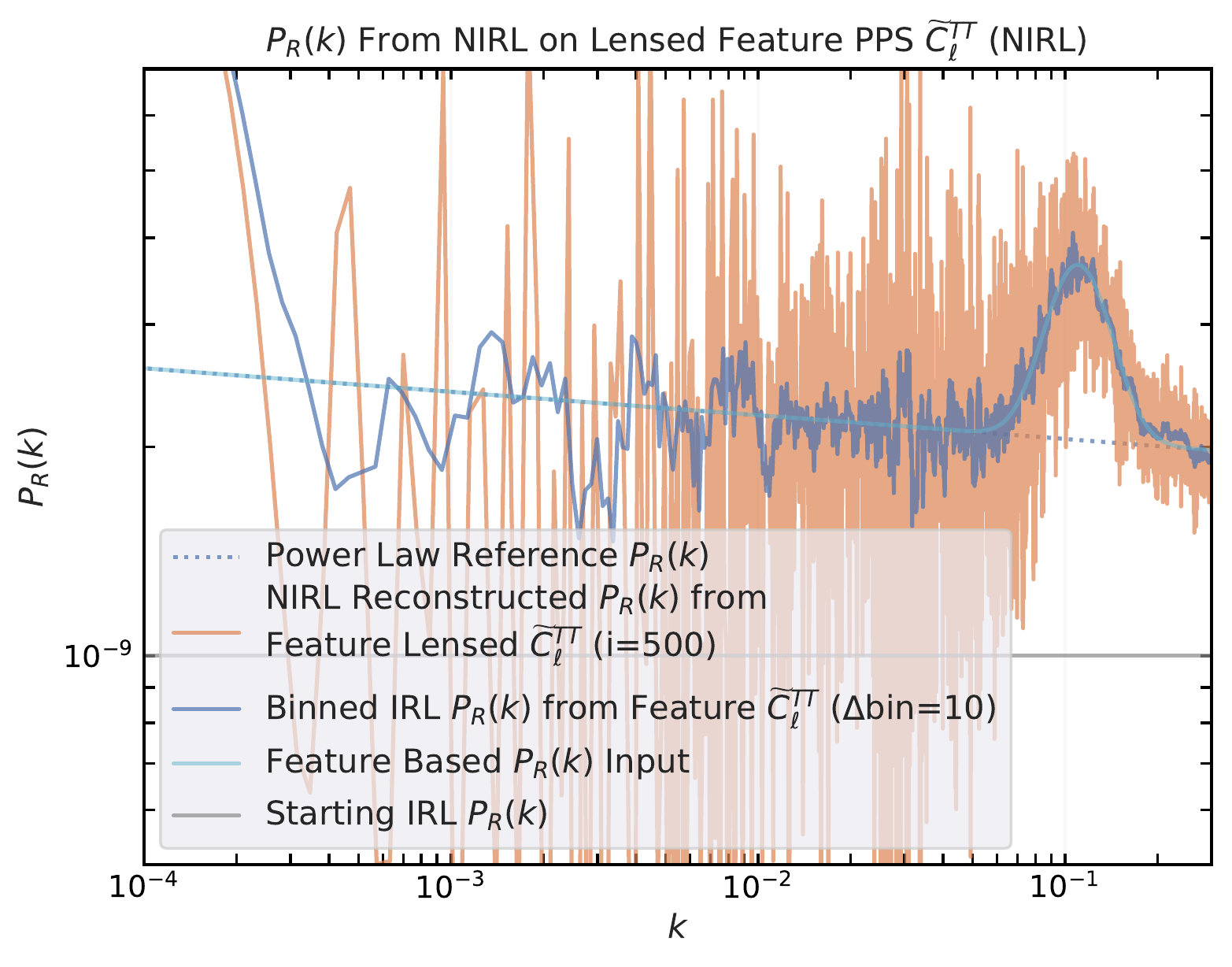}
  \caption{}
  \label{fig:pr4_feattestrlzn_reslt500_4}
\end{subfigure}
\caption{ The 4 figures show the reconstructed $P_R(k)$ from the feature based data, using 500 RL reconstruction iterations. The reference power law is given in cyan dotted lines. The reconstructed $P_R(k)$ are given in orange lines, with the binned reconstruction in blue lines. The input feature $P_R(k)$ is given in cyan solid lines.}
\label{fig:pr4_feattestrlzn_reslt500}
\end{figure}

\begin{figure}
\begin{subfigure}{1\textwidth}
\includegraphics[width=1.00\linewidth]{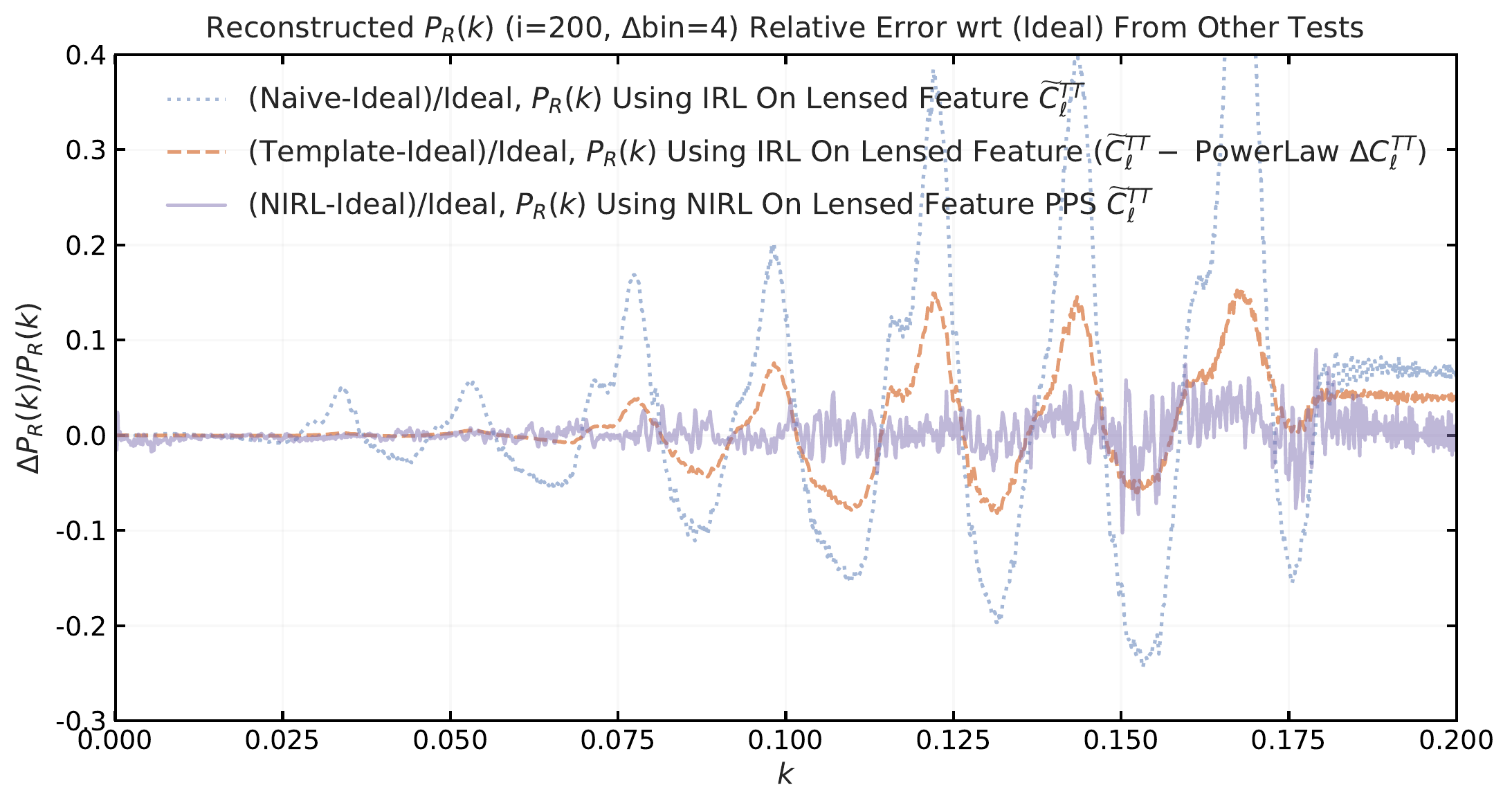}
\end{subfigure}
\caption{This figure shows the relative error of the 3 reconstructions of the $P_R(k)$ displayed in figure \ref{fig:pr4_feattestwv_reslt200_2},\ref{fig:pr4_feattestwv_reslt200_3}, and \ref{fig:pr4_feattestwv_reslt200_4} respectively, relative to the first reconstruction \ref{fig:pr4_feattestwv_reslt200_1} which is held as the ideal reference reconstruction. These are based on 200 iterations of the RL algorithm.}
\label{fig:pr4_feattestrlzn_reslt_relerr200}
\end{figure}

\begin{figure}
\begin{subfigure}{1\textwidth}
\includegraphics[width=1.00\linewidth]{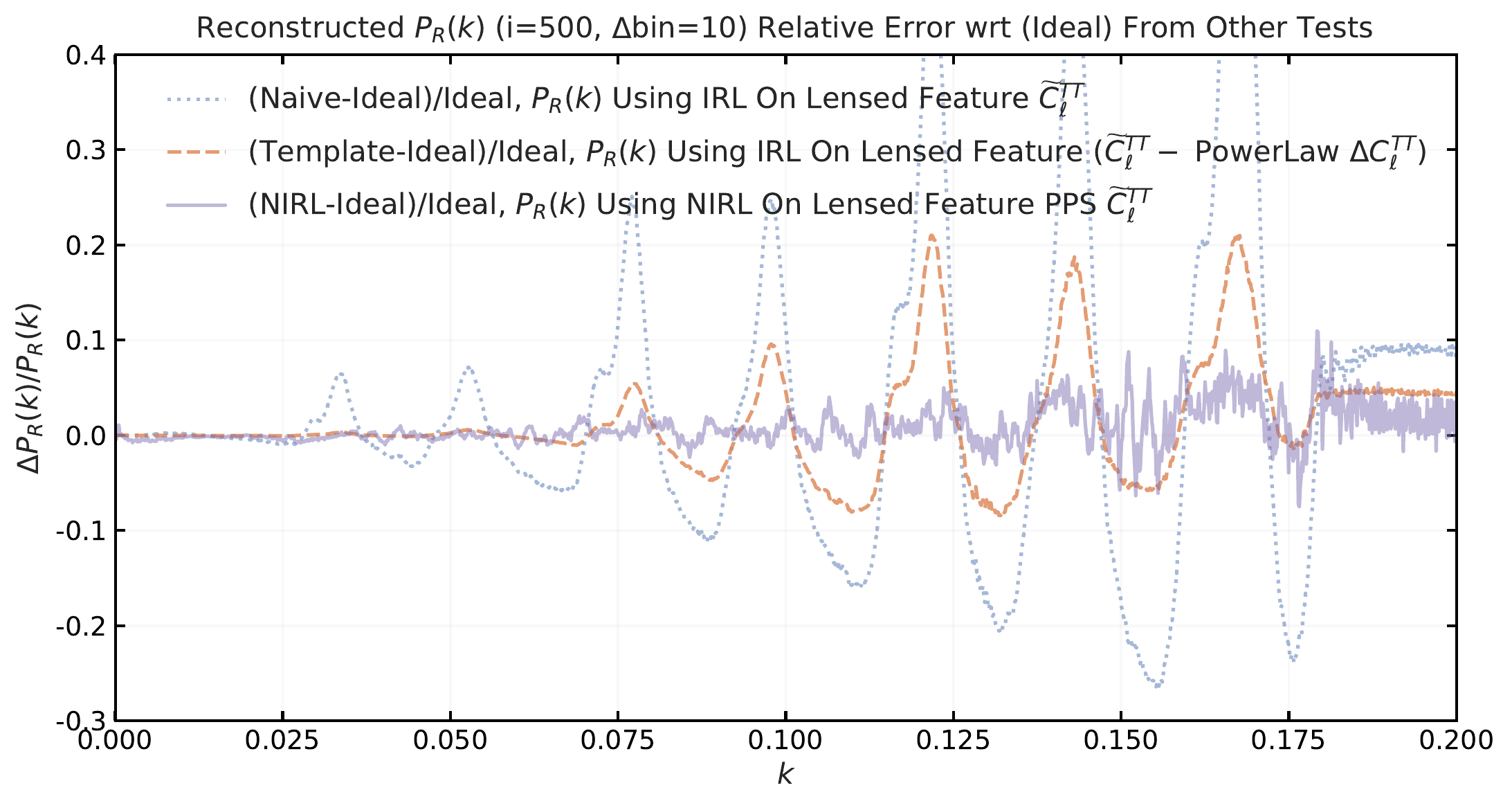}
\end{subfigure}
\caption{This figure shows the relative error of the 3 reconstructions of the $P_R(k)$ displayed in figure \ref{fig:pr4_feattestwv_reslt500_2},\ref{fig:pr4_feattestwv_reslt500_3}, and \ref{fig:pr4_feattestwv_reslt500_4} respectively, relative to the first reconstruction \ref{fig:pr4_feattestwv_reslt500_1} which is held as the ideal reference reconstruction. These are based on 500 iterations of the RL algorithm.}
\label{fig:pr4_feattestrlzn_reslt_relerr500}
\end{figure}

In the Richardson-Lucy deconvolution application we now use the error weighing term expressed in equation \ref{eq:IRL_eqns}, as well as add it to equation \ref{eq:IRL_eqns_nonlin_notanerr} and use the CV error bars for analysis. We carry out the previous set of reconstructions, Ideal, Naive, Template and NIRL, for two iteration sets of 200 and 500, as before. Post reconstruction, we also apply a simple binning function to the reconstructed $P_R(k)$ for cleaning the realisation induced noise. The results are given in figures  \ref{fig:pr4_feattestrlzn_reslt200} and \ref{fig:pr4_feattestrlzn_reslt500}. 
For the first RL reconstruction using the unlensed $C_{\ell}^{TT}$ data in figures \ref{fig:pr4_feattestrlzn_reslt200_1} and \ref{fig:pr4_feattestrlzn_reslt500_1}, we obtain the raw and binned reconstruction in orange and blue respectively, and note that the binning significantly cleans up the realisation induced 'features', and for higher iterations, a larger binning width produces equivalent results. The same is observed for the standard RL reconstruction applied to the lensed $\widetilde{C}_{\ell}^{TT}$ data in \ref{fig:pr4_feattestrlzn_reslt200_2} and \ref{fig:pr4_feattestrlzn_reslt500_2}, where the binning is successful at removing the realisation features and leaves the distinct lensing induced spikes in the reconstruction. For the third case where we subtract a power law based theoretical template from the lensed $\widetilde{C}_{\ell}^{TT}$ data prior to application of the standard RL algorithm, binning again is successful at reproducing the residual features left behind post template cleaning, as evidenced in figures \ref{fig:pr4_feattestrlzn_reslt200_3} and \ref{fig:pr4_feattestrlzn_reslt500_1}. For the NIRL reconstruction, both iteration sets and binning produce results equivalent to the Ideal case, shown in figures \ref{fig:pr4_feattestrlzn_reslt200_4} and \ref{fig:pr4_feattestrlzn_reslt500_4}.
We plot the corresponding relative error plots in figures \ref{fig:pr4_feattestrlzn_reslt_relerr200} and \ref{fig:pr4_feattestrlzn_reslt_relerr500}, where we show the relative error for the Naive, Template and NIRL reconstruction results relative to the Ideal reconstruction, with the corresponding binnings used in the previous results. As before, the NIRL reconstruction approaches the Ideal reconstruction closest, followed by the Template and then the Naive reconstructions of the $P_R(k)$. The summary of the results are given in table \ref{table:pr4_feattestrlzn_reslt_summ}. We point out that the capacity for the binning to differentiate between realisation noise and $P_R(k)$ features may be due to the lensing and feature origin contribution to the $\widetilde{C}_{\ell}^{TT}$ being more broadly distributed across the $L$ multipoles, as evidenced in figures \ref{fig:pr4_feattest_6} and \ref{fig:pr4_feattest_4}. The fluctuations induced due to the realisation noise exist on a different 'frequency' on the $P_R(k)$ and hence binning can selectively smooth those out without affecting the reconstruction features. A combination of smoothing and pre-analysis data binning techniques may be required for a comprehensive analysis that mitigates noise induced fluctuations from physical or other non random effect derived features. This would also be informed by the physical models being used, which can provide insight into the kinds of physical $P_R(k)$ features expected and the kind of smoothing/binning techniques that may be used to complement that.

\begin{table}
{\begin{center}
\begin{tabular}{ l r r l }
  \toprule
  Data & \multicolumn{2}{c}{Algorithm}{Result} \\
  \midrule
  Unlensed ${C}_{\ell}^{TT}\textbf{(Bump Feature Realisation)}$ & IRL & Exact \\
  Lensed $\widetilde{C}_{\ell}^{TT}\textbf{(Bump Feature Realisation)}$ & IRL & Poor \\
  Lensed $\widetilde{C}_{\ell}^{TT}\textbf{(Bump Feature Realisation)}-\Delta C_{\ell}^{TT}\textbf{(Power Law)}$ & IRL & Average \\
  Lensed $\widetilde{C}_{\ell}^{TT}\textbf{(Bump Feature Realisation)}$ & NIRL & Good \\
  \bottomrule
\end{tabular}
\end{center}}
\caption{This table lists the last 4 Reconstruction tests from \ref{table:feat_testswv} with their reconstruction quality.}
\label{table:pr4_feattestrlzn_reslt_summ}
\end{table}

\section{Discussion}
\label{sec:discuss}

In this paper we propose and demonstrate an algorithmic modification to the Richardson-Lucy deconvolution algorithm to address any situation of non-linear convolution based data and reconstruction of the convolved signal, where the non linear convolution process and kernel are known. We assume the system follows the underlying properties of (RL) utilization, such as being positive definite everywhere. We term this algorithm the Non-Linear Iterative Richardson-Lucy (NIRL) estimator

The NIRL algorithm consists of an iterative kernel update mechanism, where the convolution kernel is updated at each iteration to include the contribution from the secondary contribution from the previous iteration fit to the signal. This algorithm is implemented in the case of weak lensing of the CMB and we perform the kernel update using a precomputed lensing kernel shape factor term. The method is also compared to the existing template approach already utilized to deal with lensing, in order to justify the computation cost with increase in reconstruction precision.

We choose $\Lambda$CDM cosmology with Planck best-fit parameter inferences and simulated ideal lensed power spectra (using CAMB) to validate the precision of the lensing kernel shape factor and conclude it is within the cosmic variance limit. We note the possibility to improve it further by increasing the $L$ range in order to approach CAMB constructed power spectra accuracy.

We perform a null test whereby we utilize a power law $P_R(k)$ input to simulate lensed $C_{\ell}^{TT}$ spectra and compare the $P_R(k)$ reconstruction to the ideal delensed spectra reconstruction, template based delensed spectra deconstruction and the lensed spectra. We establish that the NIRL reconstruction approaches maximum reconstruction quality and is comparable to the template based delensing method, and the ideal delensed spectra, owing to all three being power laws based. We this establish that within the assumptions utilized earlier NIRL estimator converges to a stable reconstructed solution.

We perform a feature based test whereby we utilize a feature based $P_R(k)$ input to simulate the unlensed and lensed $C_{\ell}^{TT}$ spectra and compare the $P_R(k)$ reconstruction to ideal delensed spectra reconstruction, power law template based delensed spectra reconstruction and the lensed spectra. We establish that the NIRL reconstruction outperforms other methods and is comparable to the ideal delensed data derived spectra. Thus we establish that within the assumptions utilized earlier, the NIRL estimator is successful at improving upon the power law template based approach and is the optimal free-form reconstruction algorithm. We also apply this to features that are softer and produce deviations close to or below the cosmic variance limit and observe it still produces better accuracy as it is accounting for a distinct lensing systematic. We also observe that binning is capable of parsing between data/realisation induced noise and 'features' vs features produced from physical or other non-random effects. These can be augmented with other smoothing techniques.

We expect to further optimize the algorithm by implementing the sparsity algorithm developed in our previous work \cite{Chandra:2021ydm}, as well as carry out a full covariance matrix calculation for the reconstructed $P_R(k)$ as well as test existing $C_{\ell}^{TT}$ data for features of interest given the newfound accuracy of the NIRL estimator and it's ability to discriminate finely from features vs lensing artefacts without any prior assumptions.

One disadvantage of the NIRL method is that the lensing kernel matrix is a large matrix (~50GB) and needs to be read and the iterative corrections to the kernel computed during each iteration. This increases time per RL iteration from under 1 minute to upto 4 minutes, or the equivalent scaling depending on the CPU used. The time increase can be mitigated by storing the kernel matrix in RAM, as the read in time of the matrix from the hard drive, is the i time constraint during the iteration, but this would necessitate using large amounts of RAM dedicated to a single CPU. It may also be improved by using a solid state disk, but this has not been verified.

\section{Acknowledgements}
\label{sec:acknow}

This work has been carried out with the support of the Council for Science and Industrial Research (CSIR) and University Grants Commission (UGC) graduate funding at Inter-University Centre for Astronomy and Astrophysics (IUCAA), Pune for RSC. We would like to acknowledge Shabbir Shaikh for useful discussions. The work acknowledges the use of the IUCAA High Performance Computing facility. We acknowledge the use of Matplotlib and NumPy libraries.


\begin{appendices}

\section{Kernel Lensing Simulation Accuracy}

\begin{figure}
    \centering 
\begin{subfigure}{1.0\textwidth}
  \includegraphics[width=\linewidth]{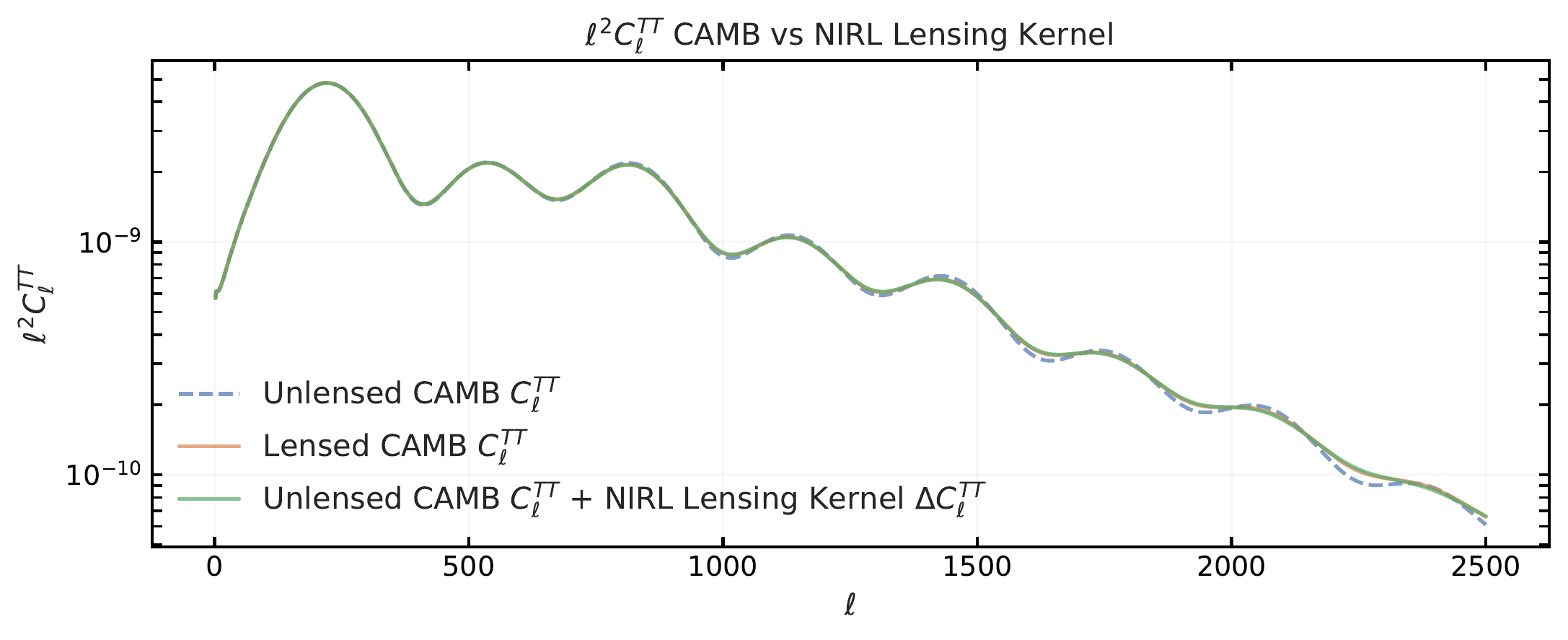}
  \caption{}
  \label{fig:pr4_kerneltest_1}
\end{subfigure}

\begin{subfigure}{0.47\textwidth}
  \includegraphics[width=\linewidth]{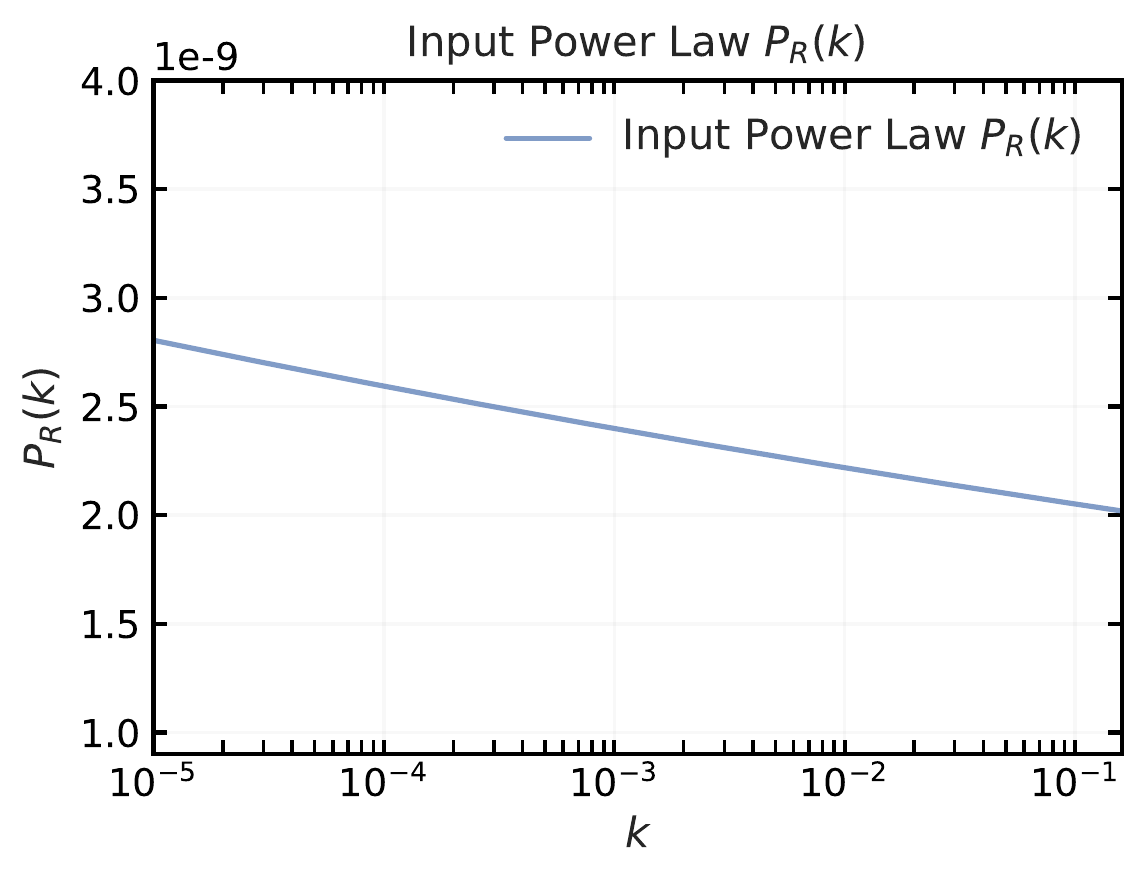}
  \caption{}
  \label{fig:pr4_kerneltest_2}
\end{subfigure}\hfil 
\begin{subfigure}{0.47\textwidth}
  \includegraphics[width=\linewidth]{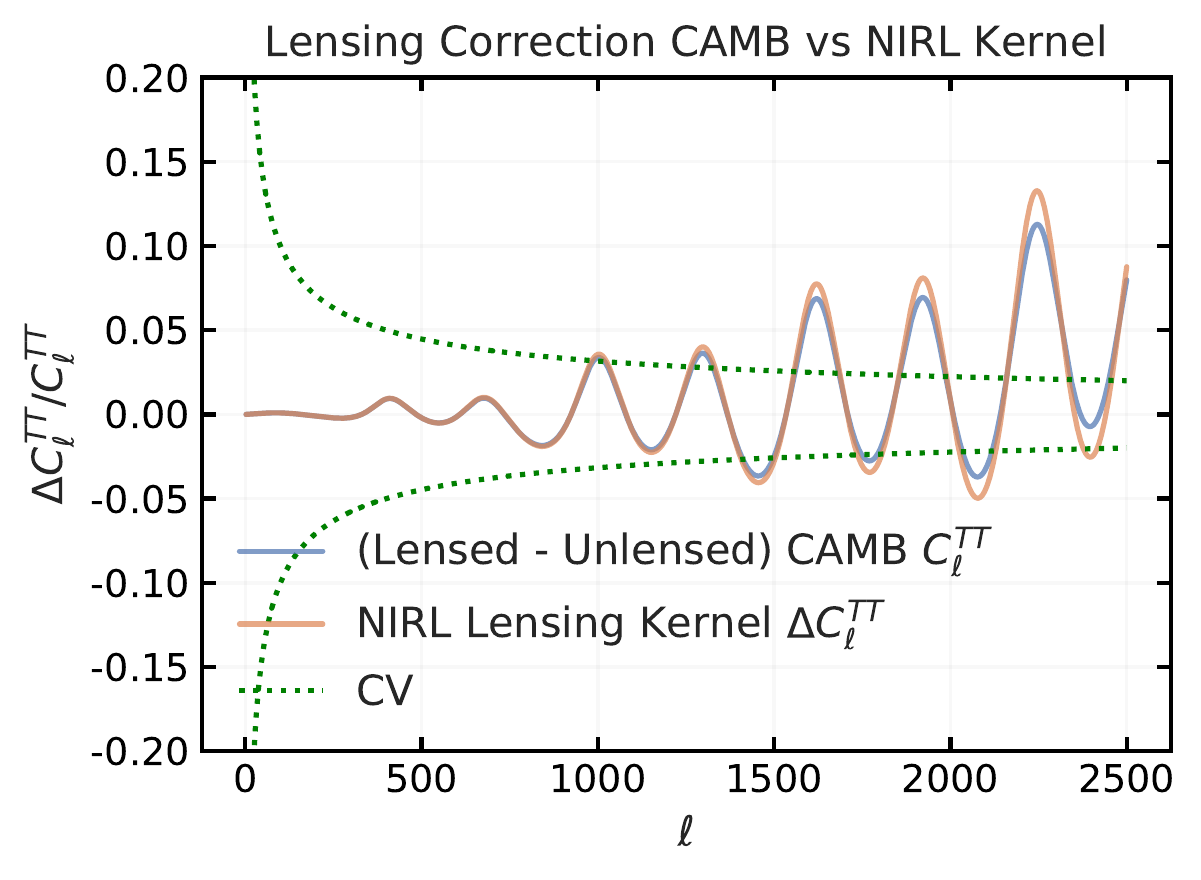}
  \caption{}
  \label{fig:pr4_kerneltest_3}
\end{subfigure}
\caption{Figure \ref{fig:pr4_kerneltest_1} shows the unlensed CAMB ${C}_{\ell}^{TT}$ in the dashed blue line. The lensed $\widetilde{C}_{\ell}^{TT}$ from CAMB and from NIRL Lensing Kernel are shown in solid orange and green lines respectively. Figure \ref{fig:pr4_kerneltest_2} shows the input power law $P_R(k)$ in blue and figure \ref{fig:pr4_kerneltest_3} shows the weak lensing contribution $\Delta C_{\ell}^{TT}$ from CAMB and NIRL Kernel in blue and orange respectively. Cosmic variance bounds are shown in green dashed lines. }
\label{fig:pr4_kerneltest}
\end{figure}

The plot for step 1. in table \ref{table:pwlw_tests} is given in figure \ref{fig:pr4_kerneltest}. The relative difference between the lensed and unlensed ${C}_{\ell}^{TT}$ from both CAMB and NIRL Kernel is given in the third plot, which clearly inspects our Kernel precision. It can be seen that at higher $\ell$, our NIRL Kernel underestimates the lensing contribution. The reason for this is mentioned in the consideration in table \ref{table:wklns_lrange}. To get an increased precision to the CAMB ${C}_{\ell}^{TT}$, we need to experiment with the extra ${\ell}$ range in the Kernel evaluation. However for the current work, the precision difference is much below cosmic variance and is relative to the CAMB model kernel used for the power spectra, hence it will not be extremely important when analysing the convergence and reconstruction capability of NIRL, which will use the lensed kernel generated data using our algorithm instead.

\end{appendices}

\end{document}